\documentclass[12pt]{article}
\usepackage{eqsection,subeqnarray,indent,amsfonts,amssymb}
\usepackage{amsmath}     
\usepackage{bm}          
\usepackage{epic,eepic}  
\usepackage{graphicx}

\def\today{July 16, 2004 \\[1mm] 
           Revised: January 18, 2005 \\[1mm]
           Final version: August 25, 2005}
\begin{document}

\bibliographystyle{plain}

\date{\today}

\title{ Transfer Matrices and Partition-Function Zeros \\
       for Antiferromagnetic Potts Models \\[5mm]
       \large\bf IV.~Chromatic polynomial with cyclic boundary conditions}

\author{
  \\
  {\small Jesper Lykke Jacobsen}                            \\[-0.2cm]
  {\small\it Laboratoire de Physique Th\'eorique et Mod\`eles Statistiques}
                                                            \\[-0.2cm]
  {\small\it Universit\'e Paris-Sud}                        \\[-0.2cm]
  {\small\it B\^atiment 100}                                \\[-0.2cm]
  {\small\it F-91405 Orsay, FRANCE }        \\[-0.2cm]
  {\small\tt JACOBSEN@IPNO.IN2P3.FR}                        \\[5mm]
  {\small Jes\'us Salas}                       \\[-0.2cm]
  {\small\it Grupo de Modelizaci\'on, Simulaci\'on Num\'erica y Matem\'atica
             Industrial}    \\[-0.2cm]
  {\small\it Universidad Carlos III de Madrid} \\[-0.2cm]
  {\small\it Avda.\  de la Universidad, 30}    \\[-0.2cm]
  {\small\it 28911 Legan\'es, SPAIN}           \\[-0.2cm]
  {\small\tt JSALAS@MATH.UC3M.ES}              \\[-0.2cm]
  {\protect\makebox[5in]{\quad}}  
  \\
}
\maketitle
\thispagestyle{empty}   

\begin{abstract} 
We study the chromatic polynomial $P_G(q)$ for $m\times n$ square- and 
triangular-lattice strips of widths $2\leq m \leq 8$ with cyclic boundary
conditions. This polynomial gives the zero-temperature limit of the 
partition function for the antiferromagnetic $q$-state Potts model 
defined on the lattice $G$. 
We show how to construct the transfer matrix in the 
Fortuin--Kasteleyn representation for such lattices and obtain the 
accumulation sets of chromatic zeros in the complex $q$-plane in the
limit $n\to\infty$. We find that the different phases that appear in this
model can be characterized by a topological parameter. We also compute
the bulk and surface free energies and the central charge. 
\end{abstract} 

\medskip
\noindent
{\bf Key Words:}
Chromatic polynomial; antiferromagnetic Potts model; triangular lattice;
square lattice; transfer matrix; Fortuin--Kasteleyn representation;
Beraha numbers; conformal field theory.

\clearpage

\newcommand{\be}{\begin{equation}}
\newcommand{\ee}{\end{equation}}
\newcommand{\<}{\langle}
\renewcommand{\>}{\rangle}
\newcommand{\widebar}{\overline}
\def\reff#1{(\protect\ref{#1})}
\def\spose#1{\hbox to 0pt{#1\hss}}
\def\ltapprox{\mathrel{\spose{\lower 3pt\hbox{$\mathchar"218$}}
 \raise 2.0pt\hbox{$\mathchar"13C$}}}
\def\gtapprox{\mathrel{\spose{\lower 3pt\hbox{$\mathchar"218$}}
 \raise 2.0pt\hbox{$\mathchar"13E$}}}
\def\textprime{${}^\prime$}
\def\proof{\par\medskip\noindent{\sc Proof.\ }}
\def\qed{\hbox{\hskip 6pt\vrule width6pt height7pt depth1pt \hskip1pt}\bigskip}
\def\proofof#1{\bigskip\noindent{\sc Proof of #1.\ }}
\def\half{ {1 \over 2} }
\def\third{ {1 \over 3} }
\def\twothird{ {2 \over 3} }
\def\smfrac#1#2{\textstyle{#1\over #2}}
\def\smhalf{ \smfrac{1}{2} }
\newcommand{\real}{\mathop{\rm Re}\nolimits}
\renewcommand{\Re}{\mathop{\rm Re}\nolimits}
\newcommand{\imag}{\mathop{\rm Im}\nolimits}
\renewcommand{\Im}{\mathop{\rm Im}\nolimits}
\newcommand{\sgn}{\mathop{\rm sgn}\nolimits}
\newcommand{\tr}{\mathop{\rm tr}\nolimits}
\newcommand{\diag}{\mathop{\rm diag}\nolimits}
\newcommand{\Gal}{\mathop{\rm Gal}\nolimits}
\newcommand{\mycup}{\mathop{\cup}}
\newcommand{\Arg}{\mathop{\rm Arg}\nolimits}
\def\hboxscript#1{ {\hbox{\scriptsize\em #1}} }
\def\zhat{ {\widehat{Z}} }
\def\phat{ {\widehat{P}} }
\def\qtilde{ {\widetilde{q}} }
\renewcommand{\mod}{\mathop{\rm mod}\nolimits}
\renewcommand{\emptyset}{\varnothing}

\def\scra{\mathcal{A}}
\def\scrb{\mathcal{B}}
\def\scrc{\mathcal{C}}
\def\scrd{\mathcal{D}}
\def\scrf{\mathcal{F}}
\def\scrg{\mathcal{G}}
\def\scrl{\mathcal{L}}
\def\scro{\mathcal{O}}
\def\scrp{\mathcal{P}}
\def\scrq{\mathcal{Q}}
\def\scrr{\mathcal{R}}
\def\scrs{\mathcal{S}}
\def\scrt{\mathcal{T}}
\def\scrv{\mathcal{V}}
\def\scrz{\mathcal{Z}}

\def\q{{\sf q}}

\def\Z{{\mathbb Z}}
\def\R{{\mathbb R}}
\def\C{{\mathbb C}}
\def\Q{{\mathbb Q}}

\def\T{{\mathsf T}}
\def\H{{\mathsf H}}
\def\V{{\mathsf V}}
\def\D{{\mathsf D}}
\def\J{{\mathsf J}}
\def\P{{\mathsf P}}
\def\QQ{{\mathsf Q}}
\def\RR{{\mathsf R}}

\def\bsigma{{\boldsymbol{\sigma}}}
\def\bone{{\mathbf 1}}
\def\vv{{\bf v}}
\def\uu{{\bf u}}
\def\w{{\bf w}}

\newtheorem{theorem}{Theorem}[section]
\newtheorem{definition}[theorem]{Definition}
\newtheorem{proposition}[theorem]{Proposition}
\newtheorem{lemma}[theorem]{Lemma}
\newtheorem{corollary}[theorem]{Corollary}
\newtheorem{conjecture}[theorem]{Conjecture}


\newenvironment{sarray}{
          \textfont0=\scriptfont0
          \scriptfont0=\scriptscriptfont0
          \textfont1=\scriptfont1
          \scriptfont1=\scriptscriptfont1
          \textfont2=\scriptfont2
          \scriptfont2=\scriptscriptfont2
          \textfont3=\scriptfont3
          \scriptfont3=\scriptscriptfont3
        \renewcommand{\arraystretch}{0.7}
        \begin{array}{l}}{\end{array}}

\newenvironment{scarray}{
          \textfont0=\scriptfont0
          \scriptfont0=\scriptscriptfont0
          \textfont1=\scriptfont1
          \scriptfont1=\scriptscriptfont1
          \textfont2=\scriptfont2
          \scriptfont2=\scriptscriptfont2
          \textfont3=\scriptfont3
          \scriptfont3=\scriptscriptfont3
        \renewcommand{\arraystretch}{0.7}
        \begin{array}{c}}{\end{array}}

\clearpage

%
%

\section{Introduction} \label{sec.intro} 

The two-dimensional Potts model conceals a very rich physics, in particular
in the antiferromagnetic regime. Let $G=(V,E)$ be a finite undirected graph
with vertex set $V$ and edge set $E$. The $q$-state Potts model is initially
defined for $q$ a positive integer in terms of spins $\sigma_x = 1,2,\ldots,q$
living on the vertices $x \in V$. Its Hamiltonian reads
\be
   H(\{\sigma\})   \;=\;
   - J \sum_{\< xy \> \in E} \delta(\sigma_x, \sigma_y) \; ,
\ee
where $\delta$ is the Kronecker delta. Fortuin and Kasteleyn
\cite{Kasteleyn} have shown that
the partition function at inverse temperature $\beta$ can be rewritten as
\be
   Z_G(q,v)   \;=\;
   \sum_{ E' \subseteq E }  q^{k(E')} v^{|E'|} \;,
 \label{FK_rep}
\ee
with $v = \text{e}^{\beta J} - 1$. The sum runs over the $2^{|E|}$ subsets
$E' \subseteq E$, with $k(E')$ being the number of connected components
(including isolated vertices) in the subgraph $(V,E')$. We now promote
\reff{FK_rep} to the definition of the model, which permits us to consider
$q$ as an arbitrary complex number.

We shall here be interested in the zero-temperature antiferromagnetic
case $J = -\infty$ (i.e., $v=-1$), whose restriction to $q$ a
positive integer can be interpreted as a coloring problem. 
In this case, $P_G(q) \equiv Z_G(q,-1)$ is known as the chromatic polynomial. 
Its evaluation for a general graph is a hard problem in the sense that  
the best algorithms currently available 
have a worst-case execution time which grows exponentially with the
number of sites of the graph. 
For strip graphs (or, more generally, recursive
families of graphs) the chromatic polynomial $P_{G_n}(q)$ of any member of the
family $G_n$ can be computed from a transfer matrix $\mathsf{T}$ and certain
boundary condition vectors $\bm{u}$ and $\bm{v}$:
\be
  P_{G_n} \;=\; \bm{u}^T \cdot \mathsf{T}^n \cdot \bm{v} \; .
 \label{PGn}
\ee
Thus, the computation time grows as a polynomial in $n$; it does however
still grow exponentially in the strip width (due to the $L$ dependence
of $\text{dim} \, \mathsf{T}$). Thus, it is necessary to introduce new
algorithms to be able to handle as large $L$ as possible. For example,
identifying symmetries may give an equivalent transfer matrix of lower
dimension (in this respect, the Fortuin-Kasteleyn representation makes
manifest the $S_q$ symmetry of the original spin Hamiltonian).

It is important to remark that the Potts spin model has a probabilistic   
interpretation (i.e., has non-negative Boltzmann weights) only when $q$ is
a positive integer and $v\geq -1$. The Fortuin--Kasteleyn random-cluster
model \reff{FK_rep}, which extends the Potts model to non-integer $q$,
has non-negative weights only when $q\geq 0$ and $v\geq 0$. In all other cases,
the model belongs to the ``unphysical'' regime (i.e., some weights are negative
or complex), and some familiar properties of statistical mechanics
need not hold. For example, even for integer $q\geq 2$ and real $v$ in the
range $-1\leq v <0$,
where the {\em spin} representation exists and has non-negative weights, the
dominant eigenvalue $\lambda_1$ of the transfer matrix in the {\em cluster}
representation 
need not be simple (i.e., the eigenvector may not be unique);
and even if simple, it may not play any role in determining
$Z_{G_n}(q,v)$ because the corresponding amplitude may vanish.
Both these behaviors are of course impossible for any
transfer matrix with {\em positive}\/ weights, 
by virtue of the Perron--Frobenius theorem.
It is important to note that the dominant eigenvalues in the cluster and
spin representations need not to be equal, because one of them may have 
a vanishing amplitude (see below).
In any case, the Potts model can thus make probabilistic sense
in the cluster representation at parameter values
where it fails to make probabilistic sense in the spin
representation, and vice versa.
It is worth mentioning that for $q = 4 \cos^2(\pi/n)$ with
integer $n=3,4,\ldots$ and {\em planar}\/ graphs $G$,
the Potts-model partition function $Z_G(q,v)$ admits
a third representation, in terms of a restricted solid-on-solid (RSOS)
model \cite{Pasquier_alone,ABF}.

But even in the most general case (i.e., with arbitrary real, or even
complex $q$ and $v$) we consider the study of the random-cluster model
\reff{FK_rep} fully legitimate. Note in this context that there is a
long story of studying statistical-mechanical models in ``unphysical''
regimes (e.g., the hard-core lattice gas at its negative-fugacity
critical point
\cite{Kurtze_79,Shapir_82,Dhar_83,Poland_84,Baram_87,Guttmann_87,%
Lai_95,Park_99,Todo_99,Brydges_03} or the Yang--Lee edge singularity
\cite{Lee_Yang,Kortman_71,Fisher_78,Alcantara_81,Parisi_81,%
Cardy_85,Itzykson_86}). In particular, Baxter \cite{Baxter_86_87}
studied the zero-temperature limit of the triangular-lattice Potts
antiferromagnet, which belongs to this ``unphysical'' regime.
Generally speaking, the ``unphysical'' regime can be understood using
the standard tools of statistical mechanics with appropriate
modifications. In particular, although a rigorous proof is still
lacking, conformal field theory (CFT) seems to apply also in the
``unphysical'' regions (see e.g., \cite{Cardy_85,Itzykson_86} for the
CFT description of the Lee-Yand edge singularity). In general,
``unphysical'' regions are described by non-unitary field theories.

It is well established 
that antiferromagnetic models in general, and the
chromatic polynomial in particular, are very sensitive to the choice of
boundary conditions. Indeed, different choices may lead to quite different
thermodynamic limits. 
Following our earlier publications \cite{transfer1,transfer2,transfer3}, 
where we studied free or periodic boundary conditions in the
transverse (space-like) direction and {\em free}\/ boundary conditions
in the longitudinal (time-like) direction,
we here subject the chromatic polynomial to {\em cyclic boundary conditions}\/,
i.e., free in the transverse direction and {\em periodic}\/
in the longitudinal direction. 
For a transfer matrix written in the Potts spin basis, cyclic boundary 
conditions can be easily implemented by setting 
$P_{G_n} = \text{Tr} \, \mathsf{T}^n$. 
The situation is, however, less straightforward in the
Fortuin--Kasteleyn representation, and we are obliged to employ a transfer 
matrix that is different from ---and more complicated than--- the ones used in 
\cite{transfer1,transfer2,transfer3}. 

Our study is of interest for understanding the general properties of the
so-called Berker-Kadanoff (BK) phase in the antiferromagnetic Potts model
\cite{Saleur_90_91}. Generically, the {\em two-dimensional} 
Potts model on a given regular lattice
possesses a curve $v_\text{FM}(q)>0$ of ferromagnetic phase transitions
which are second-order in the range $0 < q \le 4$ \cite{Baxter_73}.
Along this curve the thermal operator (which corresponds to taking $v$
away from its critical value $v_\text{FM}(q)$) is relevant and it becomes
marginal in the limit $q\to 0$ of spanning trees. The analytic continuation
of the curve $v_\text{FM}(q)$ into the antiferromagnetic regime yields
a second critical curve $v_\text{BK}(q)<0$ with $0 < q < 4$
along which the thermal operator is {\em irrelevant}. Therefore, for each
$q \in (0,4)$, the critical point $v_\text{BK}(q)$ acts as the
renormalization-group attractor of a finite range of $v$ values: this
is the BK phase.

The above scenario has been verified in detail for several regular lattices.
In particular, for the square and triangular lattices the corresponding models
are Bethe Ansatz soluble \cite{Baxter_82}\footnote{
   Baxter \cite{Baxter_82,Baxter_82b} has shown that the 
   two-dimensional Potts model is equivalent to a homogeneous 
   six-vertex model on certain curves of the
   $(q,v)$ plane, including $v_\text{FM}(q)$ and $v_\text{BK}(q)$ for both
   the square and triangular lattices. This homogeneous six-vertex model is 
   soluble using the Bethe Ansatz \cite{Baxter_82}. In addition, 
   Baxter also showed that the
   square-lattice model on the mutually dual curves $v_{\pm}(q)$ is 
   equivalent to a certain class of inhomogeneous six-vertex models that 
   can nevertheless be solved by an extension of the Bethe 
   Ansatz \cite{Baxter_82b}.
}
and the analytic forms of $v_\text{FM}(q)$ and $v_\text{BK}(q)$ 
are exactly known. The scenario is also inherent in
the Coulomb gas approach to critical phenomena \cite{Nienhuis_87}, which,
albeit less rigorous, confirms that the general properties of the
BK phase are lattice independent.

The best understood case is that of the square lattice. Here, $v_\text{FM}(q)
= +\sqrt{q}$ and $v_\text{BK}(q)=-\sqrt{q}$ \cite{Baxter_82,Saleur_90_91}.
There are also two mutually dual (and hence equivalent) curves of
antiferromagnetic transition points \cite{Baxter_82b}, $v_{\pm}(q) =
-2 \pm \sqrt{4-q}$ for $q \in [0,4]$, which are believed to form the
boundaries of the BK phase \cite{Saleur_90_91}. Exactly on
$v_{\pm}(q)$, the model exhibits a different type of critical
behavior \cite{Saleur_90_91,AFpaper}.

More precisely, for a given width $L$ and $q \in I(L)$ belonging to
a suitable interval $I(L)$, there exists mutually dual couplings 
$v_\pm(q,L)$ so that for $v_-(q,L) < v < v_+(q,L)$, the dominant eigenvalue 
$\lambda_\text{BK}(v)$ in the cluster representation is simple. 
As $L \to \infty$, one has $v_\pm(q,L) \to v_\pm(q)$ and 
$\lim_{L\to\infty} I(L)$ contains $[0,4]$. 
The finite-size scaling of $\lambda_\text{BK}(v)$,
and of its adjacent scaling levels, determines
the critical exponents of the BK phase. However, for certain special 
values of $q$ (the generalized Beraha numbers, see below) the amplitudes
of $\lambda_\text{BK}(v)$, and of many subdominant eigenvalues, vanish;
yet other eigenvalues are degenerate in pairs but have amplitudes of 
opposite sign and thus, cancel out from the partition function. 
The remaining eigenvalues correspond to those of the relevant RSOS 
representation 
(or spin representation, when $q$ is a positive integer). On the other hand, 
for $v \notin [v_-(q,L),v_+(q,L)]$ the dominant eigenvalue in the cluster
representation is again simple, but is a {\em different} analytical 
expression of $v$, and its amplitude does {\em not} vanish at the
special values of $q$.

On the triangular lattice, $v_\text{FM}(q)$ is the uppermost branch ($v>0$) of
the curve $v^3 + 3v^2 - q = 0$ \cite{Baxter_78}. We know of numerical evidence
(to be published elsewhere) that the middle and lower branches of this same
curve may be respectively $v_\text{BK}(q)$ and the lower boundary $v_-(q)$ of
the BK phase. Less is known about the upper boundary $v_+(q)$ of the BK phase.
If we assume, in analogy with the square-lattice case, that it is the first
locus of partition function zeros encountered in the antiferromagnetic regime
upon decreasing the temperature from infinity, one has $v_+(0) = 0$, and the
slope 
$\left. \frac{\text{d}}{\text{d}q} v_+(q) \right|_{q\to 0^+} \simeq -0.1753
\pm 0.0002$ is known numerically \cite{forests}. Moreover, another Bethe
Ansatz soluble case \cite{Baxter_86_87}, $v=-1$, provides evidence that
$v_+(q_0) = -1$, where $q_0$ will be defined below.

In the present publication we study the chromatic polynomial with cyclic
boundary conditions for the square and the triangular lattices. In both cases
the chromatic line $v=-1$ passes through a finite part of the BK phase, whence
our results will enable us to examine the role of cyclic boundary conditions
for this phase. We shall obtain strong evidence that with cyclic boundary
conditions the BK phase splits into an {\em infinite number of distinct
phases}, each of which has a precise topological characterization in terms of
the Fortuin-Kasteleyn clusters. However, the chromatic line only intersects a
finite number of these phases. For the square (resp.\ triangular) lattice
chromatic polynomial we thus find a total of three (resp.\ seven) phases.

In particular, we examine the loci of zeros of $P_{G_{L \times n}}(q)$, for
strips of width $L$ and length $n$, in the plane of complex $q$. When
$n\to\infty$, the accumulation points of these zeros form either isolated
limiting points (when the amplitude of the dominant eigenvalue vanishes) or
continuous limiting curves $\scrb_L$ (when two or more dominant eigenvalues
become equimodular). These latter curves constitute the boundaries between the
different phases just described. We refer to
\cite{transfer1,transfer2,transfer3} for details, and for the relation to the
Beraha--Kahane--Weiss theorem \cite{BKW}. 
The rather unusual phase diagram referred to
above is linked to a difference of the $L\to\infty$ limits of $\scrb_L$ with
free/cylindrical \cite{transfer1,transfer2,transfer3} and cyclic boundary
conditions (this work).

The curves $\scrb_L$ will in general intersect the real $q$-axis in a number
of values, and we define $q_0(L)$ to be the largest of these values. Also, let
$q_0 = \lim_{L\to\infty} q_0(L)$. Based on our earlier studies
\cite{transfer1,transfer2,transfer3} we expect $q_0$ for a given lattice to be
independent of (reasonable) boundary conditions. The physical scenario
outlined above would then imply that we should observe the BK phase for $0 < q
< q_0(L)$. We expect $q_0(\text{sq})=3$ for the square lattice (corresponding
to $v_+(3) = -1$). For the triangular lattice, Baxter \cite{Baxter_86_87}
originally predicted $q_0(\text{tri}) \approx 3.81967$. We have recently
revisited the analysis of Baxter's equations \cite{transfer3}, arguing in
particular that his prediction for $q_0(\text{tri})$ is erroneous,
the correct value being $q_0(\text{tri}) = 2 + \sqrt{3} \approx
3.73205$.

The results available in the literature concerning square- and 
triangular-lattice strips with cyclic boundary conditions are limited to
widths $L\leq 5$.
The results for the square lattice were obtained for $L=2$ 
by Biggs, Demerell and Sands \cite{Biggs_72}, for $L=3$ by Shrock and Tsai 
\cite{Shrock_00a}, and for $L=4,5$ by 
Chang and Shrock \cite{Shrock_01a,Shrock_02a}.
The results for the triangular lattice were obtained for $L=2$ by 
Shrock and Tsai \cite{Shrock_00a}, and for $L=3,4$ by Chang and Shrock 
\cite{Shrock_01c}. When writing up this paper we learned that Chang and
Shrock have extended their results to the case $L=5$ \cite{Shrock_05a}. 

In this paper we present new exact results for the chromatic polynomial of
strip graphs of widths $L=6$. We have also computed independently the exact
formula for the triangular-lattice strip graphs of width $L=5$; furthermore,
we present the computation of the limiting curve for these strips.\footnote{ 
   The interested reader can find the exact expression of the 
   eigenvalues for $L\leq 6$ in the {\sc mathematica} files available
   as part of the electronic version of this paper in the {\tt cond-mat} 
   archive. 
}
For $L=7,8$ we have {\em exactly}\/ computed some blocks of
the transfer matrix from which {\em all}\/ the eigenvalues
can be inferred (modulo some conjectures that are very plausible
though unproven, see Properties~1 and~3 in Section~\ref{sec.sq.cyclic.blocks2}).
Unfortunately some of these blocks are so large than we have been unable to 
get the analytical expressions for the eigenvalues, as we have done for 
$L\leq 6$. From the {\em exact symbolic} expression for these blocks, we can 
nevertheless obtain {\em essentially-exact} relevant physical information, 
such as numerical values for the eigenvalues and the limiting curve 
$\mathcal{B}$. The only source of error in these results is the limited 
(but very high) machine precision of the numerical computations performed 
with {\sc mathematica}.
We have taken the symbolic computations reported in this paper as far as 
possible, using several months of CPU time. However, because of the 
exponential character (in the width $L$) of this problem, 
going beyond $L=8$ will require an extremely 
large amount of CPU time and RAM memory. 

Our approach is based on a transfer-matrix formulation of the 
Fortuin--Kasteleyn representation, which is equivalent to the
methods of Shrock and collaborators 
(See e.g. \cite{Shrock_05b} and references therein).%
\footnote{The two approaches are however formulated in somewhat
   different terms, so the equivalence is only made clear by a
   detailed comparison. 
}
Large parts of our analysis has been automatized and carried out by a computer,
thus allowing to access larger widths $L$. On the other hand, in our approach 
we find a natural candidate for an essentially 
``conserved quantum number'': the number of bridges\footnote{
 The number of bridges is equivalent to what 
 Refs.~\protect\cite{Shrock_05a,Shrock_05b} call the level.
} 
of a 
given connectivity state (see Section~\ref{sec.sq.cyclic.blocks} for details). 
This number is the link to the quantum-group theory worked out by 
Pasquier and Saleur \cite{Pasquier,Saleur_90_91}. 
Finally, let us mention a different approach developed by Noy and Rib\'o 
\cite{Noy_04} and also based on a transfer-matrix formalism. They handle
cyclic boundary conditions by introducing a new type of operation: edge 
deletion. 

The outline of the paper is as follows. In Section~\ref{sec.sq.cyclic} we
present our method for obtaining the transfer matrix for a square-lattice
strip with cyclic boundary conditions. We describe the structural properties
of the transfer matrix and its block structure.
Section~\ref{sec.sq.cyclic.blocks2} also provides an alternative (and
simplified) method for constructing the transfer matrix which has proved very
useful for large-width computations. In Section~\ref{sec.sq.chromatic} we
present the results for square-lattice strips up to widths $L=8$. In the
following Section we adopt our method to triangular-lattice strips with cyclic
boundary conditions, and the results are given in
Section~\ref{sec.tri.chromatic} for widths up to $L=8$. Finally, in
Section~\ref{sec.disc} we discuss the phase diagram of the square- and
triangular-lattice Potts antiferromagnet at zero temperature. We compute the
free energy and the central charge as a function of $q$. Finally, in
Appendix~\ref{sec.sq.cyclic.alt} we show an alternative way of computing the
transfer matrix for a cyclic square-lattice strip.

%
%

\section{Transfer matrix for square-lattice strips with cyclic boundary 
         conditions} \label{sec.sq.cyclic} 

We want to build the transfer matrix for a square-lattice strip of 
width $m$ and arbitrary length $n$ with cyclic boundary conditions, i.e.,  
free along the transverse direction and periodic along the longitudinal 
direction (Figure~\ref{figure_sq_cyclic_bc} shows an example of size 
$4\times 2$).  
As in previous work \cite{transfer1,transfer2,transfer3},
 we would like to express the transfer matrix ${\mathsf T}$
as a product of two matrices ${\mathsf H}$ and ${\mathsf V}$, representing 
respectively the horizontal and vertical bonds of the lattice 
\be 
  \mathsf{T}(m) \; = \; \mathsf{V}(m) \cdot \mathsf{H}(m)
  \label{def_T}
\ee
For each lattice strip width $m$ we have a different transfer matrix.
   When free boundary conditions are used in the longitudinal direction
   (i.e., for a strip with free or cylindrical b.c.),
   the transfer matrix in the Fortuin--Kasteleyn representation
   has been explained in detail in \cite[Section 3]{transfer1}.
   The basic idea is that the state of the system is characterized
   by a connectivity pattern $\{ {\bf v}_{\cal P} \}$ of the
   sites $\{1,\ldots,m\}$ of the top row,
   and the transfer matrix $\mathsf{T}(m)$ acts by attaching one new row
   above the current top row.

   By contrast, when periodic boundary conditions are used in the
   longitudinal direction, it is insufficient to keep track of the
   connectivity of the top row alone, as we will eventually need
   to identify the top and bottom rows.
   Rather, as first sketched in \cite[Section 7.4]{transfer1},
   the idea is to keep track of the connectivity pattern of the $2m$ sites
   constituted by the current top row {\em and}\/ the bottom row.
   Let us call these sites $1,2,\ldots,m$ and $1',2',\ldots,m'$, respectively 
   (see Figure~\ref{figure_sq_cyclic_bc} for an example with $m=4$).
   Initially the top and bottom rows are identical.
   We then enlarge the lattice by adding successive new rows
   to the top of the lattice, exactly as in \cite[Section 3.2]{transfer1};
   the join and detach operations act on the sites of the top row,
   with those of the bottom row simply ``going along for the ride''.
   At the end, when we have obtained a lattice with $n+1$ rows,
   we identify the top and bottom rows.

   Let us now explain this formalism in greater detail.
   We characterize the state of the top and bottom rows by a
   connectivity pattern $\{ {\bf v}_{\cal P} \}$, which is 
associated to a partition
$\mathcal{P}=\{P_1,\ldots,P_k\}$ of the set $\{ 1,\ldots ,m,1',\ldots ,m' \}$.
It is convenient to introduce the operators
\begin{eqnarray}
   \P_x       & = & v I  \,+\,  \D_x   \\[1mm]
   \QQ_{x,y}  & = & I  \,+\,  v \J_{xy}
\end{eqnarray}
acting on $\{ \bm{v}_\mathcal{P}\}$ (see \cite{transfer1} for more
details). Here, $I$ is the identity; $\D_x$ is a detach operator that
detaches site $x$ from the block it currently belongs to,
multiplying by a factor $q$ if $x$ is currently a singleton; and
$\J_{xy}$ is a join operator that amalgamates the blocks containing $x$ and
$y$, if they were not already in the same block.

The matrices $\mathsf{H}(m)$ and $\mathsf{V}(m)$ then act on the top row as
follows:
\begin{subeqnarray}
  \mathsf{H}(m)  &=& \prod\limits_{i=1}^{m-1} \QQ_{i,i+1}  \slabel{def_H} \\
  \mathsf{V}(m)  &=& \prod\limits_{i=1}^{m} \P_{i}         \slabel{def_V}
\end{subeqnarray}
We can similarly define matrices $\mathsf{H}'(m)$ and $\mathsf{V}'(m)$ that
act on the bottom row.

The chromatic polynomial of a square-lattice strip of width $m$ and 
length $n$ with cyclic boundary conditions is given by 
\be
  P_{m\times n}(q) \; = \; \bm{u}^\text{T} \cdot \mathsf{H}(m)  
                                           \cdot \mathsf{T}(m)^n 
                                           \cdot \bm{v}_\text{id}
\label{def_P}
\ee
The left $\bm{u}$ and right $\bm{v}_\text{id}$ vectors differ from those
found for free and cylindrical boundary conditions. The right vector
$\bm{v}_\text{id}$ now denotes the partition
\be
  \bm{v}_\text{id} \; = \; \{ \{1,1'\},\{2,2'\},\ldots,\{m,m'\} \}
  \label{del_v_id}
\ee
(i.e., we start with the top and bottom rims identified).
The left vector in \reff{def_P} acts on a connectivity state 
$\bm{v}_\mathcal{P}$ as follows
\be
  \bm{u}^\text{T} \cdot \bm{v}_\mathcal{P} \; = \; q^{|\mathcal{P}'|} 
\ee
where $|\mathcal{P}'|$ denotes the number of blocks in the connectivity
$\bm{v}_{\mathcal{P}'}$ obtained from 
$\bm{v}_\mathcal{P}$ by the action of several join operators $\J_{x,y}$:
\be
  \bm{v}_{\mathcal{P}'} \; = \; \left( 
                \prod\limits_{i=1}^m \J_{i,i'} \right) \cdot \bm{v}_\mathcal{P}
 \label{def_vp}
\ee
In other words, $\bm{u}^\text{T}$ acts on $\bm{v}_{\cal P}$ by identifying
the top and bottom rows, assigning a factor of $q$ to each block in the
resulting partition.

In this paper we are mainly concerned by the chromatic-polynomial case 
($v=-1$). In this case, the matrix $\mathsf{H}$ is a projector 
(i.e., $\mathsf{H}^2 = \mathsf{H}$). Thus, we can use instead of the 
transfer matrix $\mathsf{T}$ [c.f.,~\reff{def_T}] and the usual connectivity
states $\bm{v}_\mathcal{P}$, the modified transfer matrix  
\be
  \widetilde{\mathsf{T}}(m) \; = \; 
  \mathsf{H}(m) \cdot \mathsf{V}(m) \cdot \mathsf{H}(m) 
  \label{def_Tp}
\ee
and the basis vectors 
\be
  \bm{w}_\mathcal{P} \; = \; \mathsf{H}(m) \cdot \bm{v}_\mathcal{P}
\ee
Note that $\bm{w}_\mathcal{P}=0$ if $\mathcal{P}$ has any pair of 
nearest neighbors of the top row in the same block. 
As we start with the state $\bm{w}_\text{id}$, and at the end we are 
identifying the top and bottom rows, it is useful to work directly 
on the modified connectivity basis  
\be
  \widehat{\bm{w}}_\mathcal{P} \; = \; \mathsf{H}'(m) \cdot 
                             \mathsf{H}(m) \cdot \bm{v}_\mathcal{P}
  \label{def_wP}
\ee
This choice 
implies that $\widehat{\bm{w}}_\mathcal{P}=0$ also if $\mathcal{P}$ 
contains any pair of
nearest neighbors of the bottom row in the same block.
For simplicity, we hereafter drop the tilde to denote 
the modified transfer matrix \reff{def_Tp}, and the hat to denote the modified
connectivity basis \reff{def_wP}. In this case, the 
chromatic polynomial can be written as  
\be
  P_{m\times n}(q) \; =\;  \bm{u}^\text{T} \cdot \mathsf{T}(m)^n 
                                           \cdot \bm{w}_\text{id}
\label{def_Pp}
\ee
%

%
%
\subsection{Structural properties of the chromatic polynomial}
\label{sec.sq.cyclic.struct}

The first observation is that the square lattice with cyclic boundary 
conditions is a {\em planar} graph: thus, only non--crossing partitions 
$\mathcal{P}$ of $\{1,\ldots,m,1',\ldots,m'\}$ can occur. 
As we have transverse free boundary conditions, the number of such 
connectivities is given by the Catalan number
\be
  C_{2m} \; = \; \frac{1}{2m+1} \binom{4m}{2m} 
\ee
Their asymptotic behavior is given by 
\be
  C_{2m} \; = \; 16^m m^{-3/2} (8\pi)^{-1/2} [ 1 + O(1/m)] 
\ee
These numbers are listed in the second column of
Table~\ref{table_dim_T_cyclic} up to $m=10$.
     
The fact that we are computing the chromatic polynomial $(v=-1$)
implies that the dimension of the transfer matrix should be equal to 
the number of non-crossing {\em non-nearest-neighbor} partitions  
(i.e., partitions in which no block contains a pair
of nearest neighbors). We shall denote
these numbers by TriRing$(m)$, as we expect them to be equal
to the dimension of $\mathsf{T}$ for a cyclic triangular-lattice strip 
of width $m$ (see Table~\ref{table_dim_T_cyclic}). 

A further simplification comes from considering classes of partitions 
invariant under reflections with respect to the center of the strip. The
number of such invariant classes is denoted by 
SqRing$(m)$ in Table~\ref{table_dim_T_cyclic}. Thus, we expect that these
numbers are equal to the dimension of $\mathsf{T}$ for a cyclic 
square-lattice strip of width $m$.

We have modified the {\tt perl} code that we used for free/cylindrical
boundary conditions \cite{transfer2,transfer3} to deal with the cyclic
boundary conditions. We have found that for $m\geq 3$, SqRing$(m)$
is larger than the number of {\em distinct} eigenvalues SqRing$'(m)$ found in
\cite{Shrock_00a,Shrock_01a,Shrock_01b} 
(see Table~\ref{table_dim_T_cyclic}).\footnote{
  The authors of Ref.~\cite{Shrock_01b} use the notation $N_{P,L_y,\lambda}$ 
  to denote the number of distinct eigenvalues appearing in the chromatic 
  polynomial for cyclic strips of square and triangular lattices of width 
  $L_y$.
} 
In Section \ref{sec.sq.cyclic.blocks2} below we show how the transfer matrix
can be rewritten in a form that amounts to making each distinct eigenvalue
appear with unit multiplicity; we also explain how the issue of degeneracies
is linked to an underlying quantum-group symmetry.

Chang and Shrock found a closed expression for the numbers SqRing$'(m)$ using
coloring-matrix methods \cite{Shrock_01b}:
\be
\text{SqRing}'(m) \;=\; 2(m-1)! \sum\limits_{k=0}^{\lfloor m/2 \rfloor} 
                  \frac{(m-k)!}{ (k!)^2 \, (m-2k)! } 
\label{formula_SqRing}
\ee
We notice that these numbers coincide with the sequence {\tt A025565} of 
\cite{Sloane}, and we therefore have the alternative expression\footnote{
  The equivalence between \protect\reff{formula_SqRing} and 
  \protect\reff{formula_SqRing_alt} comes from the equality of their 
  respective generating functions. In particular \protect\cite{Shrock_01b},
$$
\sqrt{ \frac{1+x}{1-3x} } -1 \;=\; \sum\limits_{m=1}^\infty 
       \text{SqRing}'(m)\, x^m
$$
}
\be
\text{SqRing}'(m) \;=\; \sum\limits_{k=0}^{\lfloor (m+1)/2 \rfloor} 
                  \binom{m-1}{k} \, \binom{m+1-k}{k+1} \; .
\label{formula_SqRing_alt}
\ee
The asymptotic behavior for large $m$ of the number of distinct eigenvalues
was found by Chang and Shrock \cite{Shrock_01b} to be
\be
\text{SqRing}'(m) \; \sim \; 3^m \, m^{-1/2} \qquad \text{as $m\to\infty$} 
\ee

\bigskip

\noindent
{\bf Remarks}: 
1) This transfer matrix is not invariant under the interchange of the top
and bottom rows (i.e. $i \leftrightarrow i'$). The reason is that our transfer
matrix only acts on the top row, leaving fixed the bottom one. Thus, we cannot
reduce the size of the transfer matrix by including only connectivity states 
that are invariant under the  transformation $i \leftrightarrow i'$.

2) With our transfer-matrix approach we have checked all previous results
by Shrock and collaborators \cite{Shrock_00a,Shrock_01a,Shrock_01c,Shrock_02a}. 

3) Even though the final dimension of the transfer matrix $\mathsf{T}(m)$ is
given by SqRing$(m)$, the computations need to be performed in the larger
space of non-crossing connectivities (of dimension $C_{2m}$). This fact 
limits the maximum width we are able to handle. 

4) When we apply the operator $\prod\limits_{i=1}^m \J_{i,i'}$ to a
non-crossing connectivity $\bm{v}_\mathcal{P}$ in \reff{def_vp}, the resulting
connectivity state $\bm{v}_{\mathcal{P}'}$ is {\em not} guaranteed to be
non-crossing. However, as this step is the last one in the computation, we do
not need to enlarge our vector space (to include all crossing connectivities).

5) We have checked our results by comparing them to the appropriate 
chromatic polynomial with cylindrical boundary conditions
\be
Z_{m_\text{F} \times n_\text{P}}(q) = Z_{n_\text{P} \times m_\text{F}}(q) \; ,
\label{eq_check_Z}
\ee
where F and P denote free and periodic boundary conditions respectively.

%
%
\subsection{Block structure of the transfer matrix}
\label{sec.sq.cyclic.blocks}

\begin{figure}
   \setlength{\unitlength}{0.5pt}
   \centering
   \begin{picture}(300,300)

     
     \Thicklines
     \put(200,250){\oval(200,50)[b]}
     \put(100,50) {\oval(200,50)[t]}

     \dashline{10}(300,50)(300,250)

     \put(0,50){\circle*{15}}
     \put(0,250){\circle*{15}}
     \put(100,50){\circle*{15}}
     \put(100,250){\circle*{15}}
     \put(200,50){\circle*{15}}
     \put(200,250){\circle*{15}}
     \put(300,50){\circle*{15}}
     \put(300,250){\circle*{15}}

     \put( -8,  0){\large \bf 1'}
     \put( 92,  0){\large \bf 2'}
     \put(192,  0){\large \bf 3'}
     \put(292,  0){\large \bf 4'}

     \put( -8,290){\large \bf 1}
     \put( 92,290){\large \bf 2}
     \put(192,290){\large \bf 3}
     \put(292,290){\large \bf 4}
   \end{picture}

   \caption{\label{figure_sq_cyclic_bc_bis}
     Connectivity state for a cyclic strip of width 4. As an example,
     we show the connectivity state ${\cal P}=\delta_{1',3'}\delta_{2,4,4'}$. 
     This state can be seen as a bottom-row connectivity state 
     ${\cal P}_\text{bottom} =\delta_{1',3'}$ (solid line), 
     a top-row connectivity state ${\cal P}_\text{top} =\delta_{2,4}$ 
     (solid line), and a bridge connecting both rows $\delta_{4,4'}$ 
     (dashed line). 
   }
\end{figure}
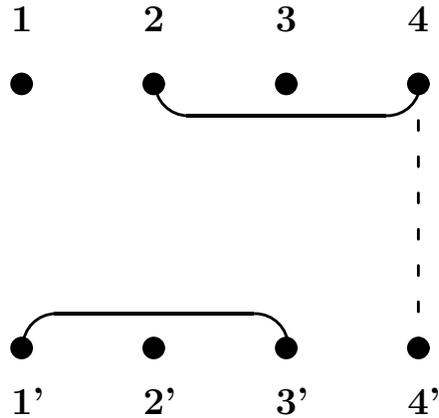

The transfer matrix for a cyclic square-lattice strip of width $m$ has
a block structure. This property greatly simplifies the computation of its
eigenvalues as we shall see below. 
Given a certain connectivity $\mathcal{P}$ of the top and bottom rows 
(see, e.g., Figure~\ref{figure_sq_cyclic_bc_bis}), one can regard it as a
top-row connectivity $\mathcal{P}_\text{top}$ and a bottom-row connectivity 
$\mathcal{P}_\text{bottom}$ joined by $\ell$ bridges. Note that each top-row 
block is connected by at most one bridge to the bottom row and vice versa. Thus,
$\ell$ can take any integer value from $0$ to $m$. 

In order to obtain the matrix element $\mathsf{T}_{\bm{a},\bm{b}}$ 
between two connectivity states $\bm{w}_\mathcal{P}$ and
$\bm{w}_{\mathcal{P}'}$, we should perform three operations 
\begin{enumerate}
   \item Apply the vertical-bond operator $\mathsf{V}$ on the initial state 
         $\bm{w}_\mathcal{P}$.
   \item Apply the horizontal-bond operator $\mathsf{H}$, 
         so we kill any nearest-neighbor connectivity state. 
   \item Project the result onto the final state $\bm{w}_{\mathcal{P}'}$. 
\end{enumerate}

Let us suppose we start from a connectivity state $\bm{w}_\mathcal{P}$ with
a certain bottom-row connectivity $\mathcal{P}_\text{bottom}$, number of 
bridges $\ell$, and position of such bridges (all these characteristics modulo
reflections with respect to the center of the strip). Then two important
observations should be made:
\begin{itemize}

  \item The bottom-row connectivity of the initial state is not modified by
        the application of either matrix $\mathsf{V}$ (in step 1) or  
        matrix $\mathsf{H}$ (in step 2). 

  \item The application of the matrices $\mathsf{V}$ (in step 1) 
        and $\mathsf{H}$ (in step 2) cannot  
        increase the number of bridges $\ell$. Thus, we can only go to 
        connectivity states $\mathcal{P}'$ with the same or smaller number 
        of bridges $\ell'\leq \ell$. Indeed, the relative position of the 
        bridges is preserved. 

\end{itemize}

These two properties imply that the transfer matrix has a lower-triangular
block form:
\be
\mathsf{T}(m) \;=\; \left( \begin{array}{cccc} 
   T_{m,m}   & 0           & \ldots & 0 \\
   T_{m-1,m} & T_{m-1,m-1} & \ldots & 0 \\
   \vdots    & \vdots      &        & \vdots\\
   T_{0,m}   & T_{0,m-1}   & \ldots & T_{0,0}
                           \end{array}
                    \right)
\label{T_blocks1}
\ee
(Note that we have arranged the connectivity states according to a
decreasing number of bridges.) Furthermore, they also imply that each diagonal
block $T_{\ell,\ell}$ has a diagonal-block form
\be
T_{\ell,\ell} \; = \; \left( \begin{array}{cccc}
   T_{\ell,\ell,1}  & 0               & \ldots & 0 \\
   0                & T_{\ell,\ell,2} & \ldots & 0 \\
   \vdots           & \vdots          &        & \vdots\\
   0                & 0               & \ldots & T_{\ell,\ell,N_\ell}
                             \end{array}
                       \right)
\label{T_blocks2}
\ee
where
each sub-block $T_{\ell,\ell,j}$ is characterized by a certain bottom-row
connectivity $\mathcal{P}_\text{bottom}$ and a position of the $\ell$ 
bridges. Its dimension is given by the number of top-row connectivities that
are compatible with $\mathcal{P}_\text{bottom}$, $\ell$, and the relative
positions of the $\ell$ bridges. The number of blocks $N_\ell$ 
($0\leq \ell\leq m$) for a square-lattice strip of width $m$ is displayed in
Table~\ref{table_blocks_sq}. 
 
In particular, this structure means that the characteristic polynomial of the
full transfer matrix $\mathsf{T}$ can be factorized as follows
\be
  \chi(\mathsf{T}) \; = \; \prod\limits_{\ell=0}^m \left[ 
          \prod\limits_{j=1}^{N_\ell} 
          \chi \left( T_{\ell,\ell,j} \right) \right] 
  \label{chi_block_T_sq}
\ee

In practice, for each strip width $2\leq m\leq 6$, we have performed 
the following procedure: 

\begin{enumerate}

\item Using a {\tt perl} script, we compute the full transfer matrix 
      $\mathsf{T}(m)$ and the left $\bm{u}$ and right vectors 
      $\bm{w}_\text{id}$. These objects are obtained in the basis of  
      non-crossing non-nearest-neighbor classes of connectivities that are 
      invariant  under reflections (with respect to the center of the strip). 
      Thus, their dimension is given by SqRing$(m)$.

\item Given the full transfer matrix $\mathsf{T}(m)$, we compute the 
      SqRing$'(m)$ distinct eigenvalues $\lambda_i(q)$. To do this, we
      split the transfer matrix into blocks $T_{\ell,\ell,j}$, each block
      characterized by a bottom-row connectivity and the number $\ell$ and 
      position of the bridges (modulo reflections). By diagonalizing these
      blocks we obtain the SqRing$'(m)$ distinct eigenvalues $\lambda_i(q)$ 
      and their multiplicities $k_i$.\footnote{
 Unfortunately, we have not found any analytical formula for the 
 multiplicities $k_i$.
} 
      Indeed,  
\be
  \sum\limits_{i=1}^{\text{SqRing}'(m)} k_i \; = \; \text{SqRing}(m)
\ee

\item The final goal is to express the chromatic polynomial in the form 
\be
  P_{m\times n}(q) \; = \; \sum\limits_{i=1}^{\text{SqRing}'(m)} 
                           \alpha_i(q) \, \lambda_i(q)^n 
  \label{def_P_eigen}
\ee
     where the $\{\lambda_i\}$ are the distinct eigenvalues of the transfer
     matrix $\mathsf{T}(m)$ and the $\{\alpha_i\}$ are some amplitudes we
     have to determine. The calculation of the amplitudes can be achieved by
     solving SqRing$'(m)$ linear equations of the type \reff{def_P_eigen}, 
     where the l.h.s.\ (i.e., the true chromatic polynomials) have been 
     obtained via Eq.~\reff{def_Pp} (i.e., by using the full transfer matrix 
     and vectors).

\end{enumerate}

We have used this procedure
to compute the transfer matrix up to width $m=6$. The next
case would involve the calculation of a symbolic matrix of dimension
SqRing$(m=7)=28940$, which is unmanageable given our computer facilities.

\bigskip

\noindent
{\bf Remark.} In the procedure described above one can obtain the eigenvalues
$\lambda_i(q)$ of the transfer matrix by considering the diagonal 
blocks $T_{\ell,\ell}$ [c.f.,~\reff{T_blocks1}].
However, we still need the full transfer matrix in order to determine the
amplitudes $\alpha_i(q)$.

%
%
\subsection{Simplified computation of the transfer matrix}
\label{sec.sq.cyclic.blocks2}

The computations can be simplified, and taken to larger strip widths, by
utilizing the following general property of the eigenvalue structure of
$\mathsf{T}(m)$:
\begin{enumerate}
 \item [1.] All distinct eigenvalues of the block $T_{\ell,\ell}$ have the
 same amplitude, henceforth denoted $\alpha^{(\ell)}$, which is given by
 Eq.~(\ref{def_alpha}) below.
\end{enumerate}
This property is an analytical result originating from the representation
theory of quantum groups \cite{Pasquier,Saleur_90_91}. Before reviewing the
arguments for it we shall pause to discuss its practical utility.

Property~1 implies that the chromatic polynomial can be written as
\be
 P_{m\times n}(q) \;=\; \sum\limits_{\ell=0}^m \alpha^{(\ell)} 
                        \sum\limits_{i=1}^{L_\ell}
                        \left(  \lambda_\ell^{(i)}  \right)^n
 \; ,
\label{P_sq_method_final}
\ee
where $L_\ell$ is the number of distinct eigenvalues in the block
$T_{\ell,\ell}$ (see Table~\ref{table_blocks_tri}). Note that
we have labeled the eigenvalues as $\lambda_{\ell}^{(i)}$, with
$\ell$ being the corresponding number of bridges.

One important consequence of Eq.~\reff{P_sq_method_final} is that in order to
obtain all the eigenvalues and amplitudes we do no longer need to compute the
full transfer matrix $\mathsf{T}(m)$. Rather, it is sufficient to diagonalize
the blocks $T_{\ell,\ell}$ for $\ell=0,1,\ldots,m$ to 
get the eigenvalues $\lambda_\ell^{(j)}$ ($j=1,\ldots,L_\ell$)
--- assuming, of course, that we can determine the amplitudes
$\alpha^{(\ell)}$ by some other method.

Property~1 and the exact expression of the amplitudes $\alpha^{(\ell)}$
follow from the presence of a quantum-group symmetry in the square-lattice
Potts model \cite{Pasquier,Saleur_90_91}. As a first step, the Potts model is
reformulated as a six-vertex model in a standard way \cite{Baxter_82}, the
cyclic boundary conditions being imposed by the insertion of an appropriate
twist operator \cite{Saleur_90_91}. It is then realized that the vertex model
transfer matrix $\mathsf{T}_\text{vertex}$ commutes with the generators of the
quantum algebra $\text{U}_{\bar{q}} \, \text{sl}(2)$
\cite{Pasquier,Saleur_90_91},
with deformation parameter $\bar{q}$ given by $\sqrt{q}=\bar{q}+1/\bar{q}$.
Decomposing the representation space into a direct sum over eigenspaces
corresponding to each value of the vertex-model spin (which in turn is
identified with $\ell$) gives Property~1, along with the expression%
\be
  \alpha^{(\ell)} \; = \; U_{2\ell} \left( \frac{\sqrt{q}}{2} \right) \; .
  \label{def_alpha}
\ee
where $U_n(x)$ is the Chebyshev polynomial of the second kind, defined by 
\cite{GR}, 
\be
  U_n(x) \;=\;
  \frac{(x + i \sqrt{1-x^2})^{n+1} - (x - i \sqrt{1-x^2})^{n+1}}
       {2i \sqrt{1-x^2}} \;=\;
  \sum_{j=0}^{\lfloor n/2 \rfloor} (-1)^j \binom{n-j}{j} (2x)^{n-2j}
  \label{def_Un}
\ee
The form \reff{def_alpha} has recently been rediscovered independently by
Chang and Shrock \cite{Shrock_01b}, using arguments based on Temperley--Lieb
algebra.  Though the reasoning in \cite{Pasquier,Saleur_90_91} and 
\cite{Shrock_01b} may not constitute fully mathematically rigorous proofs,
there is hardly any doubt that the results are correct.
We have also verified Property~1 explicitly for strip widths $m \le 6$.
The first coefficients $\alpha^{(\ell)}$ are given by 
\begin{subeqnarray}
 \alpha^{(0)} &=& 1 \\
 \alpha^{(1)} &=& q-1 \\
 \alpha^{(2)} &=& q^2 - 3 q + 1 \\
 \alpha^{(3)} &=& q^3 - 5 q^2 + 6q -1 \\
 \alpha^{(4)} &=& (q-1)(q^3 - 6q^2 + 9q -1) 
 \label{def_alphas}
\end{subeqnarray}
Another useful characterization of the amplitudes is the following 
\cite{Shrock_01b}:
\be
  \alpha^{(\ell)} \; = \; \prod\limits_{j=1}^\ell
  \left(q- B_{2\ell+1}^{(j)} \right) 
  \label{def_alpha_bis}
\ee
where $B_n^{(j)}$ is the generalized Beraha number\footnote{Note that
$q=B_n^{(j)}$ with $n,j$ integer corresponds to the quantum-group
deformation parameter $\bar{q}$ being a rational-order root of unity,
$\bar{q} = {\rm e}^{i \pi j/n}$.}
\be
 B_n^{(j)} \; = \; 4 \cos^2 \left( \frac{j\pi}{n} \right) =
 2 + 2 \cos \left( \frac{2\pi j}{n} \right)
\label{def_Bnj}
\ee
and $\ell$ ranges between $0$ and the strip width $m$. The standard
Beraha numbers are $B_n \equiv B_n^{(1)}$.

In the case of the Potts model at general temperature, and generic values of
$\bar{q}$, the representation theory for $\text{U}_{\bar{q}} \, \text{sl}(2)$
also fixes the dimension of the spin $\ell$ eigenspaces as 
$\binom{2m}{m-\ell} - \binom{2m}{m-\ell-1}$ 
\cite{Pasquier,Saleur_90_91}. This formula
should be compared with Table~II in \cite{Shrock_04}. Note however that this
is not equal to the number of distinct eigenvalues $L_\ell$ in the spin $\ell$
sector of \reff{P_sq_method_final}. There are two reasons for this: 1) we are
here considering the special case of the chromatic polynomial, for which many
eigenvalues vanish due to the non-nearest-neighbor constraint (see
Table~\ref{table_blocks_tri}), and 2) by symmetrizing with respect to
reflections of the strip we have paired eigenstates whose eigenvalues would
otherwise have been degenerate.

Let us stress that we have verified that Property~1 is in perfect agreement
with all our exact transfer matrices up to strip width $m=6$. Actually,
this empirical evidence for $2 \leq m \leq 6$ verifies a slightly stronger
property, which we state here as a conjecture:
\begin{enumerate}
 \item [2.] Two distinct blocks $T_{\ell,\ell}$ and $T_{\ell',\ell'}$
 (with $\ell \neq \ell'$), cf.~\reff{T_blocks1}, have no common
 eigenvalue. In other words, there does not exist $\ell,\ell',i,i'$ such
 that $\lambda^{(i)}_{\ell}=\lambda^{(i')}_{\ell'}$ and $\ell \neq \ell'$.
\end{enumerate}
Note that only Property~1 is necessary for
making our simplified computation of the transfer matrix work.
However, we have rather strong reasons for believing that Property~2
is actually true in general, since it essentially states that the combined
$\text{U}_{\bar{q}} \, \text{sl}(2)$ and reflection (in the case of the square
lattice) symmetries provides a {\em complete} identification of all the
symmetries in the problem. We leave the question whether Property~2 can
be proved by quantum-group techniques aside for a separate publication.
Note that Property~2 with Eq.~\reff{P_sq_method_final} would imply that
$\sum\limits_{\ell=0}^m L_\ell = \text{SqRing}'(m)$,
cf.~Eqs.~\reff{formula_SqRing}--\reff{formula_SqRing_alt}. And also note that
Property~2 with Eq.~\reff{def_alpha_bis} and the Beraha--Kahane--Weiss theorem
\cite{BKW} would imply that 
the only possible real or complex
isolated limiting points for cyclic strips of the triangular and
square lattices are odd-$n$ generalized Beraha numbers.\footnote{ 
  The position of the isolated limiting points depends on the boundary
  conditions of the strips. For square-lattice strips with free and 
  cylindrical boundary conditions \protect\cite{transfer1,transfer2} 
  we have found that (at least up to widths $m=12$) all Beraha numbers 
  up to $B_6=3$ (and only those) can be isolated limiting points. Furthermore, 
  we have found complex-conjugate pairs of complex isolated limiting points. 
}
The fact that we observe exactly this structure of isolated limiting points
also for $m=7$ lends further credibility to Property~2.

{}From the point of view of computing the transfer matrix, further efficiency
is achieved by making a further empirical observation (based on the 
strips of widths $2\leq m \leq 6$): 
\begin{enumerate}
 \item [3.] Let $T_{\ell,\ell,j_0}$ be any sub-block of $T_{\ell,\ell}$,
 cf.~\reff{T_blocks2}, whose dimension is the largest possible (there may
 be more than one such sub-block). Then the dimension of $T_{\ell,\ell,j_0}$
 is exactly $L_\ell$, and all of its eigenvalues are distinct.
\end{enumerate}
The dimensions of these sub-blocks are given by the $L_\ell$ in 
Table~\ref{table_blocks_sq}. 

The characterization of the sub-blocks $T_{\ell,\ell,j_0}$ with maximal
dimension is interesting from a computational point of view. In particular,
we have found that, for a given value of the number of bridges $\ell$, the
sub-blocks $T_{\ell,\ell,j_0}$ split into two classes depending on 
whether the bottom-row connectivity $\mathcal{P}_\text{bottom}$  and the
relative position of the $\ell$ bridges is symmetric or not under 
reflections. Let us denote those classes as $\mathcal{C}_\text{sym}$ and 
$\mathcal{C}_\text{non-sym}$, respectively. Then, we have empirically 
found for $2\leq m \leq 8$ that: 
a) All blocks belonging to class $\mathcal{C}_\text{sym}$ have the same 
dimension $L_\text{sym}$; 
b) All blocks belonging to class $\mathcal{C}_\text{non-sym}$ have the same 
dimension $L_\text{non-sym}$; and 
c) $L_\text{non-sym} > L_\text{sym}$.  
Thus, any block belonging to the non-symmetric class 
$\mathcal{C}_\text{non-sym}$ is a maximal one with $L_\text{non-sym}=L_\ell$
(see Table~\ref{table_blocks_tri}). The only exception to this rule occurs
when the non-symmetric class is empty $\mathcal{C}_\text{non-sym}=\emptyset$.
In this particular case, any block belonging to $\mathcal{C}_\text{sym}$ is
a maximal one with $L_\ell = L_\text{sym}$. 
For instance, for a strip of width $m \geq 3$, we can always choose as 
maximal blocks with $k=1,\ldots,m$ bridges those blocks characterized by 
$\mathcal{P}_\text{bottom}= 1$ (i.e., no connections among the sites of the
bottom rim) and bridges located at sites $1',\ldots,k'$. Finally, we can 
choose as maximal block with no bridges ($\ell=0$) the block with
$\mathcal{P}_\text{bottom}= \delta_{1',3'}$. 

Note that Property~3 is not very surprising at all when one recalls that
$j_0$ labels the bottom-row connectivities. Since the transfer matrix acts
on top-row connectivities only, one would not expect its spectrum to
depend on $j_0$ (apart from the reflection parity effect observed above).
To save space we refrain from turning this idea into a formal proof.

Supposing Property~3 to be true in general, it suffices to diagonalize one
maximal sub-block for each $\ell=0,1,\ldots,m$ (e.g., those described above).  
This method has allowed us to extend our results up to $m=8$.

An important question is whether this procedure is able to give {\em all}\/
the $\text{SqRing}'(m)$ distinct eigenvalues of the transfer matrix $\T(m)$. 
For the strips of widths $2\le m \le 6$, we have explicitly verified that 
this is the case, as we can handle the full transfer matrix 
(See Section~\ref{sec.sq.cyclic.blocks}). For widths $m=7,8$, the situation
is apparently less clear. In Table~\ref{table_blocks_tri} we have displayed 
the dimension $L_\ell(m)$ of the maximal block $\T_{\ell,\ell,j_0}(m)$ 
for a strip of width $m$ with $\ell$ bridges. These numbers fully agree 
with the {\em exact}\/ number of distinct eigenvalues $n_P(m,\ell)$ in the
corresponding $\ell$-bridge subspace found by Chang and Shrock 
\cite[Table~1]{Shrock_01b}. Thus, we can guarantee that for $m=7,8$ we
have not missed any transfer-matrix eigenvalue, and that our results are 
indeed exact. Furthermore, the total number of distinct eigenvalues we have
found is equal to the predicted number $\text{SqFree}'(m)$ \cite{Shrock_01b}.

\medskip

\noindent
{\bf Remark.} We stress that the results for widths $2\le m\le 6$ have been 
obtained completely {\em ab initio}, using the practical procedure 
described in Section~\ref{sec.sq.cyclic.blocks}. Thus, for these strips, 
we have not used any unproven property.
For the strips of widths $m=7,8$, we have used the simplified method
described above, which makes use of Properties~1 and~3. Property~2 is actually
not needed and used in our practical procedure. 

\medskip

%
%
\section{Square-lattice chromatic polynomials with cyclic boundary conditions}
\label{sec.sq.chromatic}

In this section we will analyze the results for the square-lattice strips of
widths $2\leq L \leq 8$. Even though the two smaller strips ($L=2,3$)
are well known in the literature, we will describe them in detail using our
approach. The other cases published in the literature $L=4,5$ will
be briefly reviewed for completeness. Finally, our new results will
be described in detail. A summary of the main characteristics of the 
limiting curves for these strips is displayed in Table~\ref{table_summary_sq}. 
Finally, in Table~\ref{zeros_sq_R} we list the real chromatic zeros
for square-lattice strips of size $L_{\rm F} \times (kL)_{\rm P}$ with 
$1\leq k\leq 10$.

For the larger $L$, we have not attempted to write the complete analytic
expressions for the eigenvalues and amplitudes in this paper, as they are
rather lengthy. The interested reader can find them in the {\sc mathematica}
file {\tt transfer4\_sq.m} available as part of the electronic version of this
paper in the {\tt cond-mat} archive. The analytic expressions for the
sub-blocks $T_{\ell,\ell,j_0}$ can be obtained by request from the authors. 

%
%
\subsection{$\bm{L=2}$}

This case was first solved by Biggs, Damerell and Sands back in 1972 
\cite{Biggs_72}. Their solution is 
\be
P_{2\times n}(q) \; = \; (q^2 -3q +3)^n + 
                         (q-1)\left[ (3-q)^n + (1-q)^n\right] + 
                         q^2 - 3q +1 
\label{Psq2}
\ee

Let us review our derivation of this result.
The transfer matrix $\mathsf{T}(2)$ has dimension four. In the basis 
$\{ \delta_{1,1'}\delta_{2,2'}, \delta_{1,1'}+\delta_{2,2'},
    \delta_{1,2'}+\delta_{2,1'}, 1 \}$, it takes the form
\be
\mathsf{T}(2) \; = \; \left( \begin{array}{c|cc|c}
 1   & 0      & 0      & 0 \\
\hline 
-1   & 2-q    & 1      & 0 \\ 
 0   & 1      & 2-q    & 0 \\ 
\hline 
 1   & 2(q-2) & 2(q-2) & q^2 -3q + 3\\ 
            \end{array}\right)
\label{Tsq2R}
\ee
where we by vertical and horizontal lines show the block structure of this
matrix (see below). We have ordered the basis elements
according to a
decreasing number of bridges so that the matrix \reff{Tsq2R} shows clearly
its lower-triangular block form. 
The right $\bm{w}_\text{id}$ and left $\bm{u}$ vectors are given by 
\begin{subeqnarray}
  \bm{u}^T          &=& (A,2A,0,A)\,, \qquad \text{with $A=q (q-1)$}\\ 
 \bm{w}_\text{id}^T &=& (1,0,0,0)  
 \label{UVsq2R}
\end{subeqnarray} 

The transfer matrix \reff{Tsq2R} can be decomposed into three blocks: 
all of them are characterized by the same bottom-row connectivity 
($\mathcal{P}_\text{bottom}=1$, i.e., there is no connection between the 
sites $1'$ and $2'$). 
In this case, the difference between blocks is solely due to the number 
of bridges $\ell$ connecting the top and bottom rows. 

The first block is one-dimensional and corresponds to the connectivity 
state $\delta_{1,1'}\delta_{2,2'}$ (thus, $\ell=2$). The eigenvalue is 
\be
\lambda_{\ell=2}  \;=\; 1 
\ee
The third block is also one-dimensional and it corresponds to the singleton
state $1$. Thus, it is characterized by zero bridges ($\ell =0$). The 
eigenvalue is 
\be
\lambda_{\ell=0} \;=\; q^2 - 3q + 3
\ee
Finally, the second block corresponds to the two remaining states, and it
is characterized by a single bridge ($\ell=1$). The eigenvalues of this 
block are
\begin{subeqnarray}
 \lambda_{\ell=1}^{(1)} &=& 3-q \\
 \lambda_{\ell=1}^{(2)} &=& 1-q 
\end{subeqnarray} 
The amplitudes associated to those eigenvalues are given by 
[c.f.~\reff{def_alphas}] 
\begin{subeqnarray}
 \alpha_{\ell = 2}       &=& \alpha^{(2)} \\ 
 \alpha_{\ell = 1}^{(j)} &=& \alpha^{(1)} \;, \quad j=1,2\\ 
 \alpha_{\ell = 0}       &=& \alpha^{(0)} = 1 
\end{subeqnarray}
where the $\alpha^{(j)}$ are given by \reff{def_alpha}, 
in full agreement with the exact result \reff{Psq2}. As discussed in 
Section~\ref{sec.sq.cyclic.blocks2}, the amplitude $\alpha^{(\ell)}$ is 
associated to every eigenvalue $\lambda_\ell^{(i)}$
coming from the block $T_{\ell,\ell}$. 

We have depicted the chromatic zeros for the square-lattice strips of width 
$L=2_\text{F}$ and lengths $n=10_\text{P},20_\text{P}$ in 
Figure~\ref{figure_sq_1}(a). We have also depicted the limiting curve 
$\mathcal{B}$ for this strip.\footnote{  
  The limiting curve for a square-lattice strip of width $L=2$ and 
  cyclic boundary conditions was first computed by Biggs, Damerell and Sands 
  \cite{Biggs_72}. We include it here to compare it to the new results
  presented in this paper.
}  
This curve was computed using the resultant method (see, e.g.,
\cite{transfer1}). It divides the complex $q$-plane into four regions, and
crosses the real $q$-axis at $q=0,2$. At this latter point it has a quadruple
point. There are also two T points at $q=2\pm \sqrt{2}$.

The isolated limiting points are very easy to locate as the amplitudes have
a simple analytic form \reff{def_alpha_bis}: there is a single isolated
limiting point at $q=1$.
  
%
%
\subsection{$\bm{L=3}$}

The solution for this strip was first found by Shrock and Tsai in 2000 
\cite{Shrock_00a}:
\begin{eqnarray}
P_{3\times n}(q) &=& (q^3 - 5 q^2 + 6 q - 1)(-1)^n +
                     (q^2 -3 q + 1)\left[ 
                         (q - 1)^n + (q - 2)^n + (q - 4)^n \right] \nonumber\\
                 & & \quad + 
                     (q - 1)\left[ \left(- (q - 2)^2\right)^n + 
                            \sum\limits_{i=6}^8 \lambda_i^n 
                            \right] + 
                     \sum\limits_{i=9}^{10} \lambda_i^n
\label{def_Poly_3R}
\end{eqnarray}
where the $\lambda_i$ for $i=9,10$ are the solutions of the equation
\be
x^2 - (q - 2)(q^2 - 3 q + 5) x + (q-1)(q^3 - 6q^2 +13q -11) \; =\;  0
\label{eq_3R_1}
\ee
and those for $i=6,7,8$ are the solutions of the third-order equation 
\begin{eqnarray}
x^3 &+& (2 q^2 - 9 q + 12) x^2 + (q^4 - 10 q^3 + 36 q^2 - 56 q + 31) x \; =
  \nonumber \\
  & & \qquad (q-1)(q^4 - 9 q^3 + 29 q^2 - 40 q + 22)
\label{eq_3R_2}
\end{eqnarray} 

We have re-derived the above result using our method. The transfer matrix is 
20-dimensional, and can be split into eight blocks. The diagonalization of
these blocks lead to the eigenvalues given in \reff{def_Poly_3R}. 
We have also confirmed that the amplitudes are precisely those given by 
\reff{def_Poly_3R}.  

There is a single block for $\ell=3$ bridges. Its eigenvalue is 
$\lambda_{\ell=3}=-1$ and its amplitude is $\alpha^{(3)}$. 
We also find two blocks with $\ell=2$. The corresponding eigenvalues are 
$\lambda_{\ell=2}^{(j)} = q-4,q-2,q-1$, and their amplitude is $\alpha^{(2)}$.  
There are three blocks with a single bridge ($\ell=1$). The eigenvalues are 
$\lambda_{\ell=1}^{(1)}=-(q-2)^2$ and $\lambda_{\ell=1}^{(j=2,3,4)}$ are 
the solutions of the third-order equation \reff{eq_3R_2}. The amplitude
is $\alpha^{(1)}$ for these four eigenvalues. Finally, there are two blocks
with no bridges $\ell=0$; their eigenvalues are the solutions of the 
second-order equation \reff{eq_3R_1}, and $\alpha^{(0)}$ is their amplitude.
 
The chromatic zeros for the square-lattice strips of width
$L=3_\text{F}$ and lengths $n=15_\text{P},30_\text{P}$ are displayed in
Figure~\ref{figure_sq_1}(b). The limiting curve $\mathcal{B}$ (computed with 
the resultant method) is also depicted in that figure. This curve was
first obtained by Shrock and Tsai \cite{Shrock_00a}.
 
The limiting curve crosses the real $q$-axis at $q=0,2$ and 
$q\approx 2.3365442725$. We find ten T points at 
$q\approx 1.7713445981 \pm  1.7900810556\, i$, 
$q\approx 2.0457165736 \pm  1.6008334033\, i$, 
$q\approx 2.2701797354 \pm  0.9704300675\, i$, 
$q\approx 2.0401211769 \pm  1.6256019349\, i$, and  
$q\approx 2.3527863293 \pm  0.3787883673\, i$. 
We also find a single isolated limiting point at $q=1$.

%
%
\subsection{$\bm{L=4}$ and $\bm{L=5}$}

The solution for $L=4$  was obtained by Chang and Shrock \cite{Shrock_01a}.
We have re-derived their result by using our approach. The
transfer matrix has dimension 94, and it can be split into 16 blocks.
After obtaining the eigenvalues of all such blocks, we arrive at the
26 distinct eigenvalues and amplitudes found by Chang and Shrock 
\cite{Shrock_01a}. These eigenvalues and amplitudes have the structure 
described in Section~\ref{sec.sq.cyclic.blocks2}. 

\begin{figure}
\centering
\begin{tabular}{cc}
   \includegraphics[width=200pt]{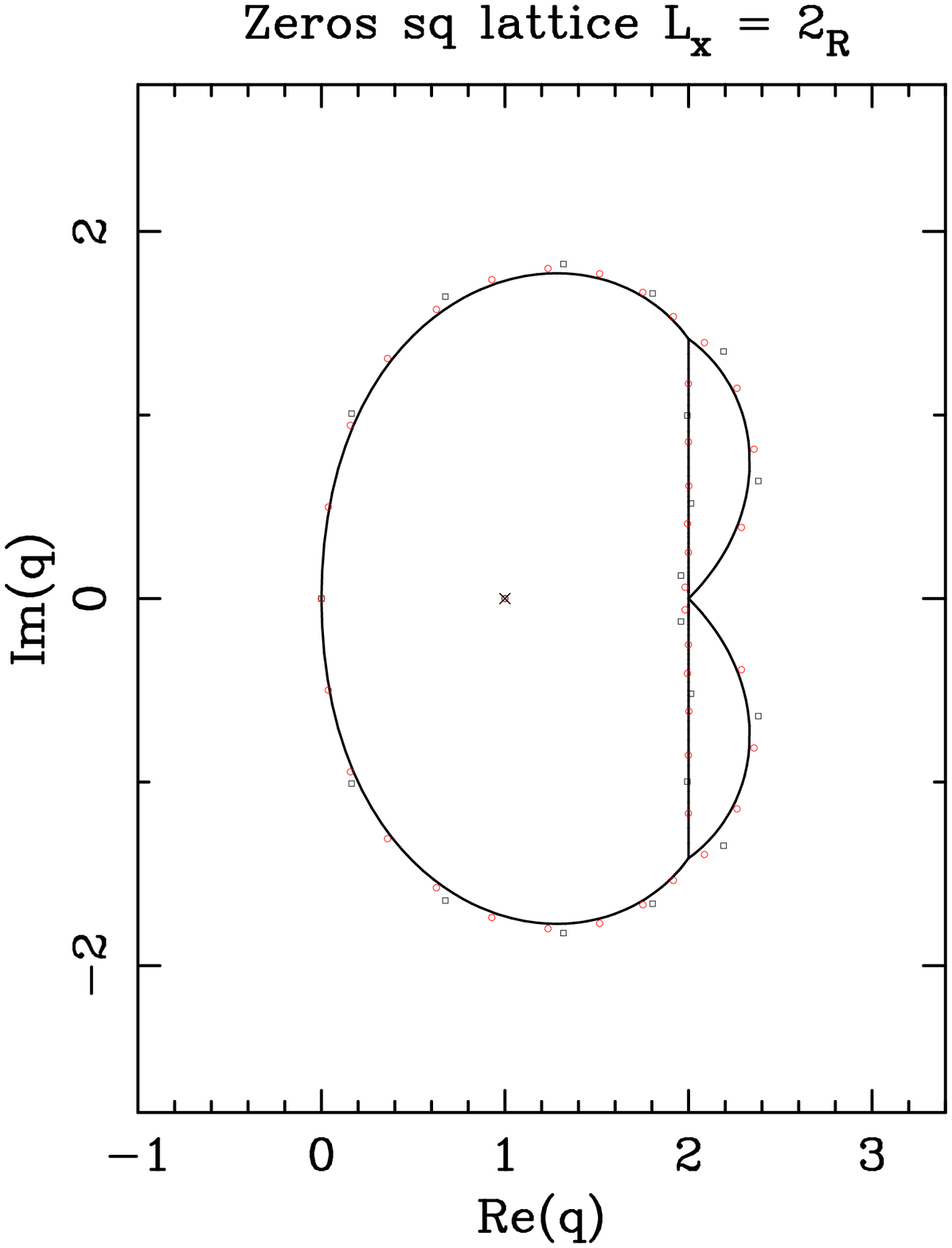} &
   \includegraphics[width=200pt]{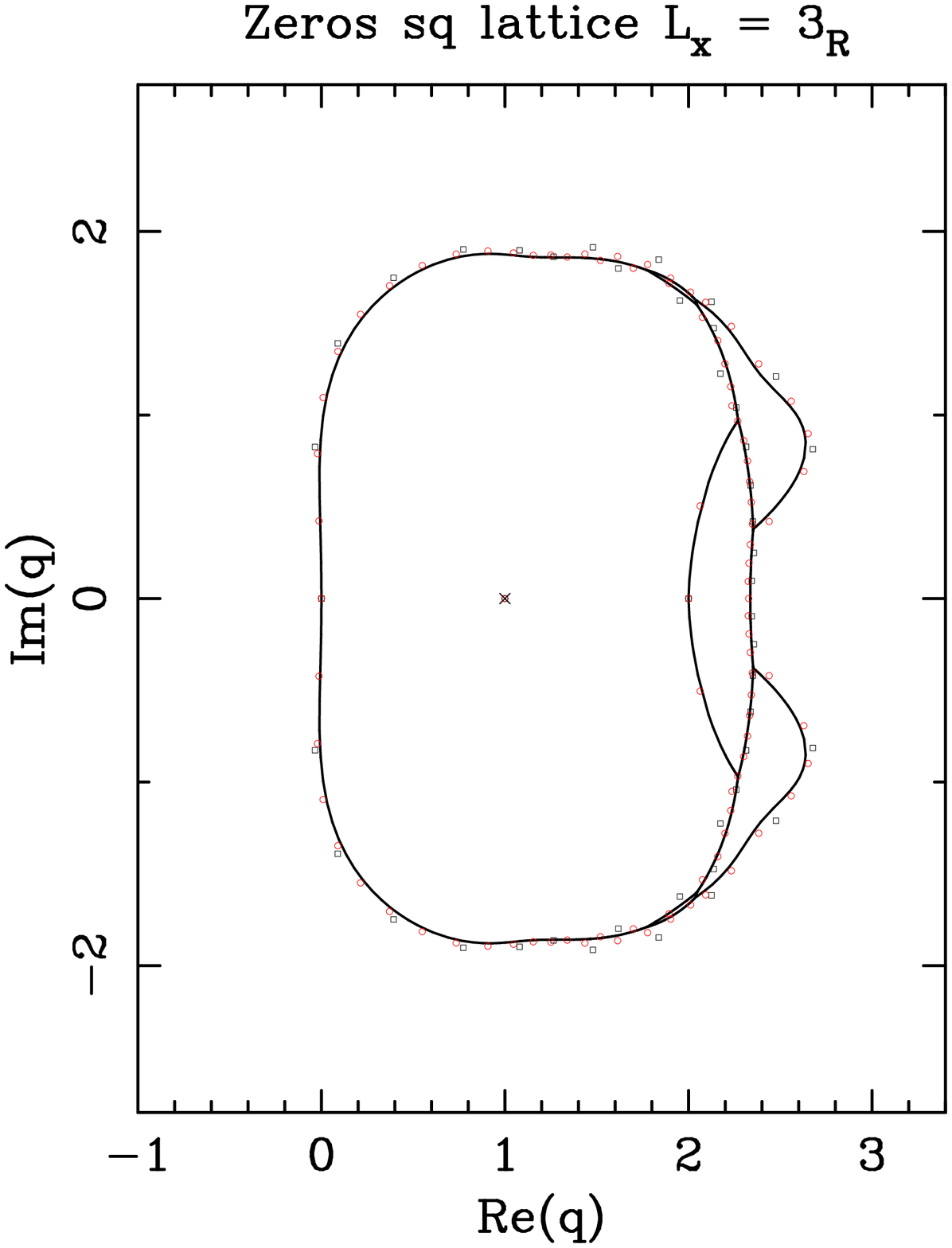}
   \\[1mm]
   \phantom{(((a)}(a)    & \phantom{(((a)}(b) \\[5mm]
   \includegraphics[width=200pt]{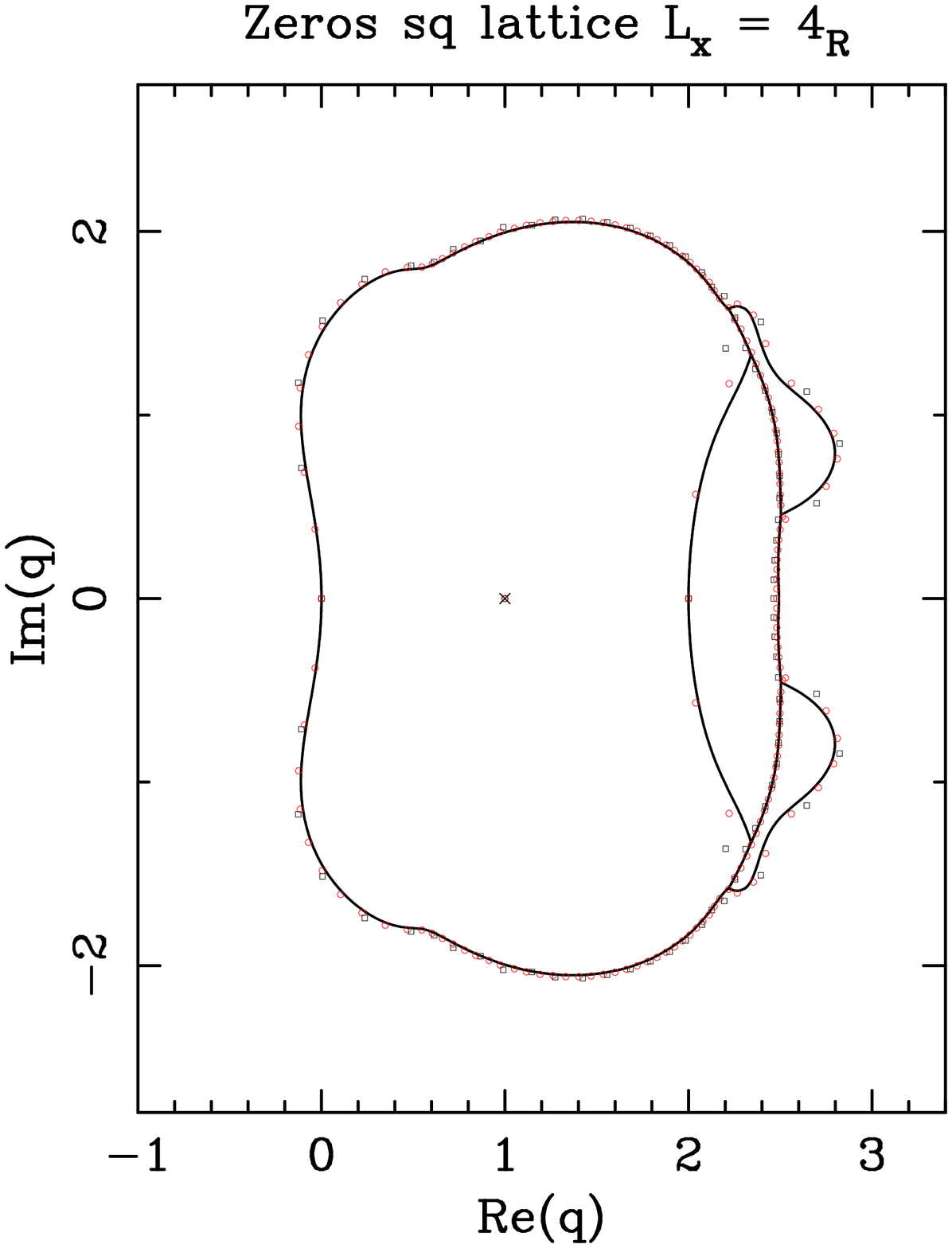} &  
   \includegraphics[width=200pt]{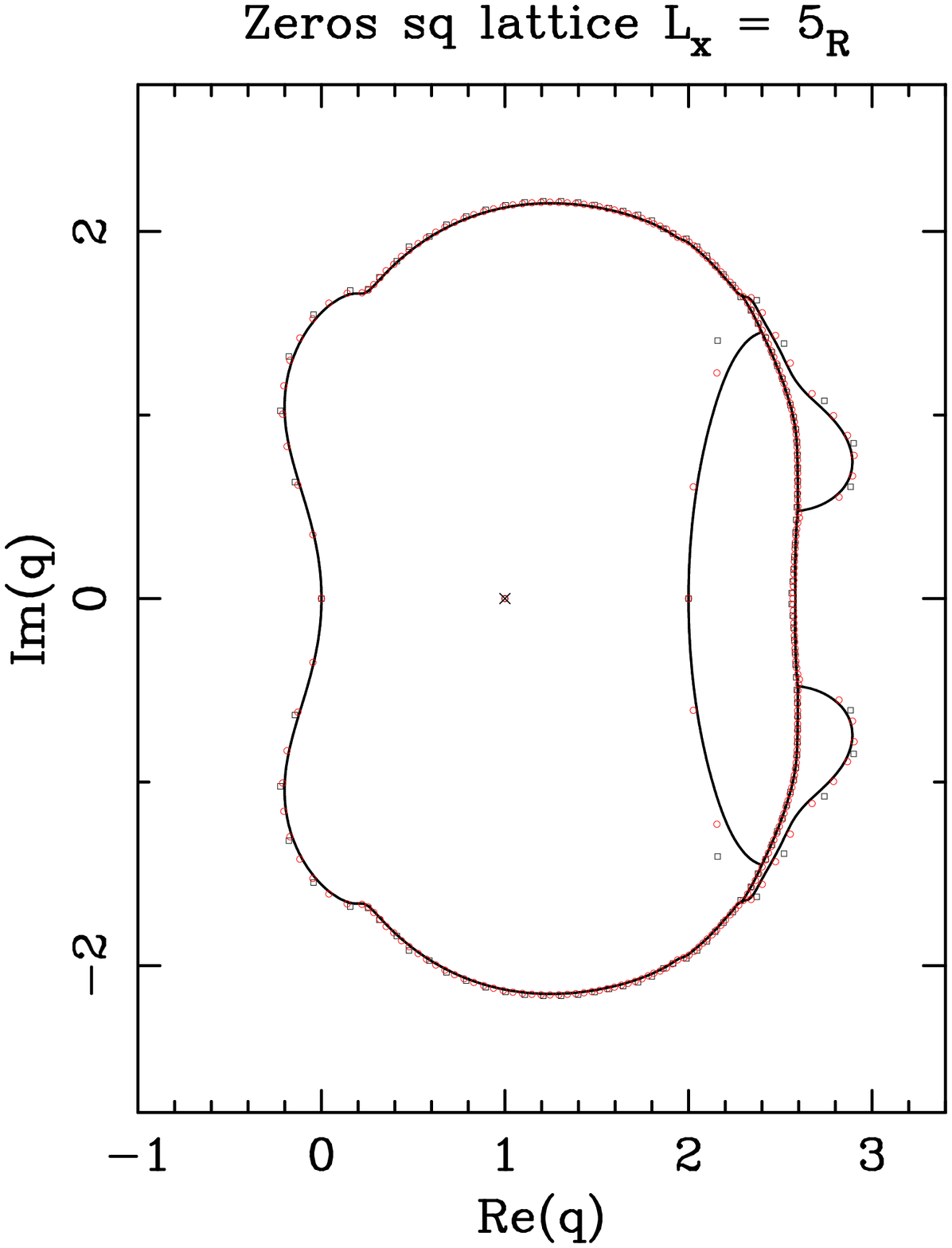} 
   \\[1mm]  
   \phantom{(((a)}(c)    & \phantom{(((a)}(d) \\
\end{tabular}
\caption{\label{figure_sq_1}
Limiting curves for square-lattice strips of width (a) $L=2$, (b) $L=3$,
(c) $L=4$, and (d) $L=5$  with cyclic boundary conditions. We also show the
zeros for the strips $L_\text{F} \times (5L)_\text{P}$ (black $\Box$) and
$L_\text{F} \times (10L)_\text{P}$ (red $\circ$) for the same values of $L$.
}
\end{figure}

In Figure~\ref{figure_sq_1}(c) we show the chromatic zeros for strips of sizes
$4_\text{F}\times 20_\text{P}$ and $4_\text{F}\times 40_\text{P}$, as well
as the limiting curve $\mathcal{B}$.\footnote{
  Although the limiting curve for this lattice was not included in the 
  original reference by Chang and Shrock \protect\cite{Shrock_01a}, it 
  appeared in the subsequent paper 
  \protect\cite{Shrock_02a} by the same authors. 
} 
The computation of this curve was done by means of the direct-search method
(see e.g. \cite{transfer1}). Thus, the number of relevant features 
reported below should be regarded as lower bounds. 
The limiting curve ${\cal B}$ crosses the real $q$-axis at three points: 
$q=0,2$, and $q\approx 2.4928455591$. We have found six T points at 
$q \approx 2.2192166698 \pm 1.5778851486\, i$, 
$q \approx 2.3408741083 \pm 1.3290891386\, i$, and 
$q \approx 2.5035997660 \pm 0.4577795448\, i$. 
As in the previous strips, there is a single isolated limiting point at $q=1$.

%
%
The solution for $L=5$ was found by Chang and Shrock \cite{Shrock_02a}.
The full transfer matrix has dimension 614, and it can be split into 
44 blocks. From them we have obtained the 70 distinct eigenvalues and
their associated amplitudes. Again we find the eigenvalue and amplitude
structure discussed in Section~\ref{sec.sq.cyclic.blocks2}.  

In Figure~\ref{figure_sq_1}(d) we depict the chromatic zeros for the 
strips of length $n=25_\text{P},50_\text{F}$. We also plot the limiting
curve $\mathcal{B}$, which was computed using the direct-search 
method.\footnote{
  This limiting curve was first obtained by Chang and Shrock 
   \protect\cite{Shrock_02a}.
}
This curve $\mathcal{B}$ crosses the real $q$-axis at $q=0,2$ and 
$q\approx  2.5823854661$. 
We have found six T points at
$q \approx 2.3986403427 \pm 1.4510688289\, i$,
$q \approx 2.5948107219 \pm 0.4767225937\, i$, and 
$q \approx 2.2911897201 \pm 1.6493441205\, i$.
Finally, there is a single isolated limiting point at $q=1$.

%
%
\subsection{$\bm{L=6}$}

The full transfer matrix $\mathsf{T}(6)$ has dimension 4028, and can be 
split into 105 blocks. After performing the diagonalization of these blocks, 
we have found 192 distinct eigenvalues (in agreement with 
Eq.~\reff{formula_SqRing}). The solution can be written in the form
\be
P_{6\times n}(q) \; = \; \sum\limits_{i=1}^{192} \alpha_i(q) \lambda_i(q)^n
\label{def_Poly_6R}
\ee

We find several simple eigenvalues: e.g. $\lambda=1$, 
$\lambda=4-q,3-q,2-q,1-q$. The other eigenvalues come from solving polynomic
equations up to order 30. We find the same structure of the eigenvalues and
amplitudes in terms of the number of bridges $\ell$ as for the other strips. 

In Figure~\ref{figure_sq_2}(a) we show the chromatic zeros for the lengths 
$30_\text{P}$ and $60_\text{P}$, and the limiting curve $\mathcal{B}$. This
curve has been computed using the direct-search method. Thus, the number
of T points and the other features discussed below should be taken as lower
bounds of the true values. 

The limiting curve ${\cal B}$ crosses the real $q$-axis at $q=0,2$ and
$q = q_0 \approx 2.6460783059$. 
We have found 10 T points at
$q \approx 2.3494422185 \pm  1.6759458630\, i$, 
$q \approx 2.3716213974 \pm  1.6415616784\, i$, 
$q \approx 2.4271173045 \pm  1.5496869588\, i$,   
$q \approx 2.4988088784 \pm  1.4046791200\, i$, and  
$q \approx 2.6565980229 \pm  0.4807158507\, i$. 
There are two small triangular-like regions delimited by the six T points 
around $q \approx 2.37 \pm  1.64\, i$. In total, there are six enclosed regions
in the complex $q$-plane; only two of them have non-void intersection with
the real $q$-axis.

There are two isolated limiting points at $q=1$ and $q=B_5$, as 
the value of $q_0$ is larger than $B_5$. This is the smallest width for 
which we find two isolated limiting points among the square-lattice 
strips considered in this paper. 

We remark that the limiting curve crosses the real $q$-axis 
at $q_0 \approx 2.6460783059 > B_5$. Thus, in principle there could be real
zeros larger than $q=B_5$ (See Table~\ref{zeros_sq_R}).\footnote{
  In refs.~\protect\cite{transfer1,transfer2} it was found that for 
  square-lattice strips with cylindrical boundary conditions and even width
  $L=8,10,12$, the limiting curve $\mathcal{B}$ does not cross the real
  $q$-axis. Indeed, for $L=4,6$, $\mathcal{B}$ crosses the real $q$-axis
  at $q_0 < B_5$. No chromatic real root greater than $B_5$ was found.
}
However, we computed the zeros of $P_{6 \times n}(q)$
up to $n=60$ and did not find any real zeros larger than $B_5$.
This point is relevant in relation to a conjecture by Woodall \cite{Woodall_97},
stating that no bipartite planar graph can have a real chromatic zero 
greater than $q=2$. This conjecture was disproved in ref.~\cite{transfer1}, 
and modified in the following way: for any bipartite planar graph $G$, 
$P_G(q) >0$ for all real $q\geq B_5$ \cite[Conjecture~7.5]{transfer1}.  

%
%
\subsection{$\bm{L=7}$}

The full transfer matrix $\mathsf{T}(7)$ has dimension 28940, and can be
split into 294 blocks (see Table~\ref{table_blocks_sq}). The 
computation of such an enormous symbolic matrix is beyond our current computer 
facilities. However, 
using the simplified method described in Section~\ref{sec.sq.cyclic.blocks2},
we have been able to obtain the diagonal blocks $T_{\ell,\ell,j_0}$ (i.e.,
for each number of bridges $\ell=0,\ldots,7$, we choose among all possible
sub-blocks of $T_{\ell,\ell}$ one with the largest possible dimension given
by $L_\ell$ in Table~\ref{table_blocks_tri}). 

The blocks $T_{\ell,\ell,j_0}$ are themselves large: their dimensions 
range from 1 to 145. This fact severely limits the applicability 
of the method: we have been 
unable to compute the characteristic polynomial of any block of dimension 
larger than 100. Thus, with our current computer facilities, we cannot 
obtain closed formulas for the eigenvalues (as in the preceding cases). 
This implies that we cannot compute the chromatic polynomial for this
strip family. However, we are able to 
compute the limiting curve (which depends only on the eigenvalues provided
none of the amplitudes $\alpha_k$ vanishes identically) 
by the direct-search method combined with {\em numerical}\/
determination of the eigenvalues at any given $q$.

\begin{figure}
\centering
\begin{tabular}{cc}
   \includegraphics[width=200pt]{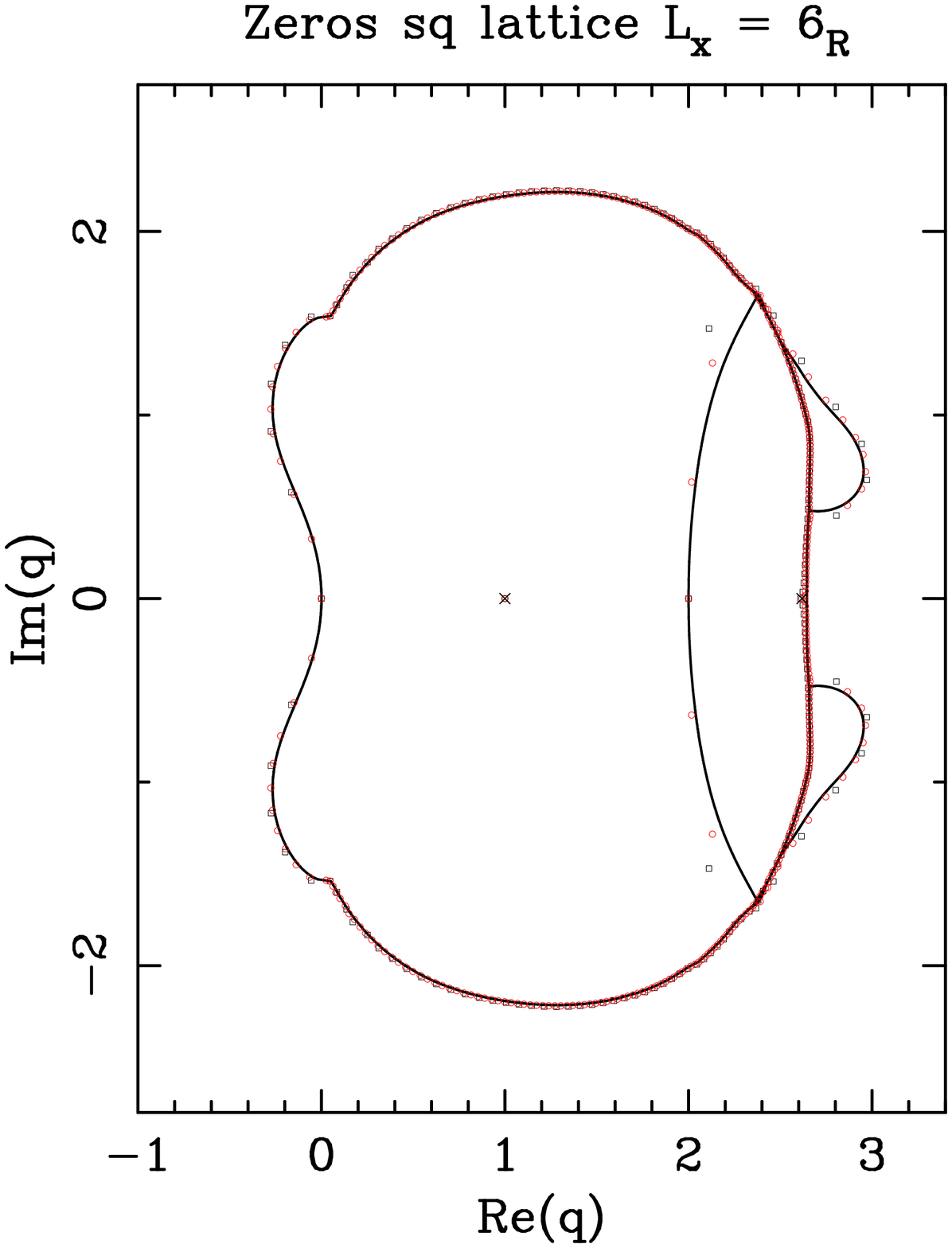} &
   \includegraphics[width=200pt]{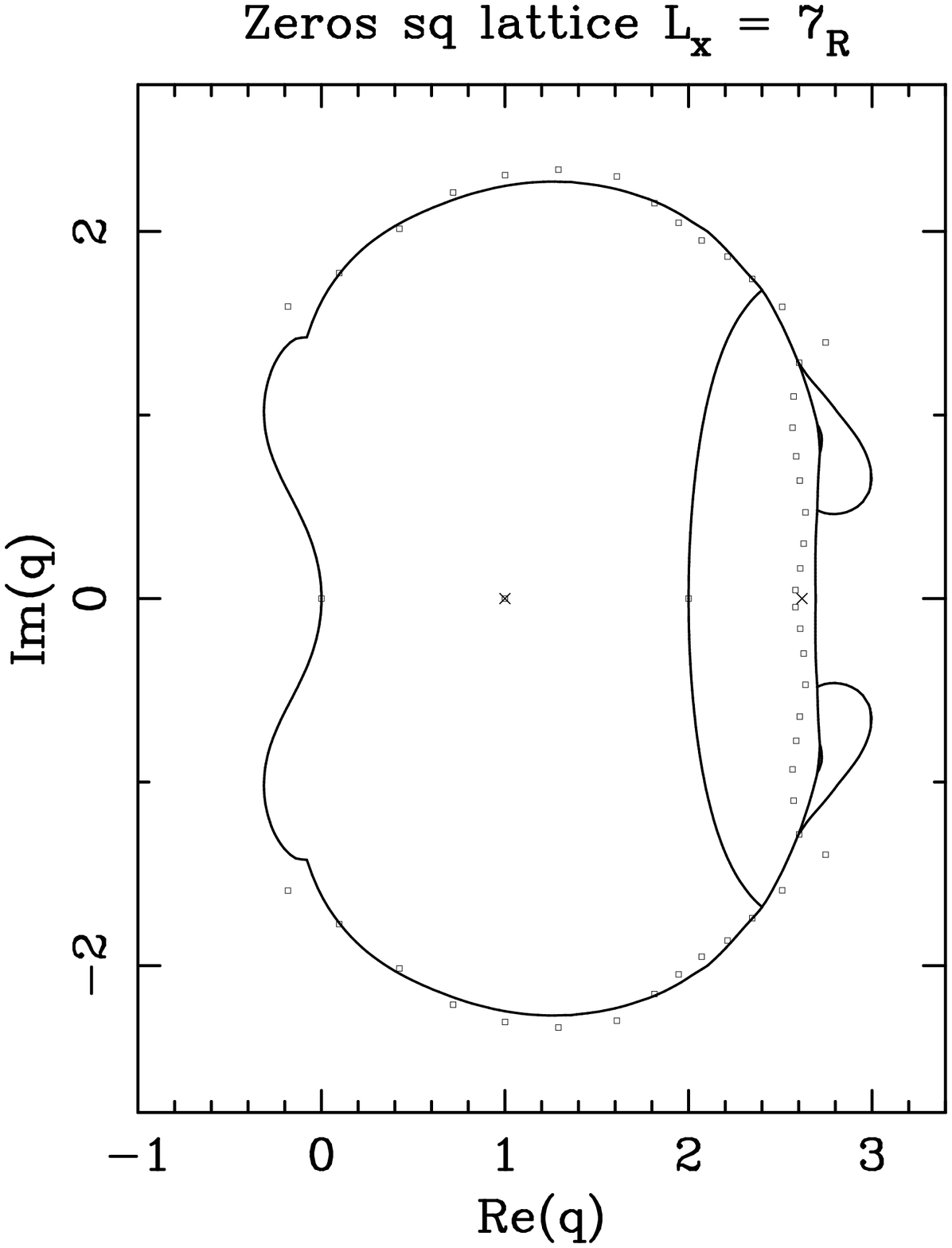}
   \\[1mm]
   \phantom{(((a)}(a)    & \phantom{(((a)}(b)\\
\end{tabular}
\caption{\label{figure_sq_2}
Limiting curves for square-lattice strips of width (a) $L=6$, and (b) $L=7$ 
with cyclic boundary conditions. In (a) we show the
zeros for the strips $6_\text{F} \times 30_\text{P}$ (black $\Box$) and
$6_\text{F} \times 60_\text{P}$ (red $\circ$). 
In (b) we show the chromatic zeros
for the strip $7_\text{F} \times 7_\text{P}$ (black $\Box$). 
}
\end{figure}

In Figure~\ref{figure_sq_2}(b) we show the zeros for a square-lattice strip
of size $7_\text{F} \times 7_\text{P}$ (obtained using Eq.~\reff{eq_check_Z} 
and previous results for cylindrical boundary conditions 
\cite{transfer1}). 
We also show the limiting curve $\mathcal{B}$; it  
has been computed using the direct-search method. The computation of this
curve is more involved than for the previous cases, as we do not have the
chromatic zeros to guide our search. Thus, the number 
of T points and the other features discussed below should be taken as lower
bounds of the true values. 

The limiting curve ${\cal B}$ crosses the real $q$-axis at $q=0,2$ and
$q = q_0 \approx 2.6927280047$. We have found 14 T points at
$q \approx  2.4008837606 \pm 1.6805209516\, i$,      
$q \approx  2.5993953677 \pm 1.2843636999\, i$,      
$q \approx  2.7015691600 \pm 0.4815451956\, i$,      
$q \approx  2.7022195651 \pm 0.9476253694\, i$,      
$q \approx  2.7126320142 \pm 0.8591968848\, i$,      
$q \approx  2.7152854276 \pm 0.7936937430\, i$, and  
$q \approx  2.7248651978 \pm 0.8452059951\, i$.      

There are two tiny complex-conjugate features of the limiting curve that are
not visible in Figure~\ref{figure_sq_2}(b). Around $q\approx 2.71 \pm 0.85$,
we find two closed regions delimited by the T points
$q \approx  2.7022195651 \pm 0.9476253694\, i$  and
$q \approx  2.7152854276 \pm 0.7936937430\, i$. These regions are split in
two by small lines connecting the T points
$q \approx  2.7248651978 \pm 0.8452059951\, i$ and
$q \approx  2.7126320142 \pm 0.8591968848\, i$. Thus, we
find four small closed regions delimited by the latter four T points
listed above.

There are two isolated limiting points at $q=1$ and $q=B_5$, as 
the value of $q_0$ is larger than $B_5$.

%
%
\subsection{$\bm{L=8}$}

The full transfer matrix $\mathsf{T}(8)$ has dimension 212370, and can be
split into 777 blocks (see Table~\ref{table_blocks_sq}). We have been unable
to compute this huge symbolic matrix. By using the method 
described in Section~\ref{sec.sq.cyclic.blocks2},
we have been able to obtain the nine diagonal blocks $T_{\ell,\ell,j_0}$. 
Several blocks $T_{\ell,\ell,j_0}$ are still very large: their dimensions 
range from 1 to 385. Again, we cannot compute the symbolic 
characteristic polynomial of the largest sub-blocks, and hence, the
chromatic polynomial. But the numerical computation of the
eigenvalues is possible, but very time consuming. This is the reason we have 
not tried to determine the whole limiting curve $\mathcal{B}$.  
The only exception is the determination of the value of $q_0$ for this
strip family: $q_0 \approx 2.7287513178$. 
Finally, we expect two isolated limiting points at $q=1$ and $q=B_5$, as 
the value of $q_0$ is larger than $B_5$.

%
%

\section{Transfer matrix for triangular-lattice strips with cyclic boundary 
         conditions} \label{sec.tri.cyclic} 

\begin{figure}
   \setlength{\unitlength}{0.5pt}
   \centering
   \begin{picture}(300,300)
     \Thicklines
     \drawline(0,50) (300,50)
     \drawline(0,150)(300,150)
     \drawline(0,250)(300,250)

     \drawline(0,50)  (0,250)
     \drawline(100,50)(100,250)
     \drawline(200,50)(200,250)
     \drawline(300,50)(300,250)

     \drawline(0,150)(100,50)
     \drawline(0,250)(200,50)
     \drawline(100,250)(300,50)
     \drawline(200,250)(300,150)

     \put(0,50){\circle*{15}}
     \put(0,150){\circle*{15}}
     \put(0,250){\circle*{15}}
     \put(100,50){\circle*{15}}
     \put(100,150){\circle*{15}}
     \put(100,250){\circle*{15}}
     \put(200,50){\circle*{15}}
     \put(200,150){\circle*{15}}
     \put(200,250){\circle*{15}}
     \put(300,50){\circle*{15}}
     \put(300,150){\circle*{15}}
     \put(300,250){\circle*{15}}

     \put( -8, 10){\large \bf 1'}
     \put( 92, 10){\large \bf 2'}
     \put(192, 10){\large \bf 3'}
     \put(292, 10){\large \bf 4'}

     \put( -8,270){\large \bf 1}
     \put( 92,270){\large \bf 2}
     \put(192,270){\large \bf 3}
     \put(292,270){\large \bf 4}
   \end{picture}
   \caption{\label{figure_tri_cyclic_bc}
     Triangular lattice with cyclic boundary conditions of size $4\times 2$. 
     This lattice is obtained from a triangular lattice with free boundary 
     conditions of size $4\times 3$ by identifying the top and bottom rows 
     $i \leftrightarrow i'$ ($i=1,\ldots,4$). 
   }
\end{figure}
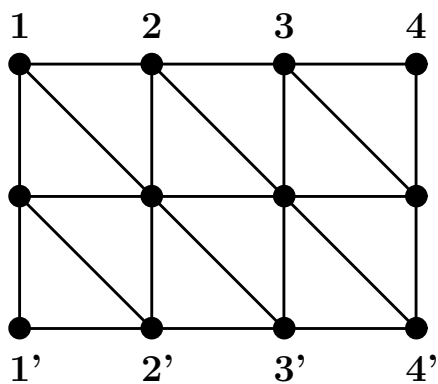

The results are very similar to those obtained for the square lattice
in Section~\ref{sec.sq.cyclic}. The transfer matrix is written as 
\be 
  \mathsf{T}(m) \; = \; \mathsf{V}(m) \cdot \mathsf{H}(m)
\ee
with the same formula for $\mathsf{H}(m)$ \reff{def_H}, but a different one 
for the ``vertical-bond'' matrix $\mathsf{V}(m)$:
\be
  \mathsf{V}(m) \; =  \; \P_m \cdot \QQ_{m,m-1} \cdot \P_{m-1} \cdot \ldots 
                             \cdot \QQ_{2,1} \cdot \P_1  
  \label{def_V_tri}
\ee
In Figure~\ref{figure_tri_cyclic_bc} we show the construction of a $4\times 2$
triangular-lattice strip. 

The chromatic polynomial is given by the same formulas as for the
square lattice \reff{def_P}/\reff{def_Pp}. Again we choose to work with
the connectivity basis 
$\widetilde{\bm{w}}_\mathcal{P} = 
\mathsf{H}(m)\cdot \mathsf{H}'(m) \cdot \bm{v}_\mathcal{P}$,
as $\widetilde{\bm{w}}_\mathcal{P}=0$ for all partitions $\mathcal{P}$ 
that contain
two nearest-neighbor sites (on either the top or the bottom rows) 
in the same block. This basis has dimension TriRing$(m)$ for a strip of
width $m$ (see Table~\ref{table_dim_T_cyclic}). 

There is no reflection symmetry for a triangular-lattice strip, so the
dimension of the transfer matrix is
TriRing$(m)$. However, the number of distinct eigenvalues should be the same
as for the square lattice, viz.\ SqRing$'(m)$ in
Table~\ref{table_dim_T_cyclic}, as the underlying $\text{U}_{\bar{q}} \,
\text{sl}(2)$ quantum-group symmetry is the same \cite{Saleur_90_91}. That
this is indeed the case has been checked explicitly by Chang and Shrock
\cite{Shrock_01b}.

Also in the triangular-lattice case, the quantum-group symmetry ensures that
the chromatic polynomial for a strip of size $m\times n$ can be written as in
\reff{P_sq_method_final}. Again, this means that only odd-$n$ generalized 
Beraha numbers can be isolated limiting points.\footnote{
   For triangular-lattice strips with free and cylindrical boundary conditions,
   we have found that {\em all} Beraha numbers up to $B_{10}$ are isolated 
   limiting points \cite{transfer3} (at least up to widths $m=12$).
   There are two different conjectures about how many Beraha numbers can 
   be isolated limiting points for this lattice. Baxter's results 
   \protect\cite{Baxter_86_87}) imply that all Beraha numbers up to 
   (and including) $B_{14}$ should be isolated limiting points for large enough
   widths $m$. The reanalysis of Baxter's solution \protect\cite{transfer3}
   led us to a new conjecture: only the Beraha numbers up to $B_{11}$ 
   (and possibly $B_{12}$) can be isolated limiting points.  
   In addition to these issues, we have found that for strips with 
   free and cylindrical there are no complex isolated limiting points up to 
   widths $m=12$. However, for ``zig-zag'' boundary conditions, we found
   complex isolated limiting points, and, more importantly, real isolated 
   limiting points that are {\em not} Beraha numbers (e.g., $q=5/2$).
} 

The block structure of $\mathsf{T}(m)$ is similar to that of the
square-lattice case 
\reff{T_blocks1}/\reff{T_blocks2}. The only difference is that the 
triangular lattice is not invariant under reflections with
respect to the center of the strip. Thus, for a given number of bridges $\ell$,
we expect that the number of blocks $N_\ell$ is larger than for the 
square case (see Table~\ref{table_blocks_tri}). Indeed,
for all $m \leq 8$, and for any $\ell$, we empirically find that
$N_\ell$ coincides with the number of distinct
eigenvalues $L_\ell$ coming from $T_{\ell,\ell}$, and that all the
$N_\ell$ blocks have the same dimension $L_\ell$
(see Table~\ref{table_blocks_tri}). 
Thus, we conclude that for $m\leq 8$
\be
\sum\limits_{\ell=0}^m L_\ell^2 \;=\; \text{TriRing}(m)
\ee
and we conjecture that this equation holds for arbitrary values of $m$.

Once again, the form \reff{P_sq_method_final} allows us to make computations
for widths $m>6$. Since all the sub-blocks $T_{\ell,\ell,j}$ are expected to
have the same dimension, we can choose any of them to perform the computations
using the simplified method described in Section~\ref{sec.sq.cyclic.blocks2}.
 
\bigskip

\noindent
{\bf Remarks}: 1. As for the square-lattice case, we have checked our results
with the help of the available results for triangular-lattice strips
with cylindrical boundary conditions [c.f., Eq.~\reff{eq_check_Z}].

2. As for the square-lattice case, the transfer matrices for the strips of 
widths $m\leq 6$ have been obtained without using any unproven property. 
For $m=7,8$ we have used Properties~1 and~3 of 
Section~\ref{sec.sq.cyclic.blocks2} to obtain {\em all}\/ the eigenvalues.
For each block $T_{\ell,\ell}$, we have obtained $L_\ell(m)$ distinct 
eigenvalues (See Table~\ref{table_blocks_tri}), where $L_\ell$ coincides with 
the predicted number of distinct eigenvalues for the sector with $\ell$ bridges 
\cite[Table1]{Shrock_05a}.

%
%
\section{Triangular-lattice chromatic polynomials with 
         cyclic boundary conditions} \label{sec.tri.chromatic}

In this section we will analyze the results for the triangular-lattice strips 
of widths $2\leq L \leq 8$. As in Section~\ref{sec.sq.chromatic}, 
we will describe our methods on the smaller widths $L=2,3$, briefly review
(for completeness) the already-known cases ($L=4$), and describe in detail
the new results presented in this section ($L\geq 5$).
In Table~\ref{zeros_tri_R} we list the real chromatic zeros
for triangular-lattice strips of size $L_{\rm F} \times (kL)_{\rm P}$ with
$1\leq k\leq 10$. A summary of the main characteristics of the 
limiting curves for these strips is displayed in Table~\ref{table_summary_tri}.

The interested reader can find the analytic expressions for the eigenvalues
and amplitudes for all lattices reported in this section  in the 
{\sc mathematica} file {\tt transfer4\_tri.m} available as part of 
the electronic version of this paper in the {\tt cond-mat} archive.   
The analytic expressions for the
sub-blocks $T_{\ell,\ell,j_0}$ can be obtained from the authors.

%
%
\subsection{$\bm{L=2}$} 

This case was first solved by Tsai and Shrock \cite{Shrock_00a}. 
The solution is  
\be
P_{2\times n}(q) \; = \; (q^2 -3q +1) + (q-2)^{2m} +  
                         (q-1)\left[ \lambda_3^n + \lambda_4^n \right] 
\label{Ptri2}
\ee
where $\lambda_{3,4}$ are the solutions of the equation
\be
(q-2)^2 + (2q-5) x + x^2 \; = \; 0
\label{eq_eigen_tri_2R}
\ee

Let us review our derivation of this result.  
The transfer matrix has dimension six. In the basis 
$\{\delta_{1,1'}\delta_{2,2'}, \delta_{1,2'}, \delta_{2,2'},
   \delta_{1,1'}, \delta_{2,1'}, 1 \}$, it takes the form
\be
\mathsf{T}(2) \; = \; \left( \begin{array}{c|cc|cc|c}
       1 &  0   & 0    &   0   &  0   & 0   \\
\hline
       0 &  2-q & 1    &   0   &  0   & 0   \\
      -1 &  2-q & 3-q  &   0   &  0   & 0 \\ 
\hline
      -1 &  0   & 0    &   2-q &  1   & 0 \\
      -1 &  0   & 0    &   2-q &  3-q & 0 \\
\hline
       1 &  q-2 & q-3  &   q-2 &  q-3 & (q-2)^2 
            \end{array}\right)
\ee
where the vertical and horizontal lines show the block structure of this
matrix (see below).
The right $\bm{w}_\text{id}$ and left $\bm{u}$ vectors are given by 
\begin{subeqnarray}
  \bm{u}^T           &=& (A, 0, A, A, 0, A)\,, \qquad \text{with $A=q(q-1)$} \\ 
 \bm{w}_\text{id}^T  &=& (1,0,0,0,0,0) 
\end{subeqnarray} 
Again, we have arranged the basis elements in decreasing order of number of 
bridges $\ell$, so the lower-triangular block form of $\mathsf{T}(2)$ becomes
apparent.

The full transfer matrix can be decomposed into four blocks: as in the 
square-lattice case with $L=2$, all of them are characterized by the 
same bottom-row connectivity $\mathcal{P}_\text{bottom}=1$, and differ only by 
the number of bridges $\ell$. 
The first block is one-dimensional and it corresponds to the 
connectivity state with $\ell=2$ bridges. Its eigenvalue is
\be
\lambda_{\ell=2} \;=\; 1 
\ee
The fourth block is also 
one-dimensional and corresponds to $\ell=0$ bridges. Its eigenvalue is
\be
\lambda_{\ell=0} \;=\; (q-2)^2 
\ee
The second and third blocks are identical two-dimensional blocks 
corresponding to a single bridge $\ell=1$: for the former block it 
emerges at site $2'$, and for the latter one it emerges at site $1'$. 
Their eigenvalues $\lambda_{\ell=1}^{(j)}$ ($j=1,2$) are given 
by the solutions of the second-order equation \reff{eq_eigen_tri_2R}.  
Please note that the block $T_{1,1}$ is block diagonal as discussed in
Section~\ref{sec.sq.cyclic.blocks}.

We have computed the amplitudes in the same way as for the square lattice. The
results are given by [c.f.~\reff{def_alphas}] 
\begin{subeqnarray}
 \alpha_{\ell=2}       &=& \alpha^{(2)} \\ 
 \alpha_{\ell=1}^{(j)} &=& \alpha^{(1)} \,, \qquad j=1,2\\ 
 \alpha_{\ell=0}       &=& \alpha^{(0)} \;=\; 1
\end{subeqnarray}
These formulas agree with the already known exact results 
\reff{Ptri2}/\reff{eq_eigen_tri_2R}. Again, the amplitude associated to 
each eigenvalue $\lambda_{\ell}^{(j)}$ coming from block $T_{\ell,\ell}$ is
$\alpha^{(\ell)}$.

\begin{figure}
\centering
\begin{tabular}{cc}
   \includegraphics[width=200pt]{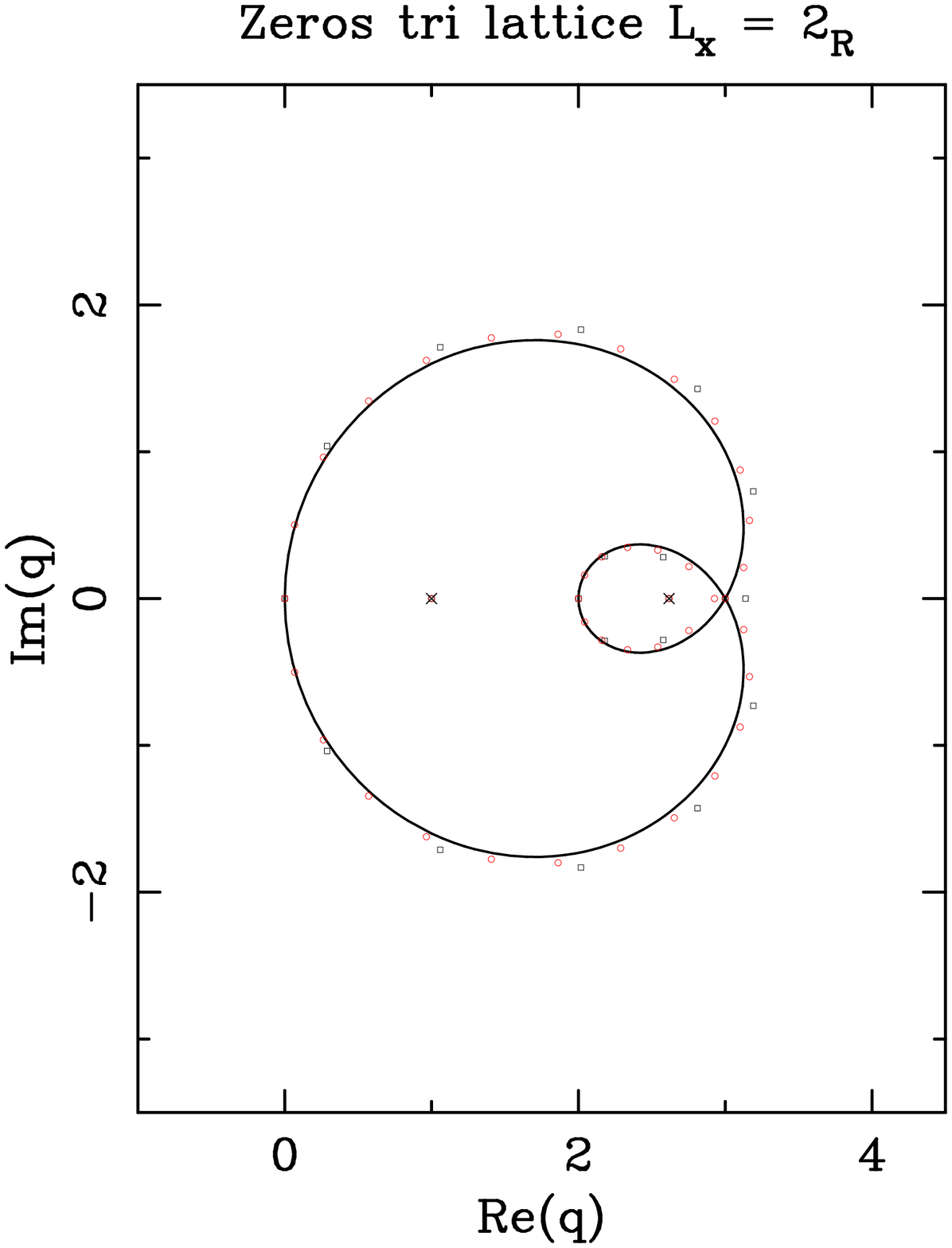} &
   \includegraphics[width=200pt]{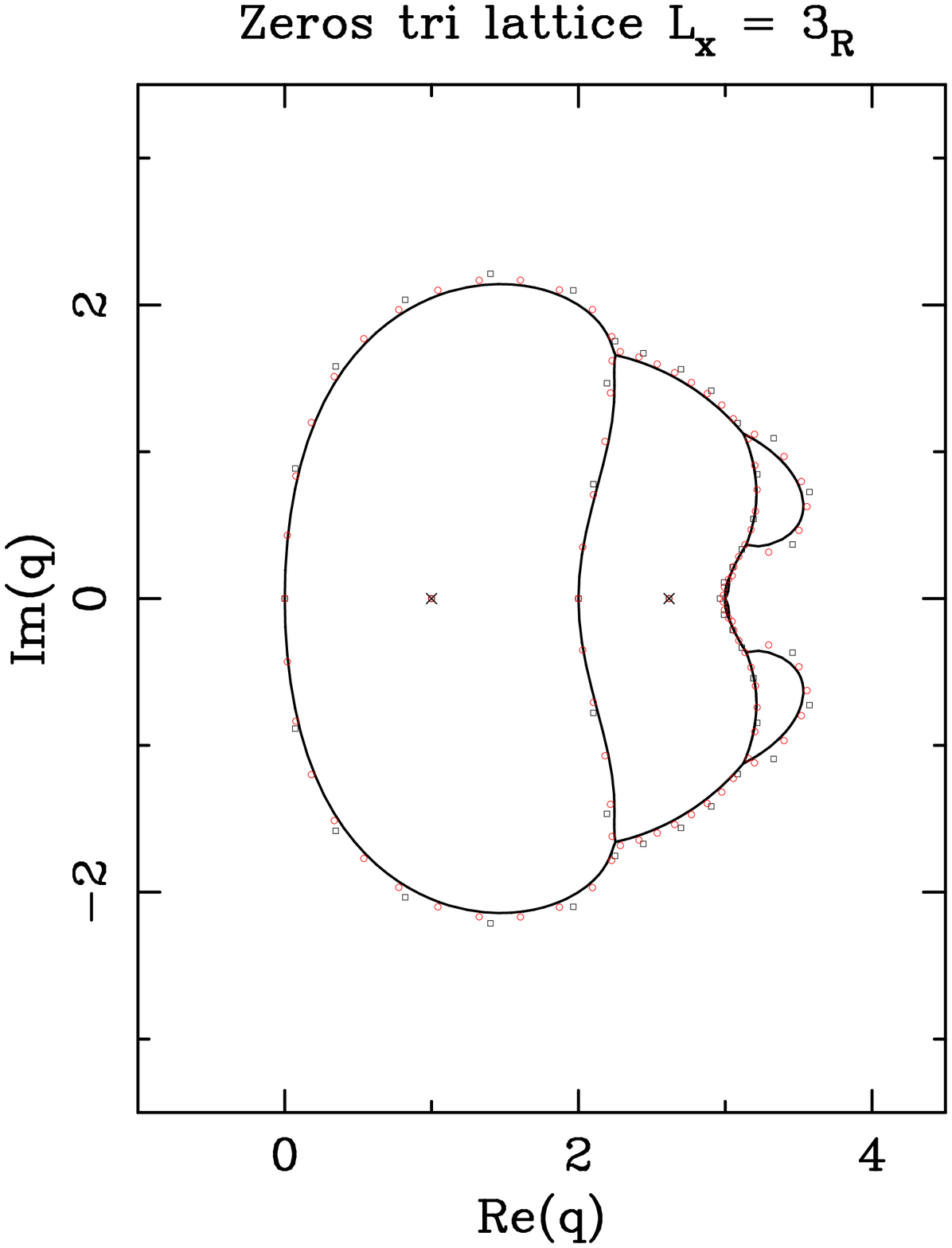}
   \\[1mm]
   \phantom{(((a)}(a)    & \phantom{(((a)}(b) \\[5mm]
   \multicolumn{2}{c}{\includegraphics[width=200pt]{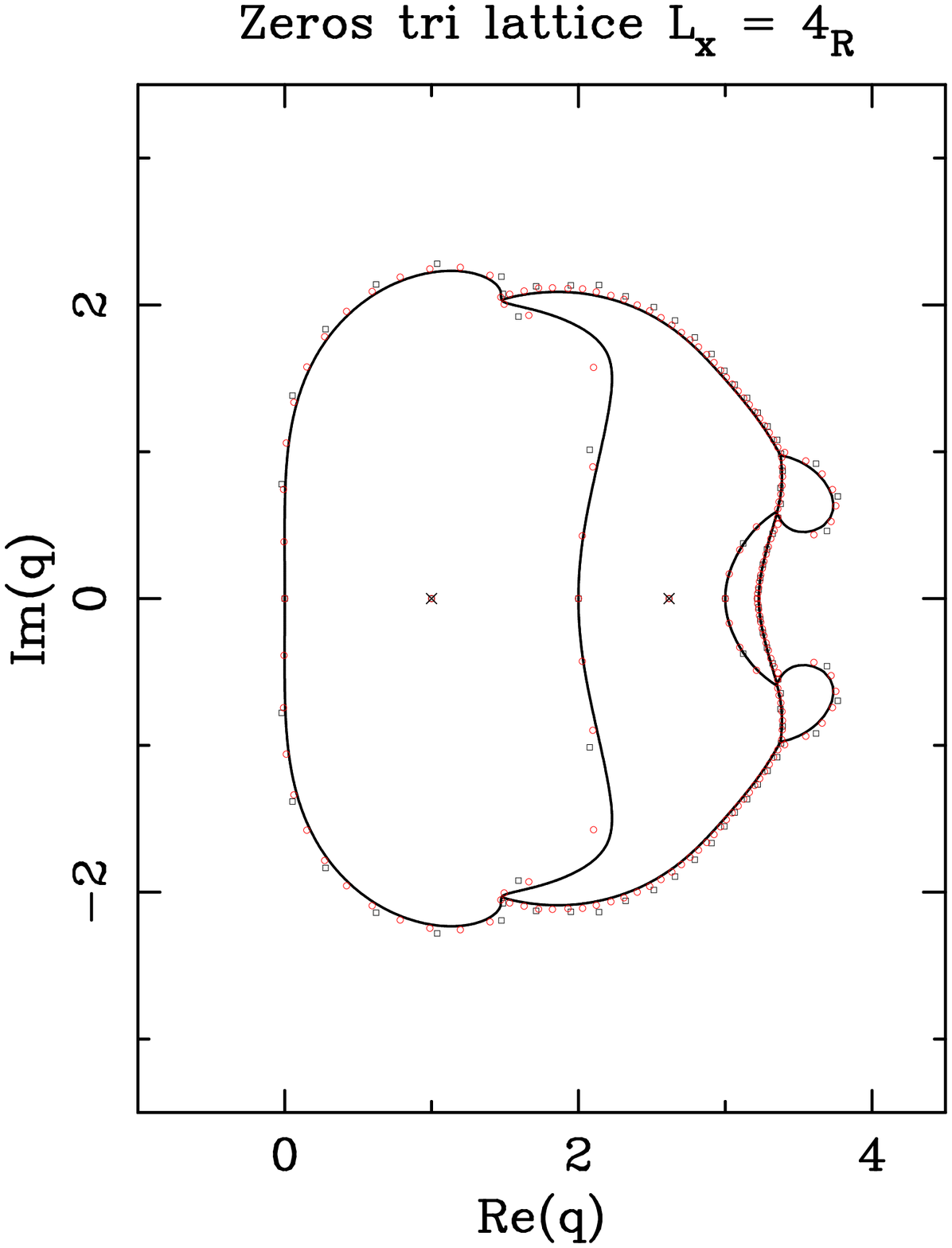} }\\[1mm] 
   \multicolumn{2}{c}{\phantom{(((a)}(c) }\\
\end{tabular}
\caption{\label{figure_tri_1}
Limiting curves for triangular-lattice strips of width (a) $L=2$, (b) $L=3$,
and (c) $L=4$ with cyclic boundary conditions. We also show the
zeros for the strips $L_\text{F} \times (5L)_\text{P}$ (black $\Box$) and
$L_\text{F} \times (10L)_\text{P}$ (red $\circ$) for the same values of $L$.
}
\end{figure}

We have depicted the chromatic zeros for the triangular-lattice strips of width
$L=2_\text{F}$ and lengths $n=10_\text{P},20_\text{P}$ in
Figure~\ref{figure_tri_1}(a). The limiting curve $\mathcal{B}$
for this strip is also shown. This curve has been obtained using the 
resultant method.\footnote{
  This limiting curve was first obtained by Shrock and Tsai 
  \protect\cite{Shrock_00a}.
} 
It crosses the real $q$-axis at $q=0,2,3$. At this
latter point it has a quadruple point. The limiting curve divides 
the complex $q$-plane into three closed regions. Finally, 
there are two isolated limiting points at $q=1$ and $q=B_5$.

%
%
\subsection{$\bm{L=3}$}

This case was first solved by Chang and Shrock \cite{Shrock_01c}, who
also computed the limiting curve. We have re-derived their result using our 
method. The full transfer matrix has dimension 30. It can be split into ten 
different blocks. These blocks provide the ten distinct eigenvalues of the 
transfer matrix. Following a similar procedure as for the square-lattice 
strips we obtained the corresponding amplitudes. 

There is a single block with $\ell=3$ bridges: this is one-dimensional and it 
corresponds to the eigenvalue 
\be
\lambda_{\ell=3} \;=\; -1
\ee
and its amplitude is $\alpha^{(3)}$. 
The block $T_{2,2}$ contains three identical sub-blocks, each of them leading
to three distinct eigenvalues $\lambda_{\ell=2}^{(j=1,2,3)}$. The simplest 
one is
\be
\lambda_{\ell=2}^{(1)} \;=\; q-2 \; ,
\ee
while the other two are the solutions of the second-order equation 
\be
x^2 - x(2q - 7) + q^2 -5q + 6  \; = \; 0 \; .
\ee
The amplitude associated to these eigenvalues is $\alpha^{(2)}$.
We have also found four (identical) sub-blocks with a single bridge ($\ell=1$). 
Each of them has four eigenvalues $\lambda_{\ell=3}^{(j)}$. The simplest one is 
\be
\lambda_{\ell=1}^{(1)} \;=\; -q^2 + 5 q -6  
\ee
and the other three are the solutions of the third-order equation 
\begin{eqnarray}
0 &=& x^3 + (2 q^2 -12 q + 19) x^2 + (q^4 - 11 q^3 + 46 q^2 - 85 q +58) x
      \nonumber \\
  & & \qquad -q^5 + 11 q^4 - 48 q^3 + 104 q^2 - 112 q + 48
\end{eqnarray}
The amplitude $\alpha^{(1)}$ corresponds to all of them.
Finally, there are two (identical) sub-blocks characterized by no bridges 
($\ell=0$). Their eigenvalues are the solutions of the second-order equation
\be
x^2 - x(q^3 - 7 q^2 +18 q -17) + q^4 - 9 q^3   + 30 q^2 - 44 q + 24 
                \;   = \; 0
\ee
and their amplitudes are $\alpha^{(0)}=1$. 

In Figure~\ref{figure_tri_1}(b) we show the chromatic zeros for the 
strips of sizes $3_\text{F}\times 15_\text{P}$ and 
$3_\text{F}\times 15_\text{P}$, and the limiting curve $\mathcal{B}$. This
latter curve was obtained using the resultant method.\footnote{
  The limiting curve for this strip was first computed
  by Chang and Shrock \protect\cite{Shrock_01c}.
} 
It crosses the real $q$-axis at $q=0,2,3$. At this latter point
the curve has a quadruple point. There are eight T points at
$q\approx 2.2532978656 \pm  1.6576481039\, i$,  
$q\approx 3.1233767820 \pm  1.1226454338\, i$, 
$q\approx 3.1427903492 \pm  0.3662742838\, i$, and
$q\approx 3.0244195343 \pm  0.1322645045\, i$. 
Finally, we find two isolated limiting points at $q=1$ and $q=B_5$.

%
%
\subsection{$\bm{L=4}$}

This case was first solved by Chang and Shrock \cite{Shrock_01c}, who
also computed the limiting curve. We have re-derived their result by using
our general procedure. The full transfer matrix has dimension 178, 
and it can be split into 26 blocks. 

In Figure~\ref{figure_tri_1}(c) we show the chromatic zeros for the
strips of sizes $4_\text{F}\times 20_\text{P}$ and 
$4_\text{F}\times 40_\text{P}$, and the limiting curve $\mathcal{B}$.\footnote{
  The limiting curve for this strip was first computed
  by Chang and Shrock \protect\cite{Shrock_01c}.
} 
This latter curve was obtained using the direct-search method. Thus, the 
number of endpoints and other features reported below should be interpreted
as lower bounds of the true values.

The limiting curve ${\cal B}$ crosses the real $q$-axis at $q=0,2,3$ and
$q\approx 3.2281262261$. We have found eight T points at 
$q\approx 3.3762587794 \pm 0.9748801876\, i$, 
$q\approx 3.3525265561 \pm 0.5771942145\, i$, 
$q\approx 3.3542356970 \pm 0.5945094569\, i$, and 
$q\approx 1.4763063347 \pm 2.0348131808\, i$.  
There are two isolated limiting points at $q=1,B_5$.

%
%
\subsection{$\bm{L=5}$}

After we completed the computation of this case we learned that
it had been independently solved by Chang and Shrock 
\cite{Shrock_05a}. Our result based on our transfer-matrix method 
fully agrees with theirs. 

The full transfer matrix is 1158 dimensional, and can be split into 70 blocks.
After the diagonalization of such blocks we find 70 different eigenvalues.
Again, we find the eigenvalue/amplitude structure discussed in 
Section~\ref{sec.sq.cyclic.blocks2}.

\begin{figure}
\centering
\begin{tabular}{cc}
   \includegraphics[width=200pt]{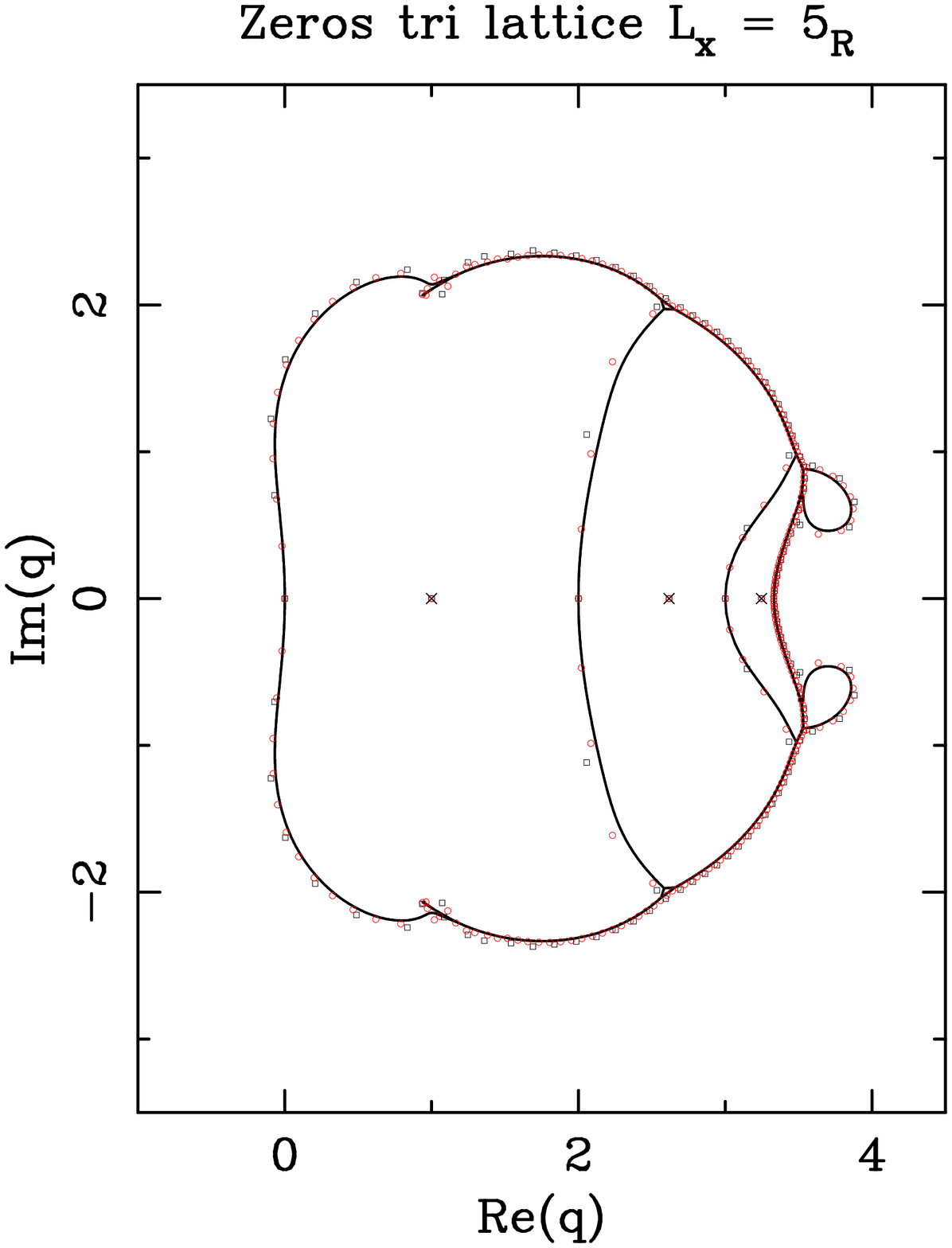} &
   \includegraphics[width=200pt]{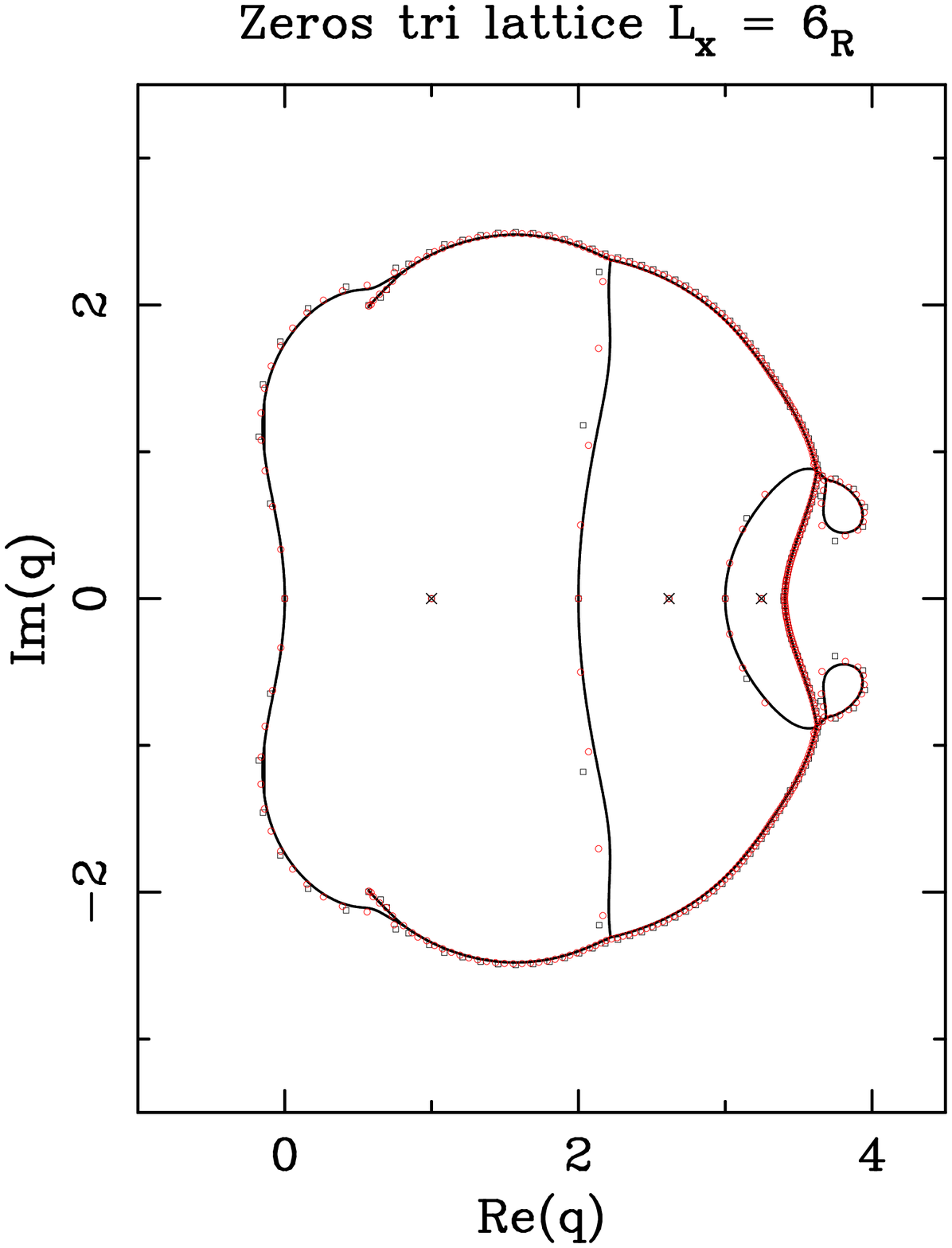}
   \\[1mm]
   \phantom{(((a)}(a)    & \phantom{(((a)}(b) \\[5mm]
   \multicolumn{2}{c}{\includegraphics[width=200pt]{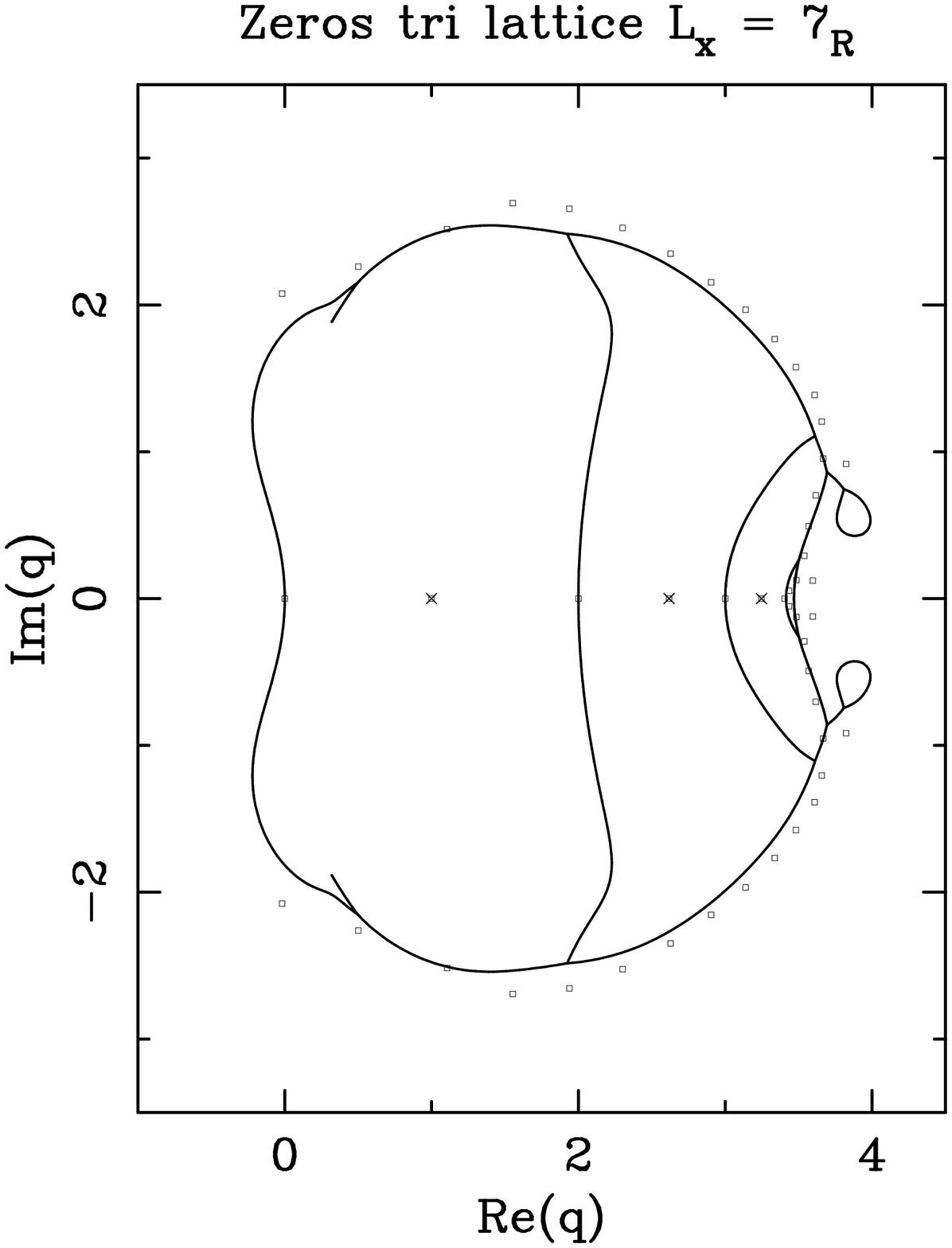}} \\[1mm] 
   \multicolumn{2}{c}{\phantom{(((a)}(c)} \\
\end{tabular}
\caption{\label{figure_tri_2}
Limiting curves for triangular-lattice strips of width (a) $L=5$, (b) $L=6$,
and (c) $L=7$
with cyclic boundary conditions. We also show the
zeros for the strips $L_\text{F} \times (5L)_\text{P}$ (black $\Box$) and
$L_\text{F} \times (10L)_\text{P}$ (red $\circ$) for $L=4,5$. For $L=7$ we
show the chromatic zeros for the strip $7_\text{F} \times 7_\text{P}$ 
(black $\Box$). 
}
\end{figure}

In Figure~\ref{figure_tri_2}(a) we show the chromatic zeros for the
strips of sizes $5_\text{F}\times 25_\text{P}$ and 
$5_\text{F}\times 50_\text{P}$, and the limiting curve $\mathcal{B}$. This
latter curve was computed via the direct-search method.\footnote{
  The computation of this limiting curve is new, as it was not published in 
  the (first) preprint version of Ref.~\protect\cite{Shrock_05a}.
} 
The curve ${\cal B}$ has a more involved shape than for the other 
triangular-lattice strips with smaller width. It crosses the real 
$q$-axis at $q=0,2,3$ and $q\approx 3.3324469422$. 
We have found 16 T points at 
$q\approx 1.0114299869 \pm 2.1132511328\, i$,      
$q\approx 1.1643009748 \pm 2.2002545351\, i$,      
$q\approx 2.5640974594 \pm 2.0365055771\, i$,      
$q\approx 2.5843109937 \pm 1.9725597103\, i$,      
$q\approx 2.6549519688 \pm 1.9693915215\, i$,      
$q\approx 3.4850579943 \pm 0.9796767236\, i$,      
$q\approx 3.5231420699 \pm 0.7322113419\, i$,  and 
$q\approx 3.5309588080 \pm 0.8827974592\, i$.      
The unusual features are a) two tiny bulb-like regions protruding from 
the T points at $q\approx 1.0114299869 \pm 2.1132511328\, i$; and b)
two small triangular-shaped regions delimited by the six T points around 
$q\approx 2.60\pm 2.00\, i$. 

Finally, there are three isolated limiting points at $q=1,B_5$ and $B_7$.
This is the triangular-lattice strip with smallest width for which we 
find an isolated limiting point at $q=B_7$.
 
%
%
\subsection{$\bm{L=6}$}

The full transfer matrix is 7986 dimensional, and can be split into 192 blocks.
The diagonalization of these blocks leads to the 192 different eigenvalues,
in agreement with the formula \reff{formula_SqRing}. The simplest eigenvalue
is $\lambda=1$ which comes from the block $T_{6,6}$; thus, its amplitude
is $\alpha^{(6)}$. The other eigenvalues are given by the solution of 
polynomic equations up to order 55. We find the block structure of 
eigenvalues and amplitudes described in Section~\ref{sec.sq.cyclic.blocks2}. 

In Figure~\ref{figure_tri_2}(b) we show the chromatic zeros for the
strips of sizes $6_\text{F}\times 30_\text{P}$ and 
$6_\text{F}\times 60_\text{P}$, and the limiting curve $\mathcal{B}$. This
latter curve was computed via the direct-search method. 
This curve crosses the real $q$-axis at $q=0,2,3$ and 
$q\approx 3.4062849655$. 
We have found 10 T points at 
$q\approx 0.8028018425 \pm 2.2224309039\, i$, 
$q\approx 2.2195798331 \pm 2.3066154729\, i$, 
$q\approx 3.6219823536 \pm 0.8757610205\, i$, 
$q\approx 3.6246351106 \pm 0.8660035657\, i$, and 
$q\approx 3.6870064797 \pm 0.8100610005\, i$.  
There are two bulb-like regions emerging from the last T points 
($q\approx 3.6870064797 \pm 0.8100610005\, i$). We have also found
two endpoints at\footnote{
  Unfortunately we have been unable to use the resultant method to verify 
  the existence of these endpoints. Instead, we have used the direct-search 
  method to locate them. The endpoint character of these points is
  suggested by the following observation. 
  If $\lambda_1$ and $\lambda_2$ are two (equimodular) dominant eigenvalues, 
  we can define the (real) parameter $t = \tan(\theta/2)$, where $\theta$ 
  is given by $\lambda_1/\lambda_2 = e^{i\theta}$ 
  (see \protect\cite[Section 4.1.1]{transfer1}). Then, an endpoint is 
  characterized by $t=0$. At the points 
  $q\approx 0.5716823545 \pm 1.9896139773\, i$, we have found that
  $t\approx 1.8\times 10^{-6}$. This computation strongly suggest that these 
  points are indeed endpoints. 
}
$q\approx 0.5716823545 \pm 1.9896139773\, i$. These points correspond to two
complex-conjugate small branches protruding from the T points at
$q\approx 0.8028018425 \pm 2.2224309039\, i$. This is first triangular-lattice
strip for which we find endpoints.
Finally, there are three isolated limiting points at $q=1,B_5$ and $B_7$.

%
%
\subsection{$\bm{L=7}$}

The full transfer matrix $\mathsf{T}(7)$ has dimension 57346, and can be
split into 534 blocks (see Table~\ref{table_blocks_tri}). By making use of
the simplified method described in Section~\ref{sec.sq.cyclic.blocks2}, 
we have obtained the relevant sub-blocks 
$T_{\ell,\ell,j_0}$ from which all distinct eigenvalues can be computed. 

Unfortunately, the sub-blocks $T_{\ell,\ell,j_0}$ have rather large dimensions,
ranging from 1 to 145. As for the square-lattice strip with $L=7$, we have
been unable to compute symbolically the characteristic polynomial of some of
these blocks. Thus, the computation of the chromatic polynomial for finite
strips of width $L=7$ is beyond our computer facilities. However, as we can
compute the eigenvalues and we know the amplitudes, we have been able to
obtain the limiting curve.

In Figure~\ref{figure_tri_2}(c) we show the zeros for a triangular-lattice 
strip of size $7_\text{F} \times 7_\text{P}$ (obtained using 
Eq.~\reff{eq_check_Z} and previous results for cylindrical boundary conditions 
\cite{transfer3}). We also show the limiting curve $\mathcal{B}$; it  
has been computed using the direct-search method. The same observations as
for the square lattice with $L=7$ hold in this case. 

The limiting curve ${\cal B}$ crosses the real $q$-axis at $q=0,2,3,B_8$ and
$q = q_0 \approx  3.4682618071$. This is the first case for which we find
five different ``phases'' on the real $q$-axis.
We have found 12 T points at 
$q \approx 0.4951745217 \pm 2.1501328665\, i$,
$q \approx 1.9229108754 \pm 2.4842273215\, i$,
$q \approx 3.5014960312 \pm 0.2613477419\, i$,
$q \approx 3.6121905140 \pm 1.1057902185\, i$,
$q \approx 3.6937370146 \pm 0.8598201628\, i$, and  
$q \approx 3.8081077493 \pm 0.7447122915\, i$. 
As for the $L=6$ case, we find two complex-conjugate endpoints at
$q\approx  0.3195792189 \pm 1.8836441569$.\footnote{
  In this case, the parameter $t = \tan(\theta/2)$ (with $\theta$
  given by $\lambda_1/\lambda_2 = e^{i\theta}$) takes the value
  $t\approx 3.2\times 10^{-6}$. This small value suggests that these
  points are indeed endpoints.
}
There are 3 isolated limiting points at $q=1$, $q=B_5$, and $q=B_7$, as
the value of $q_0$ is smaller than $B_9$.

%
%
\subsection{$\bm{L=8}$}

The full transfer matrix $\mathsf{T}(8)$ has dimension 424206, and can be
split into 1500 blocks (see Table~\ref{table_blocks_tri}). We have computed
the relevant nine blocks $T_{\ell,\ell,j_0}$ from which all distinct eigenvalues
can be obtained. 
As for $L=7$, several of these blocks have large dimensions (up to 385).
Thus, we have been unable to obtain the symbolic form of their characteristic
polynomial. We can nevertheless compute the numerical values of the
corresponding eigenvalues, and thus the limiting curve $\mathcal{B}$. 
The computation of this curve is very time consuming, whence we have focused
on obtaining the point $q_0$ with great precision. Our result is
$q_0 \approx 3.5134782609$. As this value is greater than $B_8$, we can 
observe a ``phase'' characterized by $\ell=4$ bridges.

Finally, we expect 3 isolated limiting points at $q=1$, $q=B_5$, and $q=B_7$,
as the value of $q_0$ is smaller than $B_9$.

%
%

\section{Discussion} \label{sec.disc}

We have empirically 
found an interesting structure of the eigenvalues and amplitudes of
the transfer matrix in terms of the number of bridges of the connectivity
states $\ell$. First, the transfer matrix has a lower-triangular block form
\reff{T_blocks1}. The blocks on the diagonal $T_{\ell,\ell}$ in turn have a
diagonal block form \reff{T_blocks2} in terms of the sub-blocks
$T_{\ell,\ell,j}$. The sub-blocks $T_{\ell,\ell,j}$ are characterized by the
bottom-row connectivity state and the number $\ell$ and position of the
bridges (modulo reflection symmetry with respect to the center of the strip
for the square-lattice case). Second, all the eigenvalues of the transfer
matrix can be obtained by diagonalizing, for each $\ell$, only the largest of
the sub-blocks. The final set of distinct eigenvalues can be then split into
disjoint classes labeled solely by the number of bridges $\ell$ (i.e., if a
given eigenvalue appears in several sub-blocks, all of these sub-blocks
belong to block $T_{\ell,\ell}$ with $\ell$ bridges).
Finally, all the eigenvalues belonging to the class of $\ell$ bridges have
the common amplitude $\alpha^{(\ell)}$.

\subsection{Phase diagram} \label{disc.phase.diagram} 

Motivated by this structure we have tried to give a topological meaning to the
closed regions into which the limiting curve divides the complex $q$-plane. In
other words, in each region of the $q$-plane we want to know to which block
$T_{\ell,\ell}$ the dominant eigenvalue belongs. 

We first discuss the regions having a non-empty intersection $I$ with the real
$q$-axis. We invariably find that:
\begin{enumerate}
 \item [A.] $I$ is a simple interval in $\R^* \equiv \R \cup \{\infty\}$,
  which is either $I_n \equiv (B_{2n},B_{2n+2})$ for
  $n=1,2,\ldots,n_\text{max}-1$, where $n_\text{max}$ is determined by
  $B_{2n_\text{max}} \le q_0(L) < B_{2n_\text{max}+2}$, or
  $I_{n_\text{max}} \equiv (B_{2n_\text{max}},q_0(L))$,
  or $I_0 \equiv \R^* \setminus [0,q_0(L)]$.
 \item [B.] In the region containing $I_n$, with $n=0,1,\ldots,n_\text{max}$,
  the dominant eigenvalue comes from the block with $\ell = n$ bridges.
  Its amplitude is thus $\alpha^{(n)}$.
\end{enumerate} 

\begin{figure}
  \centering
  \includegraphics[width=400pt]{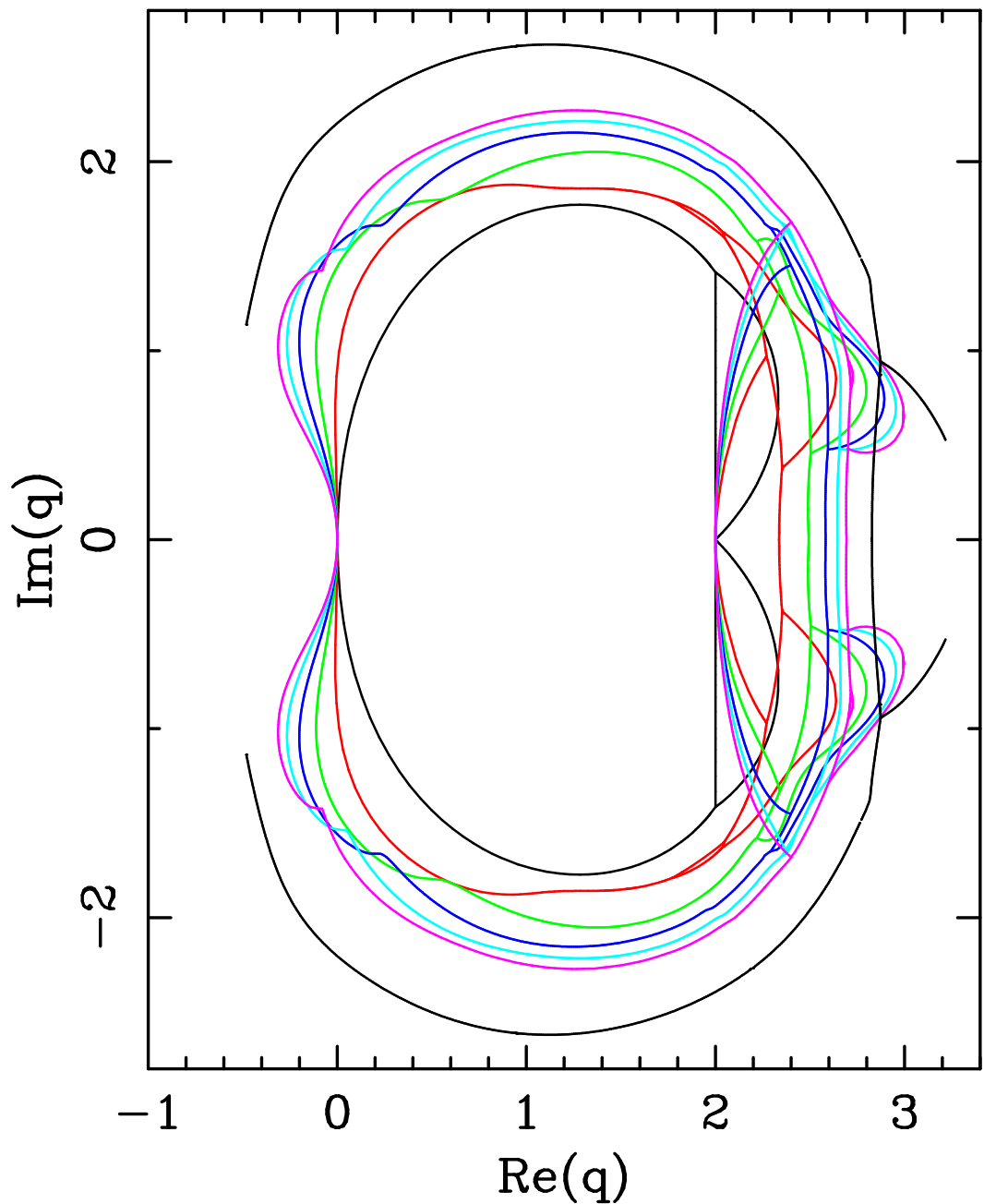}
  \caption{\label{figure_sq_allR}
   Limiting curves for square-lattice strips of width $L$ and cyclic
   boundary conditions. We show the results for $L=2$ (black), $L=3$ (red), 
   $L=4$ (green), $L=5$ (blue), $L=6$ (light blue), and $L=7$ (pink). 
   The outermost (black) curve
   shows the limiting curve for a square-lattice strip of width $L=10$ 
   with cylindrical boundary conditions \protect\cite{transfer2}.  
  }
\end{figure}

\begin{figure}
  \centering
  \includegraphics[width=400pt]{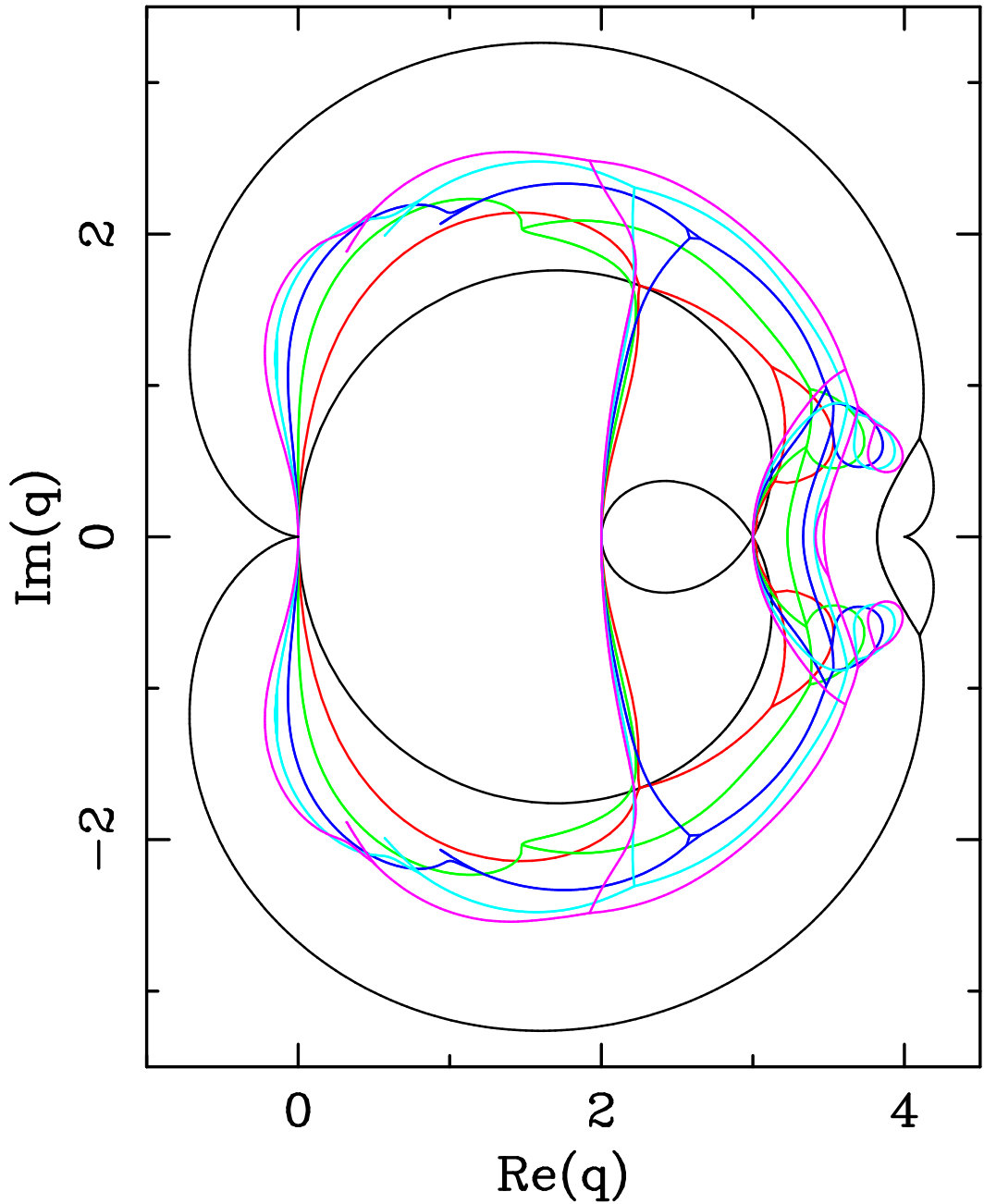}
  \caption{\label{figure_tri_allR}
   Limiting curves for triangular-lattice strips of width $L$ and cyclic
   boundary conditions. We show the results for $L=2$ (black), $L=3$ (red), 
   $L=4$ (green), $L=5$ (blue), and $L=6$ (light blue), and $L=7$ (pink).  
   The outermost (black) curve
   shows the infinite-volume limiting curve obtained by Baxter 
   \protect\cite{Baxter_86_87}. See \protect\cite{transfer3} for 
   a critical discussion on Baxter's limiting curve.
  }
\end{figure}

We have verified the Properties~A--B above for all the cyclic strips of the
square and triangular lattices considered in Sections~\ref{sec.sq.chromatic}
and \ref{sec.tri.chromatic}. We conjecture that they hold true for any strip
width $L$, for the appropriate value of $q_0(L)$, and in the thermodynamic
limit with $q_0(L)$ replaced by $q_0$. [For a discussion on the values of
$q_0$, see the Introduction. See also
Figures~\ref{figure_sq_allR}--\ref{figure_tri_allR} for plots of the limiting
curves $\mathcal{B}_L$ with $2\leq L\leq 7$.]

In Ref.~\cite{Saleur_90_91} a field theoretical mechanism was exhibited
according to which the special role of the Beraha numbers in the case of
{\em free} boundary conditions actually applies to the whole BK
phase. We therefore finally conjecture that Properties~A--B above hold
within the entire BK phase, with $q_0(L)$ replaced by the boundary
of the BK phase (in finite size, or in the $L \to \infty$ limit).
[For the extent of the BK phase in the thermodynamic limit,
see the Introduction].

Property~B should be compared with Saleur's study of the square-lattice
Potts model along the curve $v=-\sqrt{q}$, with $0 < q < 4$, i.e., the
so-called unphysical self-dual line. In this case one has
$n_\text{max}=\infty$, and it can be proved that the dominant eigenvalue for
$q \in I_n$ indeed occurs in the spin $n$ sector \cite{Saleur_90_91}. By
renormalization group arguments one would expect this to extend to the entire
BK phase. Another argument, asymptotically valid for $L \to \infty$, is that
for $q \in I_n$ the effective central charge is maximal in the $n$-bridge
sector (see below).

As to the regions inside the limiting curve that have an empty
intersection with the real $q$-axis, we have found that in most cases
(with $3\leq L \leq 7$) the dominant eigenvalue corresponds to the
class $\ell=1$. There are however some exceptions to this rule:
\begin{itemize}
  \item For the square lattice: For $L=7$, the eigenvalues which are
        dominant inside
        the four tiny closed regions around $q\approx 2.71 \pm 0.85$
        are characterized by either $\ell=0$ or $\ell=2$.
  \item For the triangular lattice: For $L=5$, the tiny bulb-like regions
        emerging from T points at 
        $q\approx 1.0114299869 \pm 2.1132511328\, i$, are characterized by
        $\ell=2$ (note that those regions are completely surrounded
        by an $\ell=1$ phase). Furthermore, the two small triangular-shaped
        regions around $q\approx 2.60 \pm 2.00\,i$ are characterized by
        $\ell=0$. For $L=6$, the two
        bulb-like regions emerging from T points at
        $q\approx 3.6870064797 \pm 0.810061005\,i$ are characterized
        by $\ell=0$. 
\end{itemize}
Therefore, we find that most of the limiting curves $\mathcal{B}$ corresponds 
to crossings (in modulus) of dominant eigenvalues coming from {\em different}\/
blocks $T_{\ell,\ell}$. 
However, there are some exceptions to this general rule: we have found 
(in most cases, small) parts of $\mathcal{B}$ that correspond to 
crossings of dominant eigenvalues coming from the {\em same}\/ block. 
For instance, the limiting curve for the square-lattice strip of width 
$L=2$ contains a vertical segment which corresponds to crossings of 
two eigenvalues belonging to the $\ell=1$ block. Smaller arcs with the same
origin are also found for square-lattice strips with $L\leq 3$ (except $L=6$),
and for triangular-lattice strips with $L\geq 5$.

\subsection{Free energy and central charge} \label{disc.free.central} 

In order to gain further insight into the phase diagram we have computed the
free energy per spin $f_L(q) = \frac{1}{L} \, \log \lambda_\star (q)$ for
real $q$ for 
strips of width $L$ and infinite length, where $\lambda_\star$ denotes the
dominant eigenvalue of the transfer matrix.

For $L \le 8$ this was done using the symbolic technique described in
Sections~\ref{sec.sq.cyclic}--\ref{sec.tri.chromatic}. Another
technique enabled us to extend these results to the range $L \le 11$.
It consists in numerically diagonalizing the transfer matrix for a strip with
fully free boundary conditions, but imposing the constraint that $\ell$
disjoint clusters propagate along the transfer direction. For each $q$,
$\lambda_\star (q)$ was selected as the numerically largest among the $L+1$
eigenvalues which are dominant with respect to each possible $\ell$-sector
($\ell=0,1,\ldots,L$). We have verified that for $L \le 8$ the two techniques
give identical results.

\begin{figure}
\centering
\begin{tabular}{cc}
   \includegraphics[width=200pt]{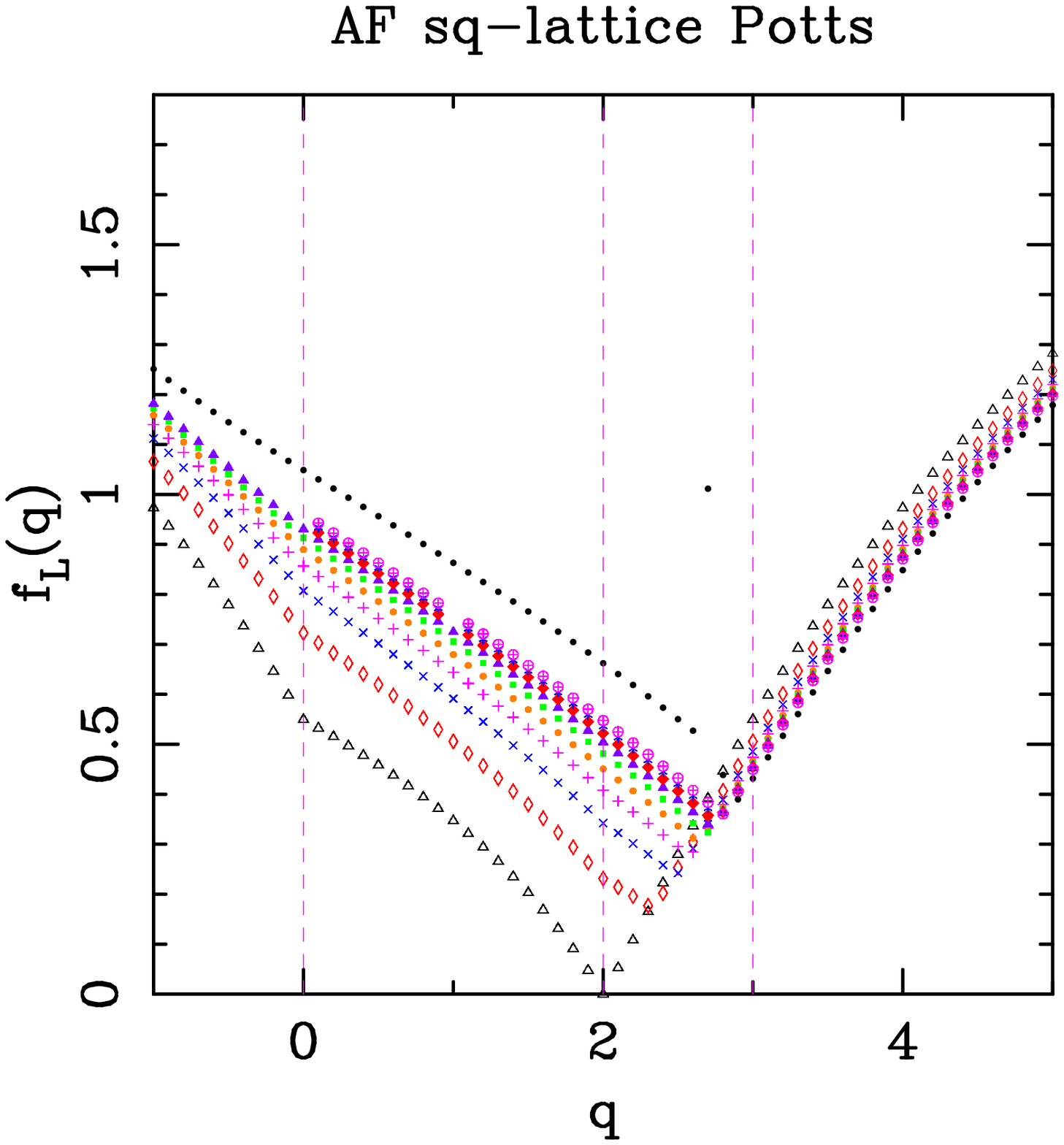} &
   \includegraphics[width=200pt]{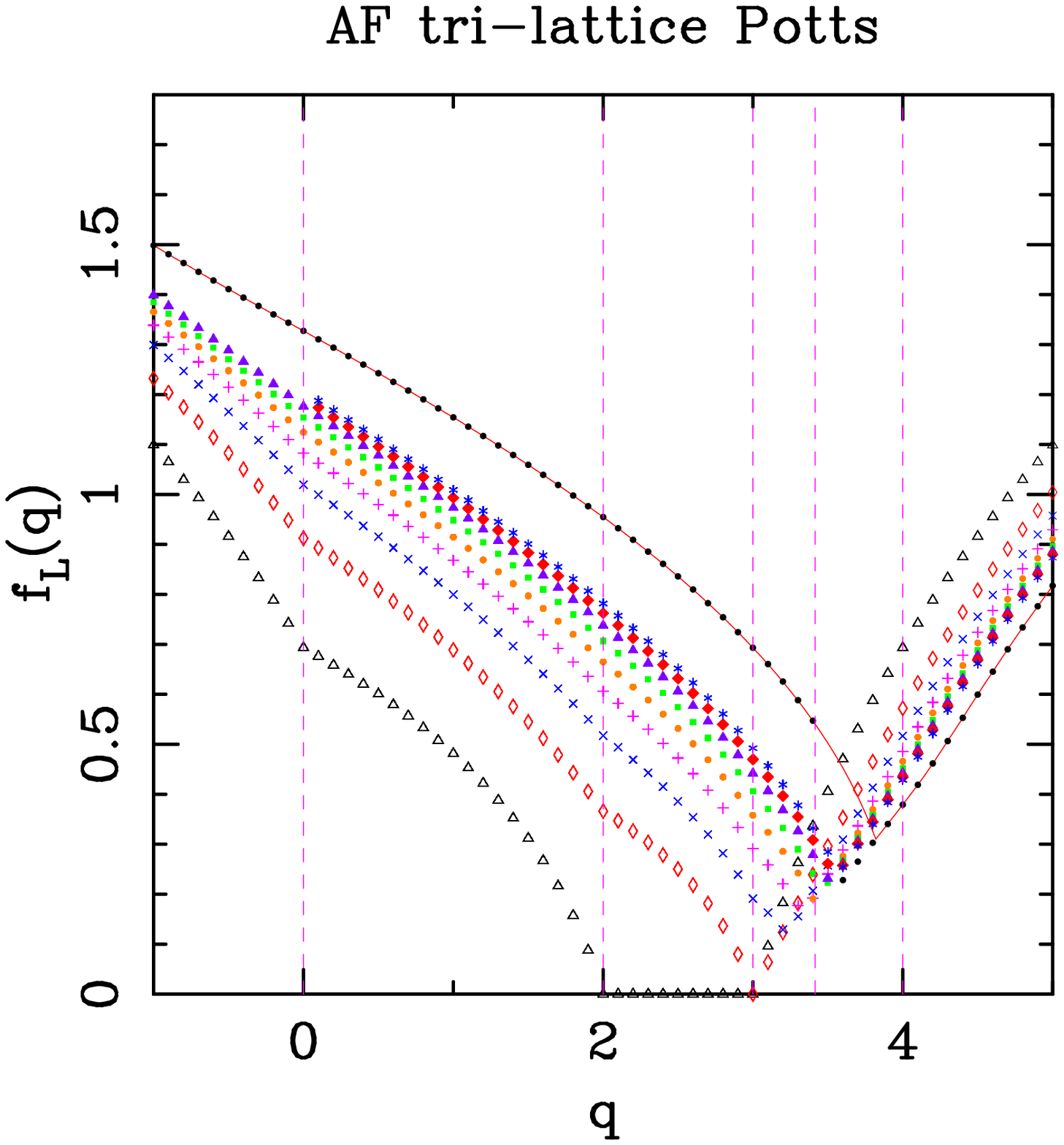}
   \\[1mm]
   \phantom{(((a)}(a)    & \phantom{(((a)}(b) \\[5mm]
\end{tabular}
\caption{\label{figure_F}
Estimates of the free energy for the square- (a) and triangular-lattice (b)
Potts model for real $q$ in the range $-1\leq q \leq 5$.
In each plot we show the values of the free energy  
obtained from the leading eigenvalue of the transfer matrix for strip widths
$L=2$ (black $\triangle$),       $3$ (red $\diamond$),   
  $4$ (blue $\times$),           $5$ (pink $+$),
  $6$ (orange $\bullet$),        $7$ (green $\blacksquare$), 
  $8$ (violet $\blacktriangle$), $9$ (red $\blacklozenge$), 
 $10$ (blue $*$),           and $11$ (pink $\oplus$). 
We also show the extrapolated value of the free energy in the thermodynamic 
limit (black $\bullet$) obtained by performing the fit  
\protect\reff{fit_free_energy} for $L_\text{min}=7$ (resp.\ $L_\text{min}=6$) 
for the square (resp.\  triangular) lattice. In (b) the solid (red) line 
shows the exact value of the triangular-lattice free energy found by
Baxter \protect\cite{Baxter_86_87}. The vertical (pink) dashed lines show even
Beraha numbers $q=B_{2n}=B_{2n}^{(1)}$ \protect\reff{def_Bnj}. In (a) we show
$1\leq n\leq 3$, and in (b), $1\leq n \leq 4$ and the limit $n\to\infty$ 
(i.e., $q=4$). 
}
\end{figure}

\begin{figure}
\centering
\begin{tabular}{cc}
   \includegraphics[width=200pt]{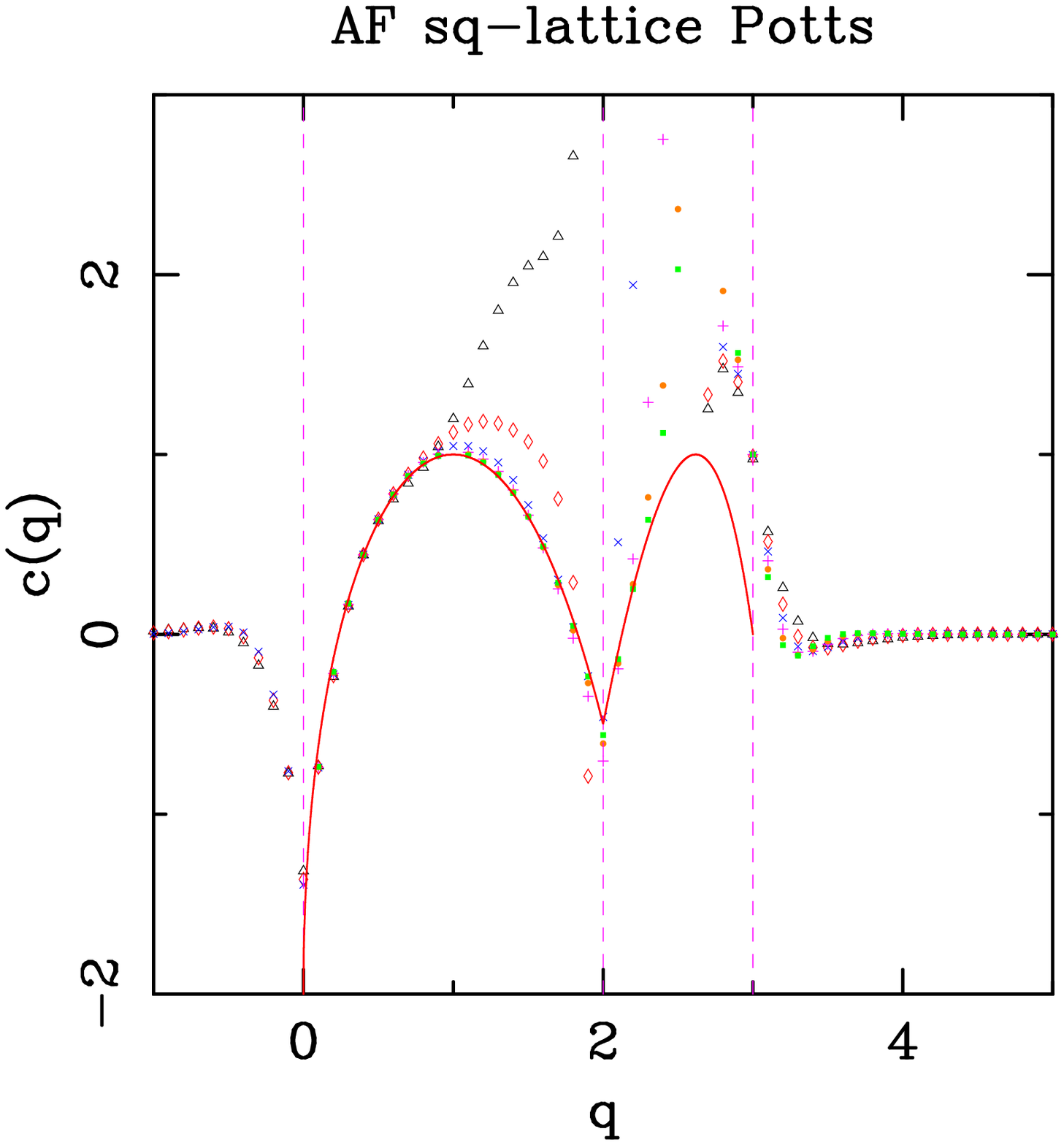} &
   \includegraphics[width=200pt]{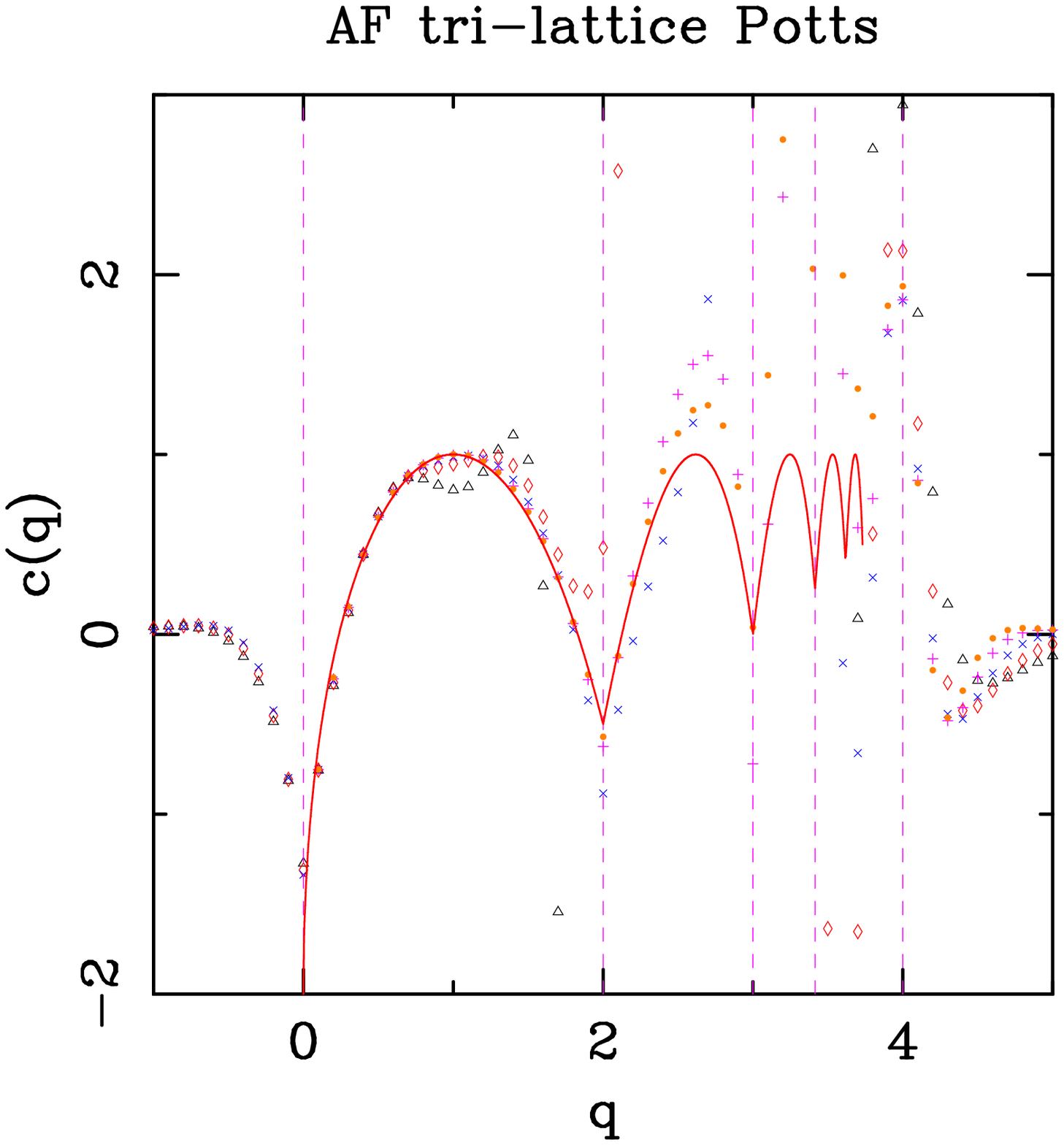}
   \\[1mm]
   \phantom{(((a)}(a)    & \phantom{(((a)}(b) \\[5mm]
\end{tabular}
\caption{\label{figure_C}
Estimates of the central charge for the zero-temperature Potts 
antiferromagnet on the square (a) and triangular (b) lattices
for real $q$ in the range $-1\leq q \leq 5$.  
In each plot we show the values of the central
charge obtained from the fits to the Ansatz \protect\reff{fit_free_energy}
for $L_\text{min}=2$ (black $\triangle$), 
                 $3$ (red $\diamond$), 
                 $4$ (blue $\times$), 
                 $5$ (pink $+$), 
                 $6$ (orange $\bullet$), and 
                 $7$ (green $\blacksquare$).
The solid (red) line shows the theoretical prediction 
for the effective central charge $c_\text{eff}$ \protect\reff{c_theor}. 
The vertical (pink) dashed lines are as in Figure~\protect\ref{figure_F}.
}
\end{figure}

\begin{figure}
\centering
\begin{tabular}{cc}
   \includegraphics[width=200pt]{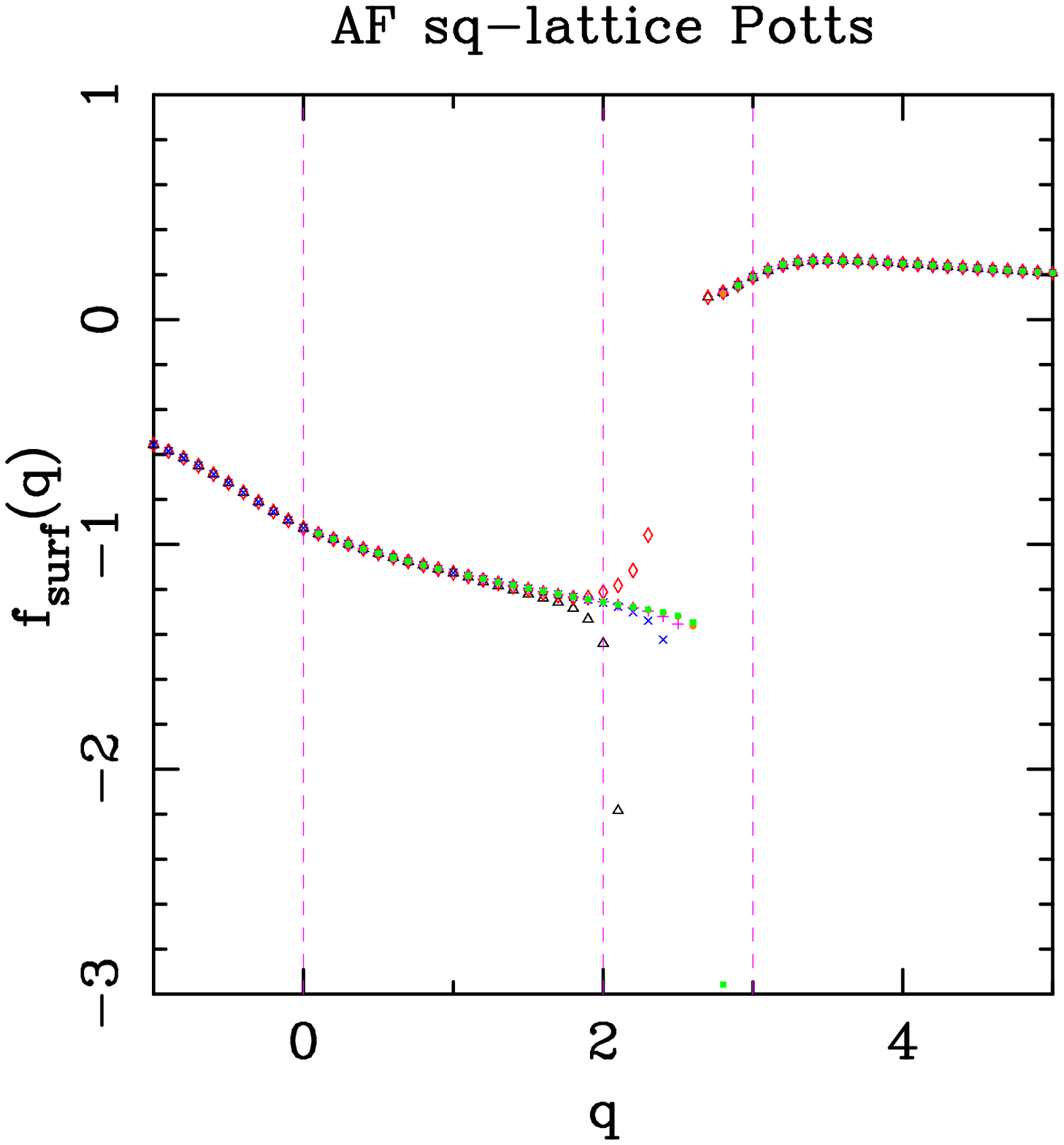} &
   \includegraphics[width=200pt]{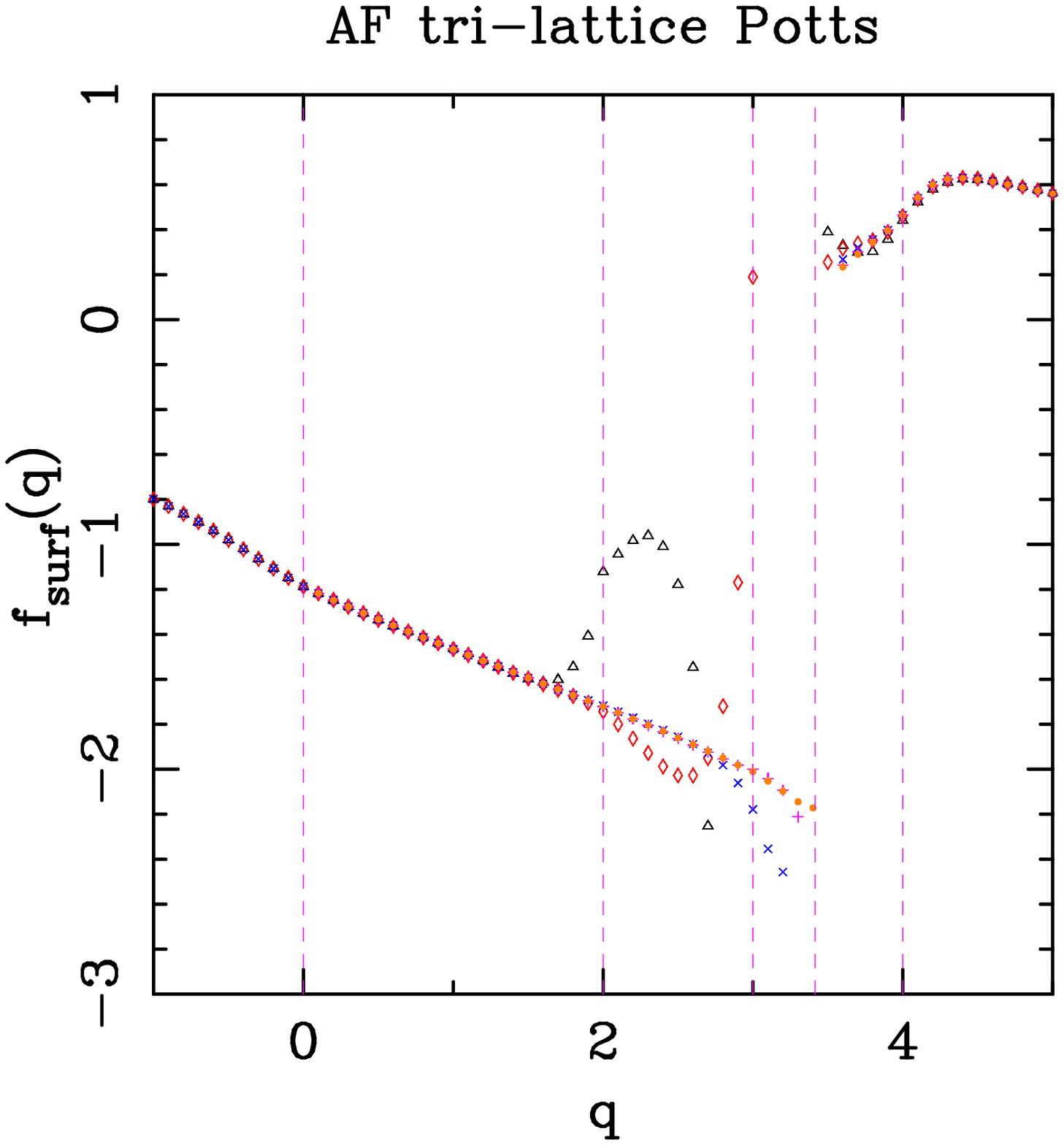} 
   \\[1mm]
   \phantom{(((a)}(a)    & \phantom{(((a)}(b) \\[5mm]
\end{tabular}
\caption{\label{figure_Fs}
Estimates of the surface free energy $f_\text{surf}(q)$ 
for the zero-temperature Potts 
antiferromagnet on the square (a) and triangular (b) lattices
for real $q$ in the range $-1\leq q \leq 5$.  
In each plot we show the values of the surface free energy  
obtained from the fits to the Ansatz \protect\reff{fit_free_energy}
for $L_\text{min}=2$ (black $\triangle$), 
                 $3$ (red $\diamond$), 
                 $4$ (blue $\times$), 
                 $5$ (pink $+$), 
                 $6$ (orange $\bullet$), and 
                 $7$ (green $\blacksquare$).
The vertical (pink) dashed lines are as in Figure~\protect\ref{figure_F}.
}
\end{figure}

\begin{figure}
\centering
\begin{tabular}{cc}
   \includegraphics[width=200pt]{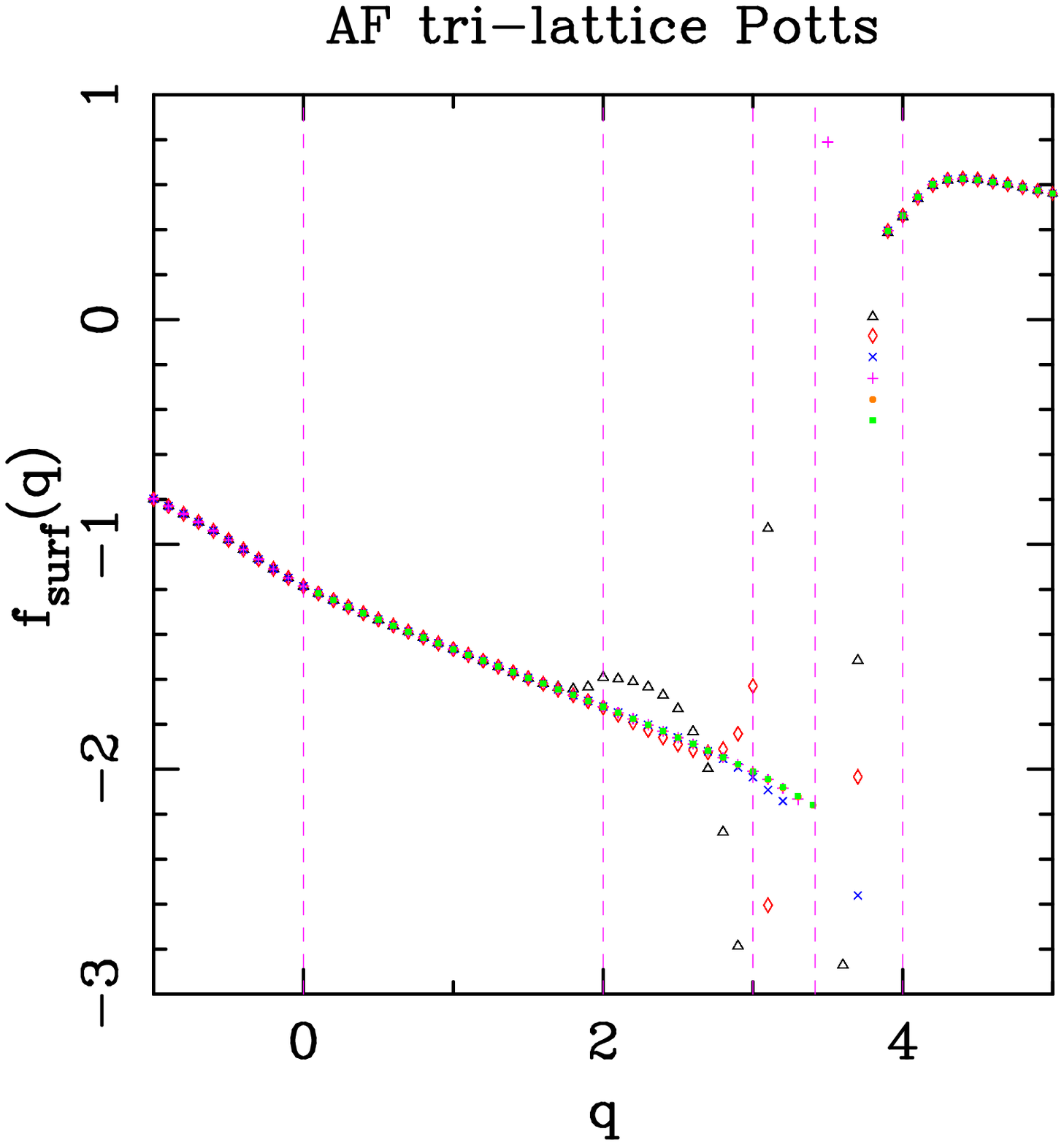} &
   \includegraphics[width=200pt]{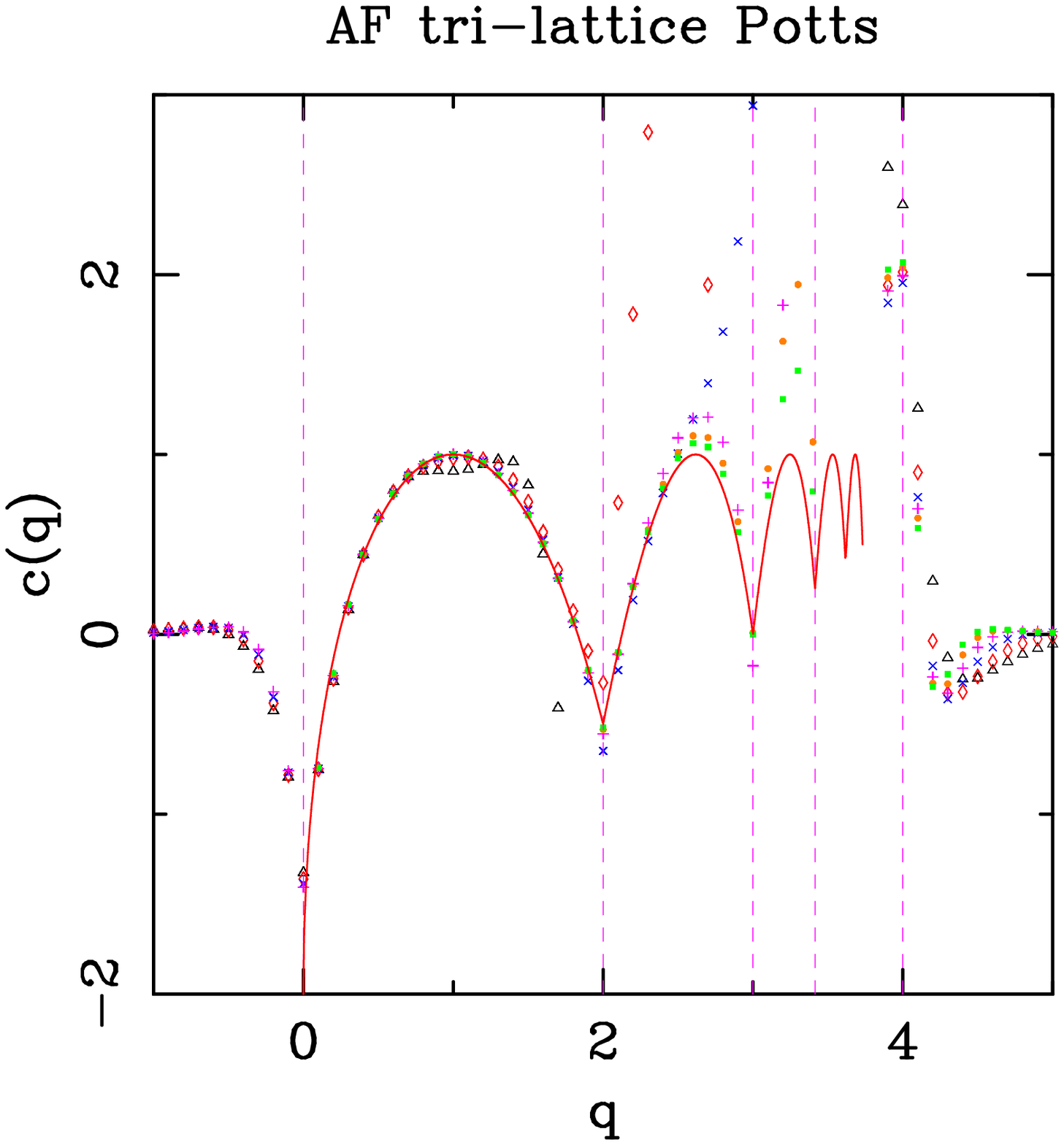}
   \\[1mm]
   \phantom{(((a)}(a)    & \phantom{(((a)}(b) \\[5mm]
\end{tabular}
\caption{\label{figure_C_tri}
Estimates of the surface free energy $f_\text{surf}(q)$ (a) and the central 
charge $c(q)$ (b) on the triangular lattice when 
$f_\text{bulk}(q)$ in the Ansatz \protect\reff{fit_free_energy} 
is fixed to its analytically known value \protect\cite{Baxter_86_87}.
We show the estimates of $f_\text{surf}(q)$ and $c(q)$ for real $q$
in the range $-1\leq q \leq 5$ and different values of 
$L_\text{min} = 2$ (black $\triangle$), 
               $3$ (red $\diamond$),
               $4$ (blue $\times$), 
               $5$ (pink $+$), 
               $6$ (orange $\bullet$), and 
               $7$ (green $\blacksquare$).
The solid (red) line in (b) shows the theoretical prediction for the effective
central charge $c_\text{eff}$ \protect\reff{c_theor}. 
The vertical (pink) dashed lines are as in Figure~\protect\ref{figure_F}.
}
\end{figure}

In Figure~\ref{figure_F}(a) [resp.\ (b)] we plot $f_L(q)$ as a function of $q$ 
for square-lattice (resp.\  triangular-lattice) strips with $2\leq L \leq 11$ 
(resp.\ $2\leq L \leq 10$). It is also useful to use instead of $q$, 
the (real) parameter $n$ defined by
\be
q \;= \; 4 \cos\left( \frac{\pi}{n} \right)^2 \;, \qquad n\geq 2 \; .
\label{def_n}
\ee
Special behavior (phase transitions, vanishing amplitudes, level crossings) is
expected when $n$ takes rational values. This is due to the special form 
\reff{def_alpha_bis} of the amplitudes $\alpha^{(\ell)}$, and to the
particularities of the representation theory of $\text{U}_{\bar{q}} \,
\text{sl}(2)$ when $\bar{q}$ is a root of unity \cite{Pasquier,Saleur_90_91}.
The vertical dashed lines in Figures~\ref{figure_F}--\ref{figure_C_tri} 
correspond to even Beraha numbers $q=B_{2n}$.

For a critical system, conformal field theory predicts the following
finite-size scaling behavior with free transverse boundary conditions 
\cite{Bloete_86,Cardy_88}:\footnote{
 Formula~\protect\reff{fit_free_energy0} is valid for two-dimensional 
 systems only for manifolds with Euler number $\chi=0$ \protect\cite{Cardy_88}. 
 If $\chi\neq 0$, then an additional term $\sim L^{-2}\log L$ appears 
 \protect\cite{Cardy_88}.
}
\be
f_L(q) \;= \; f_\text{bulk}(q) + \frac{f_\text{surf}(q)}{L} + 
                                 \frac{\pi G c(q)}{24 L^2}  + 
                                 o\!\left(L^{-2}\right),
\label{fit_free_energy0}
\ee
where $f_\text{bulk}(q)$ and $f_\text{surf}(q)$ are respectively the bulk and
surface contributions to the free energy in the thermodynamic limit, $c(q)$ is
the central charge, and the geometrical factor $G=1$ (resp.\ $G=2/\sqrt{3}$)
for the square (resp.\ triangular \cite{Indekeu_86}) lattice.\footnote{
  Note that $f_L(q)$ is the free energy per spin, whereas the standard CFT
  formulae require normalization per unit area. To redraw
  Figure~\ref{figure_tri_cyclic_bc} in an undistorted way
  (i.e., with each triangular plaquette being equilateral)
  the time slices must make an angle
  $\alpha = 2\pi/3$ with the transfer direction. The projected width
  of an $L$-spin strip is thus $L \, \sin \alpha = L/G$ lattice spacings.
}
Neglecting all higher-order
corrections on the right-hand side, one can obtain approximations to
$f_\text{bulk}(q)$, $f_\text{surf}(q)$, and $c(q)$ from fits involving three
system sizes: $L=L_\text{min},L_\text{min}+1,L_\text{min}+2$. However,
we find that the $o(L^{-2})$ terms are important and our estimates for 
$c(q)$ have strong finite-size-scaling effects.

In order to obtain accurate estimators for $c(q)$ we need to include the
first correction-to-scaling terms in \reff{fit_free_energy0}. We have used the
following improved Ansatz:
\be
f_L(q) \;= \; f_\text{bulk}(q) + \frac{f_\text{surf}(q)}{L} + 
                                 \frac{\pi G c(q)}{24 L^2}  + 
                                 \frac{A}{L^3} + \frac{B}{L^4} \; .
\label{fit_free_energy}
\ee
The $L^{-3}$ term is the expected (non-universal) contribution of the
non-singular part of the free energy due to the free transverse boundary
conditions (see e.g., the discussion in Privman's review \cite{Privman}). The
last $L^{-4}$ term is a non-universal correction predicted
by conformal field theory.\footnote{ 
  There are indeed many more non-universal corrections terms due to 
  irrelevant operators and some of them might be more relevant than the
  $L^{-4}$ term included in the Ansatz \protect\reff{fit_free_energy}.
  However, we expect that the $L^{-4}$ term can mimic the effect of such
  higher-order terms. 
}

We have obtained approximations to 
$f_\text{bulk}(q)$, $f_\text{surf}(q)$, and $c(q)$ from fits involving five 
system sizes: $L=L_\text{min}+k$ with $0\leq k \leq 4$.
The best approximations for $f_\text{bulk}(q)$, with $L_\text{min}=7$ (resp.\
$L_\text{min}=6$) for the square (resp.\  triangular) lattice,
are displayed as (black) solid circles in Figure~\ref{figure_F}. 
The corresponding estimates for $c(q)$ are shown in Figure~\ref{figure_C},
and for $f_{\rm surf}(q)$ in Figure~\ref{figure_Fs}.
Finally, for the triangular lattice we have also tried a
four-parameter Ansatz, obtained from \reff{fit_free_energy} by fixing
$f_{\rm bulk}(q)$ to its analytically known value \cite{Baxter_86_87}.
The resulting estimates for $f_{\rm surf}(q)$ and $c(q)$
are shown in Figure~\ref{figure_C_tri}.

The estimates for the bulk free energy $f_\text{bulk}(q)$ shown in 
Figure~\ref{figure_F} are rather smooth as a function of $q$, except close to
$q_0$. In the triangular-lattice case
(b), we have also shown the exact value of $f_\text{bulk}(q)$ with 
{\em cylindrical} boundary conditions, as obtained by Baxter
\cite{Baxter_86_87}. The very good agreement supports the idea that
$f_\text{bulk}(q)$ is independent of (reasonable) boundary conditions.
However, for $q\approx q_0$ this agreement is not longer good, due
to large correction-to-scaling effects. In particular, 
for the triangular lattice
our numerical results for $3.5 \ltapprox q \ltapprox 4$ 
do not shed light on the question of which of Baxter's
eigenvalues is dominant where, and hence on the value of $q_0(\text{tri})$
(See \cite{transfer3} for a more comprehensive discussion on this issue). 

The results for $c(q)$ make sense within the critical BK phase, i.e., for 
$0 < q < q_0$, where $q_0(\text{sq})=3$ and $q_0(\text{tri})=2+\sqrt{3}$ 
(but see the Introduction for a discussion of the latter value). 
Estimates for both
lattice types, and for different values of $2\leq L_\text{min}\leq 7$, 
are displayed in Figure~\ref{figure_C}. The variation with respect to 
$L_\text{min}$ is a measure of the size of the residual corrections to scaling.

It should be noted that the effective central charge $c_\text{eff}$
(i.e., the one observed on the above plots) depends on both the ``real''
central charge $c_\text{real}$ (i.e., the one computed from field theory) and 
the critical exponents $\tilde{x}_\ell$ corresponding to the insertion of
$\ell$ bridges. The latter are actually surface critical exponents, since
the transverse boundary conditions are free. We have for the BK phase
\cite{Dup_Sal_87}
\begin{eqnarray}
 c_\text{real}  & = & 1 - \frac{6(n-1)^2}{n}     \label{def_c_real} \\
 \tilde{x}_\ell & = & \frac{\ell^2 - (n-1) \ell}{n} \label{def_x_ell}
\end{eqnarray}
and since $\ell = \lfloor n/2 \rfloor$ from Property~A of 
Section~\ref{disc.phase.diagram}, we find
\be
 c_\text{eff}(q) \;=\; c - 24 \tilde{x}_\ell \;=\; 
 1-6 \frac{(n-1)^2 - 4 \left( \lfloor \frac{n}{2} \rfloor^2 - (n-1)
   \lfloor \frac{n}{2} \rfloor \right)}{n} \; .
\label{c_theor}
\ee
This continuous oscillatory function is shown in 
Figures~\ref{figure_C} and~\ref{figure_C_tri}(b). 
Note that for $n$ an even integer one has $c_\text{eff}=1-6/n$, while for
$n$ an odd integer one has $c_\text{eff} = 1$ independently of $n$.
However, exactly for integer $n$ we do not expect \reff{c_theor} to 
hold (see below). Note that for free longitudinal boundary conditions 
by contrast, only
the $\ell=0$ sector contributes, and thus the effective central charge 
\reff{c_theor} is equal to the BK value \reff{def_c_real}. 

Finally, for values of $q$ outside the limiting curve we generally find that 
the effective central charge is close to zero, consistent with non-critical
behavior. Indeed, with a finite bulk correlation length $\xi$, the scaling
form \reff{fit_free_energy} must be replaced by
\be
f_L(q) \;= \; f_\text{bulk}(q) + \frac{f_\text{surf}(q)}{L} + 
          O\left( e^{-L/\xi} \right) \; .
\label{fit_free_energy_non_critical}
\ee
Note however that for the triangular lattice, we expect the regime
$q_0(\text{tri}) < q \le 4$ to be critical \cite{Nienhuis_82,Dup_Sal_87}.
In particular the case $q=4$ is equivalent to the three-coloring of the
bonds of the hexagonal lattice \cite{Baxter_70}, a critical model with
$c = 2$ \cite{Batchelor_94}.
In addition, the 3-state zero-temperature square-lattice Potts 
antiferromagnetic model can be mapped to the F-model ({\em alias} the
equal-weighted six-vertex model) \cite{Lenard,Baxter_70b}, which is a
critical $c=1$ theory.

In the case of the triangular lattice, we have also tried a four-parameter
Ansatz, obtained from \reff{fit_free_energy} by fixing $f_\text{bulk}(q)$
to its analytically known value \cite{Baxter_86_87}.
In Figure~\ref{figure_C_tri}(a) we have plotted the corresponding estimates for
$f_\text{surf}(q)$, and in Figure~\ref{figure_C_tri}(b) the
estimates for the central charge. While we find a smooth behavior of 
$f_\text{surf}(q)$ for $q\ltapprox 3$ and $q\gtapprox 4$, there 
is a rather abrupt behavior in the regime $3 \ltapprox q \ltapprox 4$. 
In that region we should expect strong finite-size-scaling corrections
due to the proximity of several phase transitions. We conjecture that 
$f_\text{surf}(q)$ has a jump discontinuity at $q=q_0$, while 
$f_\text{bulk}(q)$ is a continuous function of $q$ 
(Baxter \cite{Baxter_86_87} explicitly showed that $f_\text{bulk}(q)$ is 
continuous for the triangular lattice).

Our best numerical results for the central charge contained in 
Figures~\ref{figure_C}(a) (square lattice) and~\ref{figure_C_tri}(b)
(triangular lattice), 
show a remarkable agreement with the theoretical prediction 
for $c_\text{eff}(q)$ \reff{c_theor} in the interval $q\in(0,2)$. 
We also find good agreement with the prediction $c=0$ in the non-critical 
region for $q\ltapprox -0.4$, and $q\gtapprox 3.5$ (resp.\  $q\gtapprox 4.5$) 
for the square (resp.\  triangular) lattice. For $q\ltapprox 0$ and 
$q\gtapprox q_0$ there are strong corrections to scaling because the
central charge is discontinuous at those points. Finally, we remark that 
at $q = q_0$ we find excellent agreement with the theoretical predictions:
$c = 1.00(1)$ for the $q=3$ model on the square lattice, and $c=2.01(5)$ for
the $q=4$ model on the triangular lattice.\footnote{
  At $q=q_0$ we find that the dominant sector is $\ell=0$, thus there
  are no corrections due to the exponents $\widetilde{x}_\ell$
  \protect\reff{def_x_ell}. In addition, when performing the fits close to 
  $q=4$ for the triangular lattice (with the exact value of $f_\text{bulk}(q)$),
  we have found noticeable even--odd oscillations. The estimate $c=2.01(5)$ 
  for $q=4$ has been obtained by fitting data points with $L=4,6,8,10$. 
} 

\subsection{Isolated limiting points and real crossings of the limiting curve}

We recall that isolated limiting points occur when the amplitude of the
dominant eigenvalue vanishes. In Tables~\ref{zeros_sq_R} and~\ref{zeros_tri_R}
we show the real chromatic zeros for strips of size 
$L_{\rm F}\times (kL)_{\rm P}$ with $1\leq k\leq 10$. As explained in 
Ref.~\cite{transfer1}, the convergence to isolated limiting points 
is exponentially fast. 

First note that $\alpha^{(0)}=1$ never vanishes.
Concerning the region containing $I_n$ (with $n=1,2,\ldots,n_\text{max}$),
we remark that as $q$ moves through the interval $[B_{2n},B_{2n+2}]$, 
$\alpha^{(n)}$ increases monotonically from $-1$ to $1$, passing through $0$
when $q=B_{2n+1}$. (This statement can be easily proved from the first
of the definitions in \reff{def_Un}.) In conjunction with the fact that
all zeros of the Chebyshev polynomials of the second kind are real,
the Properties~A--B of Section~\ref{disc.phase.diagram} give the corollary:
\begin{itemize}
 \item An isolated limiting point must necessarily be real.
 \item No isolated limiting point exists in the region containing
 $I_0$. For $n=1,2,\ldots,n_\text{max}-1$, the region containing $I_n$
 has exactly one isolated limiting point which is $B_{2n+1}$. The region
 containing $I_{n_\text{max}}$ has the isolated limiting point
 $B_{2n_\text{max}+1}$ provided that $B_{2n_\text{max}+1}<q_0(L)$.
\end{itemize}
As before, we conjecture the validity of these properties for any $L$, and
for the entire BK phase.

By definition, the limiting curve crosses the real $q$-axis in $q_0(L)$.
Further crossings are possible for $0 \le q < q_0(L)$, provided that
the two dominant eigenvalues become degenerate and that the corresponding
amplitudes differ only by a sign. From \reff{def_Un} it is easy to see 
that the last condition is satisfied for $q=B_{2n}$ with
$n=1,2,\ldots,n_\text{max}$. That the first condition is also satisfied
follows from the $\text{U}_{\bar{q}} \, \text{sl}(2)$ symmetry
\cite{Saleur_90_91}, and we have also verified this directly for all
strips studied in this paper.

In the thermodynamic limit we have $q_0(\text{sq})=3$ and
$q_0(\text{tri})=B_{12}=2+\sqrt{3}$ (as discussed in the Introduction). For
the square lattice chromatic polynomial, we thus predict that the limiting
curve crosses the real $q$-axis three times (at $B_2$, $B_4$ and $B_6$),
dividing the real $q$-axis into three different phases (corresponding to
$\ell=0,1,2$). Furthermore, there are
two isolated limiting points (at $B_3$ and $B_5$). For the triangular lattice,
the limiting curve crosses the real $q$-axis seven times (at
$B_2,B_4,\ldots,B_{12}$, and at $B_\infty$), dividing the real $q$-axis
into seven different phases (corresponding to $\ell=0,\ldots,6$). 
Furthermore, there are five isolated limiting
points (at $B_3,B_5,\ldots,B_{11}$).

\subsection{Cancellations for $\bm{q=B_n}$, $\bm{n}$ integer}

One important point about the BK phase is that it is {\em not} defined
for $q = B_n$ with $n$ integer \cite{Saleur_90_91}. In these cases there
are cancellations among eigenvalues that give rise to a different 
physics.

The most trivial case corresponds to $q=B_2=0$. At this value the
coefficients $\alpha^{(\ell)}$ \reff{def_alpha} take the form
\be
\alpha^{(\ell)}(q=0) \;=\; \begin{cases}
                       1 & \qquad \text{if $\mod(\ell,2)=0$} \\
                      -1 & \qquad \text{otherwise}
                         \end{cases}
\ee
We have seen that the eigenvalues become degenerate in pairs, one belonging 
to an odd-$\ell$ sector, and the other one to an even-$\ell$ sector. Thus
there is a complete cancellation of all eigenvalues and the result is
as expected, $Z_{G_n}(0) = 0$.

At $q=B_3=1$, the coefficients take the form 
\be
\alpha^{(\ell)}(q=1) \;=\; \begin{cases}
                       1 & \qquad \text{if $\mod(\ell,3)=0$} \\
                       0 & \qquad \text{if $\mod(\ell,3)=1$} \\
                      -1 & \qquad \text{if $\mod(\ell,3)=2$} \\
                         \end{cases}
\ee
The (non-zero) eigenvalues with non-zero amplitudes become degenerate 
forming $2n$-tuples. Exactly $n$ of them have $\alpha=1$, 
and the other half, $\alpha=-1$. Thus, there is an exact cancellation of
all contributions, giving $Z_{G_n}(q=1)=0$. 

For $q=B_4=2$, the coefficients take the form
\be
\alpha^{(\ell)}(q=2) \;=\; \begin{cases}
                   1 & \qquad \text{if $\mod(\lfloor \ell/2 \rfloor,2)=0$}\\
                  -1 & \qquad \text{otherwise}
                     \end{cases}
\ee
In the triangular-lattice case we find a similar cancellation as for the
$q=1$ case: the non-zero eigenvalues with non-zero amplitude form $2n$-tuples
whose net contribution is zero. Thus, in this case we also have the expected
result $Z_{G_n}(q=2)=0$. For the square lattice we find a non-critical 
behavior: there is an exact cancellation of all non-zero eigenvalues, 
except for those taking values $\lambda=\pm 1$. The net contribution
for a $L\times n$ lattice is always of the form 
$Z_{L\times n} = 2$ if $n$ is even, and $Z_{L\times n} = 0$ if $n$ is odd.

The last trivial case corresponds to $q=B_6=3$ for the triangular lattice.
This is very similar to $q=2$ for the square lattice: all non-zero eigenvalues
cancel exactly, except those with $\lambda=\pm 1$. Again, for each triangular
lattice of size $L\times n$ the net contribution is given by
$Z_{L\times n} = 6$ if $\mod(n,3)=0$, and $Z_{L\times n} = 0$ otherwise. 

For the square lattice the only non-trivial case is $q=B_5$.\footnote{
  The value $q=B_6=3$ is also non-trivial, as we expect a critical point
  for the 3-state antiferromagnet at zero temperature. However, we have 
  found that the dominant eigenvalue (for all widths $L\leq 8$) belongs to 
  the $\ell=0$ sector, and $\alpha^{(0)}\neq 0$.  
}
Up to $L=5$ we find that the dominant eigenvalue (belonging to the $\ell=0$ 
sector) actually contributes to the partition function. But for 
$6\leq L\leq 8$, we find that the dominant eigenvalue belongs to the 
$\ell=2$ sector,
and $\alpha^{(2)}(q=B_5)=0$. The first non-zero contribution corresponds to
a singlet belonging to $\lambda=0$. The numerical results are given in 
Table~\ref{table_F_qBn}. 

For the triangular lattice, the value $q=B_5$ is also non-trivial. The 
dominant eigenvalue belongs to the $\ell=2$ sector, but again 
$\alpha^{(2)}(q=B_5)=0$. In this case, as $L$ increases the first eigenvalue
contributing to the partition function is deeper into the spectrum of the
transfer matrix. For even $L$, this eigenvalue is complex and belongs to 
$\ell=1$, and for odd $L$, it is real and belongs to $\ell=0$. See
Table~\ref{table_F_qBn} for the numerical results.

For $q=B_7$, the dominant eigenvalue has $\ell=0$ for $2\leq L\leq 4$,
and $\ell=3$ for $5\leq L\leq 8$. This latter case is the interesting one, as
the corresponding amplitude vanishes. We find that the first eigenvalue that
gives a net contribution to the partition function belongs to $\ell=0$, and
as $L$ increases, its position in the transfer-matrix spectrum lowers
rapidly. See Table~\ref{table_F_qBn} for the numerical results. 

These results show that exactly at $q=B_n$ the physics of the system
is not described by the BK phase. In some cases $q=0,1,2$ (and $q=3$ for the
triangular lattice) the partition function is trivial. In other cases,
$q=B_5$ and $q=B_7$ (the latter one only for the triangular lattice), 
the free energy comes from a subdominant eigenvalue of the transfer matrix.
We refrain from extracting the effective central charge from the results 
displayed in Table~\ref{table_F_qBn}; as the data shows strong parity effects
and the number of data points is rather limited. 

\appendix
%
%
\section{Alternative construction of the transfer matrix for square-lattice 
         strips with cyclic boundary conditions} \label{sec.sq.cyclic.alt}

We have already discussed in Section~\ref{sec.sq.cyclic} 
how to build the transfer matrix for a square-lattice strip with 
cyclic boundary conditions: the idea was to 
leave the bottom row fixed and act only on the top row. This procedure leads
to fairly large transfer matrices with a block structure. Each block is 
labeled by the bottom-row connectivity and the number and position of the 
bridges joining the top and bottom rows (modulo reflection symmetry) 
This transfer matrix is {\em not} invariant under the operation of 
interchanging the top and bottom rows. This latter 
operation can be viewed as a reflection with respect to the middle of the
strip (in the longitudinal direction). For simplicity, we will call it 
a $t$-reflection. 

Smaller transfer matrices invariant under $t$-reflections can be built as 
follows. The idea is to obtain a transfer matrix acting on both the top and
bottom rims. Thus, the matrices $\mathsf{H}$ and $\mathsf{V}$ take  
the form
\begin{subeqnarray}
  \widehat{\mathsf{H}}(m)  &=& \prod\limits_{i=1}^{m-1} \QQ_{i,i+1} \cdot 
                               \prod\limits_{i=1}^{m-1} \QQ_{i',(i+1)'} 
                               \slabel{def_Hbis} \\
  \widehat{\mathsf{V}}(m)  &=& \prod\limits_{i=1}^{m} \P_{i} \cdot
                               \prod\limits_{i=1}^{m} \P_{i'} \cdot       
                               \slabel{def_Vbis}
  \label{def_HVbis}
\end{subeqnarray}
These two matrices act on the top and bottom rows simultaneously, and are 
by construction $t$-invariant (i.e. invariant under a $t$-reflection 
$i \leftrightarrow i'$). The $t$-invariant transfer matrix is defined as
\be
\widehat{\mathsf{T}}(m) \;=\; \widehat{\mathsf{V}}(m) \cdot 
                              \widehat{\mathsf{H}}(m) \; .
\label{def_Tbis}
\ee
Note that as we are acting on 
both rims simultaneously, we can only obtain the partition function for 
strips of even length. Then, the partition function for a square-lattice
strip of size $m\times 2n$ is given by
\be
  P_{m\times 2n}(q) \; = \; \bm{u}^\text{T} \cdot \widehat{\mathsf{H}}(m)
                                            \cdot \widehat{\mathsf{T}}(m)^n 
                                            \cdot \bm{v}_\text{id}
\label{def_Pbis}
\ee
where the matrices $\widehat{\mathsf{H}}$ and $\widehat{\mathsf{T}}$ are 
given by \reff{def_Hbis}/\reff{def_Tbis}. 

As we are considering the chromatic--polynomial case, the matrix 
$\widehat{\mathsf{H}}$ is a projector, and then it is more convenient 
to use the modified transfer matrix (see Section~\ref{sec.sq.cyclic})
\be
\widehat{\mathsf{T}}'(m) \;=\; \widehat{\mathsf{H}}(m) \cdot
                               \widehat{\mathsf{V}}(m) \cdot 
                               \widehat{\mathsf{H}}(m)
\ee
and the connectivity-state basis 
\be
   \bm{w}_\mathcal{P} \;=\; \widehat{\mathsf{H}}(m) \cdot \bm{v}_\mathcal{P}
\ee
As in Section~\ref{sec.sq.cyclic}, we will consider hereafter this modified 
transfer matrix with the prime dropped. 

The fact the transfer matrix \reff{def_Tbis} is now $t$-invariant  
can be used to obtain a smaller transfer matrix. We can pass to a new basis 
consisting on connectivities that are either even or odd under 
$t$-reflection.\footnote{ 
  Let us denote $\{\bm{v}_i\}$ the connectivity basis. Some of these elements
  are $t$-invariant; the rest can be grouped into pairs $(\bm{v}_a,\bm{v}_b)$ 
  that map into each other under a $t$-reflection. A basis for the 
  even (i.e., $t$-reflection invariant) subspace is then given
  by the elements of the first set together with the combinations 
  $\bm{v}_a + \bm{v}_b$ from the second set. A basis for the 
  odd subspace is given by the combinations 
  $\bm{v}_a - \bm{v}_b$ from the second set.
} 
In this new basis the transfer matrix takes the block-diagonal form
\be
\widehat{\mathsf{T}}(m) \;=\; \left( \begin{array}{cc} 
                \widehat{\mathsf{T}}_{+}(m) & 0 \\
                            0               & \widehat{\mathsf{T}}_{-}(m) 
                            \end{array} 
                    \right)
\ee
where $\widehat{\mathsf{T}}_{+}$ (resp.\  $\widehat{\mathsf{T}}_{-}$) 
corresponds to the subspace which is even (resp.\ odd) under
a $t$-reflection. The left and right vectors take the form 
\begin{subeqnarray}
\bm{w}_\text{id}^T &=& \left( \bm{w}_{\text{id},+}^T,0 \right) \\
\bm{u}^T           &=& \left( \bm{u}_{+}^T          ,0 \right) \,, 
\end{subeqnarray} 
as the start vector $\bm{w}_\text{id}$ is clearly $t$-reflection invariant,
and the action of $\bm{u}^T$ on any $t$-reflection-odd connectivity is 
identically zero. Thus, the contribution of the $t$-reflection-odd subspace to
the partition function vanishes, and we can work entirely within the 
$t$-reflection-invariant subspace. The dimension of this subspace for a 
square-lattice of width $m$ is denoted by $\text{SqRingT}(m)$ in 
Table~\ref{table_dim_T_cyclic_blocks_TB}. We expect that for large
enough width $m$, the dimension of the $t$-reflection-invariant subspace will
be approximately one half of the original dimension.

This transfer matrix also has a block structure. In this case, due to its
invariance under $t$-reflection, the blocks are labeled merely by the number 
of bridges $\ell$ joining the top and bottom rows. Thus, for a strip of width
$m$, the transfer matrix $\widehat{\mathsf{T}}(m)$ can be split into 
$\ell=m+1$ blocks, as in eq.~\reff{T_blocks1}: 
\be
\widehat{\mathsf{T}}(m) \;=\; \left( \begin{array}{cccc}
   \widehat{T}_{m,m}   & 0                     & \ldots & 0 \\
   \widehat{T}_{m-1,m} & \widehat{T}_{m-1,m-1} & \ldots & 0 \\
   \vdots              & \vdots                &        & \vdots\\
   \widehat{T}_{0,m}   & \widehat{T}_{0,m-1}   & \ldots & \widehat{T}_{0,0}
                           \end{array}
                    \right)
\label{That_blocks1}
\ee
The difference with the method discussed in Section~\ref{sec.sq.cyclic} is 
that now each diagonal block $\widehat{T}_{\ell,\ell}$ does not have 
in general any sub-block structure, like in \ref{T_blocks2}. 
The dimension of the largest block of  $\widehat{\mathsf{T}}$ is in general 
greater than the dimension of the largest
block of $\mathsf{T}$ (see Tables~\ref{table_dim_T_cyclic_blocks_TB} 
and~\ref{table_blocks_sq}). As an example, for width $m=5$,
the largest block of the matrix $\mathsf{T}(5)$ has dimension 21; but the
dimension of the largest block of $\widehat{\mathsf{T}}(5)$ is 127. 
The fact that the dimension of the largest block for $\widehat{\mathsf{T}}$ 
grows very quickly with the strip width, makes the symbolic computation of its 
characteristic polynomial (and hence, of its eigenvalues) a difficult task.

The simplest case is the strip of width $m=2$. The basis \reff{Tsq2R} we used
to construct the matrix $\mathsf{T}(2)$ is already 
$t$-reflection invariant. In this basis, the matrix $\widehat{\mathsf{T}}(2)$
takes the form 
\be
\widehat{\mathsf{T}}(2) \; = \; \left( \begin{array}{c|cc|c}
 1        & 0          & 0          & 0 \\
\hline
q-3       & q^2 -4q+5  & 4-2q       & 0 \\
-1        & 4-2q       & q^2 -4q+5  & 0 \\
\hline
q^2 -5q+8 & T_{42}     & T_{42}     & (q^2 -3q + 3)^2\\
            \end{array}\right)
\label{Tsq2Rbis}
\ee
with 
\be
T_{42} \;=\; 2(q-2)(q^2-4q+6)
\ee
The right $\bm{w}_\text{id}$ and left $\bm{u}$ vectors are equal to those
given by \reff{UVsq2R}. It is easy to verify that   
\be
\widehat{\mathsf{T}}(2) \;=\; \mathsf{T}(2)^2
\ee
This result is expected as both matrices are written in the same basis and
each time we act with $\widehat{\mathsf{T}}(2)$ we add two layers to the strip. 

The first non-trivial case corresponds to the strip of width $m=3$. In this 
case the $t$-reflection-invariant basis has dimension 15. We have found
four blocks labeled by the number of bridges $\ell$: 

\begin{itemize}
\item  $\ell=3$: This block is one-dimensional and the eigenvalue is
       $\lambda=1$.

\item  $\ell=2$: This block is four-dimensional and the eigenvalues are
       $\lambda=(q-4)^2, (q-1)(q-4), (q-2)^2$, and $(q-1)^2$.  

\item  $\ell=1$: This block has dimension 10. One eigenvalue is 
       $\lambda=(q-1)(q^3-6q^2+13q-11)$ and the other ones come from
       solving two different equations of third order. 

\item  $\ell=0$: This block is three-dimensional. One eigenvalue is
       $\lambda=(q-2)^4$ and the other two are the solutions of a second-order
       equation. 
\end{itemize}

Thus, we find 15 distinct eigenvalues. This number is larger than the 
number of distinct eigenvalues for this strip (see eq.~\reff{def_Poly_3R}).  
Indeed, we have verified that the amplitudes
corresponding to five of the above eigenvalues are identically zero. 
As a matter of fact, the zero amplitudes correspond to the eigenvalues that
are not positive definite for real $q$ (e.g., $\lambda=(q-1)(q-4)$ and
$(q-1)(q^3-6q^2+13q-11)$). This is expected as the transfer matrix
$\widehat{\mathsf{T}}$ is physically equivalent to $\mathsf{T}^2$: hence,
the former must contain the square of the eigenvalues of the latter. 

We have checked that this procedure works also for the square-lattice 
strip of width $m=4$ and the result agrees with the one presented in the
text.

\bigskip

\noindent
{\bf Remark:} This procedure can also be implemented for the triangular
lattice. In this case, the relevant symmetry is a $t$-reflection followed
by a (standard) reflection with respect to the center of the strip. This 
second step is needed to bring the diagonal bonds into the right position.

%
%
\section*{Acknowledgments}

We wish to thank 
Jean--Fran\c{c}ois Richard for some interesting comments on an earlier
version of this paper, 
Hubert Saleur for sharing his insight in quantum groups, and
Alan Sokal for discussions and his collaborations on related projects.
We also thank an anonymous referee for many useful suggestions that 
improved the presentation of our results.
This research was partially supported by U.S. National Science 
Foundation grant PHY-0116590 and by CICyT (Spain) grant 
MTM2004-01728 (J.S.).

\clearpage
%
%

\clearpage
%
%
%
%
\begin{table}
\centering 
\begin{tabular}{|r|r|r|r|r|}
\hline\hline
 $m$ & $C_{2m}$
     & TriRing  
     & SqRing     
     & SqRing$'$ \\ 
\hline\hline
  1  &          2 &       2 &       2 &     2 \\
  2  &         14 &       6 &       4 &     4 \\
  3  &        132 &      30 &      20 &    10 \\
  4  &       1430 &     178 &      94 &    26 \\
  5  &      16796 &    1158 &     614 &    70 \\
  6  &     208012 &    7986 &    4028 &   192 \\
  7  &    2674440 &   57346 &   28940 &   534 \\
  8  &   35357670 &  424206 &  212370 &  1500 \\
  9  &  477638700 & 3210246 & 1607246 &  4246 \\
 10  & 6564120420 &         &         & 12092 \\
\hline
\end{tabular}
\caption{\label{table_dim_T_cyclic}
   Dimensionality of the transfer matrix for cyclic boundary conditions.
   For each square-lattice width $m$ we give the number of
   non-crossing partitions $C_{2m}$,
   non-crossing non-nearest-neighbor partitions TriRing,
   the number of classes of partitions invariant under reflections SqRing,
   and the number of distinct eigenvalues SqRing$'$ \cite{Shrock_01b}.
}
\end{table}

%
%
\begin{table}
\centering 
\begin{tabular}{|r|rrrrrrrrrr|r|}
\hline\hline
$m$  &$N_0$& $N_1$& $N_2$& $N_3$& $N_4$& $N_5$& $N_6$& $N_7$ & $N_8$ & $N_9$ &
     \# blocks \\ 
\hline\hline
  2  &   1 &    1 &    1 &      &      &      &      &     &   &   &    3 \\
  3  &   2 &    3 &    2 &    1 &      &      &      &     &   &   &    8 \\
  4  &   3 &    5 &    5 &    2 &   1  &      &      &     &   &   &   16 \\
  5  &   7 &   13 &   12 &    8 &   3  &    1 &      &     &   &   &   44 \\
  6  &  13 &   27 &   30 &   20 &  11  &    3 &    1 &     &   &   &  105 \\
  7  &  32 &   70 &   77 &   61 &  34  &   15 &    4 &  1  &   &   &  294 \\
  8  &  70 &  166 &  199 &  163 & 106  &   49 &   19 &  5  & 1 &   &  777 \\
  9  & 179 &  435 &  528 &  468 & 318  &  174 &   72 & 24  & 5 & 1 & 2204 \\
\hline
\end{tabular}
\caption{\label{table_blocks_sq}
  Block structure of the transfer matrix of a cyclic square-lattice
  strip of width $m$ as a function of number of bridges $\ell$. 
  For each strip width $m$, we quote the number of
  blocks $N_\ell$ for a given number of bridges $\ell$, and
  the total number of blocks (\# blocks).
}
\end{table}

\clearpage
%
%
\begin{table}
\centering
\begin{tabular}{|r|rrrrrrrrrr|r|}
\hline\hline
$m$  &$L_0$& $L_1$& $L_2$& $L_3$& $L_4$& $L_5$& $L_6$& $L_7$& $L_8$ & $L_9$ &
      SqRing$'$ \\
\hline\hline
  2  &   1 &    2 &    1 &      &      &      &      &     &   &    &    4 \\
  3  &   2 &    4 &    3 &    1 &      &      &      &     &   &    &   10 \\
  4  &   4 &    9 &    8 &    4 &   1  &      &      &     &   &    &   26 \\
  5  &   9 &   21 &   21 &   13 &   5  &    1 &      &     &   &    &   70 \\
  6  &  21 &   51 &   55 &   39 &  19  &    6 &    1 &     &   &    &  192 \\
  7  &  51 &  127 &  145 &  113 &  64  &   26 &    7 &  1  &   &    &  534 \\
  8  & 127 &  323 &  385 &  322 & 203  &   97 &   34 &  8  & 1 &    & 1500 \\
  9  & 323 &  835 & 1030 &  910 & 622  &  334 &  139 & 43  & 9 &  1 & 4246 \\
\hline
\end{tabular}
\caption{\label{table_blocks_tri}
  Block structure of the transfer matrix of cyclic square- and
  triangular-lattice
  strips of width $m$ as a function of number of bridges $\ell$. 
  For each strip width $m$, we quote the number of
  different eigenvalues $L_\ell$ for a given number of bridges $\ell$, and
  the total number of distinct eigenvalues SqRing$'(m)$. For the triangular
  lattice, the number of blocks $N_\ell$ for a given value of $\ell$ coincide
  with the number of distinct eigenvalues $L_\ell$ for the same value of
  $\ell$.
}
\end{table}

%
%
\def\kk{\phantom{$1$}}
\begin{table}
\centering
\begin{tabular}{|r||c|c|c|c|c|l||c|}
\cline{2-8}
\multicolumn{1}{c||}{\mbox{}}&
\multicolumn{6}{|c||}{Eigenvalue-Crossing Curves $\mathcal{B}$} &
\multicolumn{1}{|c|}{Isolated Points}\\
\hline\hline
Lattice    & \# C & \# E & \# T & \# D & \# ER &\multicolumn{1}{|c||}{$q_0$}
           & \# RI \\
\hline\hline
$2_\text{R}$&  1 & 0 &\kk 2 & 1 & 3 & 2            &  1 \\
$3_\text{R}$&  1 & 0 &   10 & 0 & 6 & 2.3365442725 &  1 \\
$4_\text{R}$&  1 & 0 &\kk 6 & 0 & 4 & 2.4928455591 &  1 \\
$5_\text{R}$&    $1^\dagger$ 
            &    $0^\dagger$
            &\kk $6^\dagger$
            &    $0^\dagger$   
            &    $4^\dagger$
            &    2.5823854661 &  1 \\
$6_\text{F}$&    $1^\dagger$
            &    $0^\dagger$
            &   $10^\dagger$
            &    $0^\dagger$  
            &    $6^\dagger$ 
            &    2.6460783059 &  2 \\
$7_\text{F}$&    $1^\dagger$
            &    $0^\dagger$
            &   $14^\dagger$
            &    $0^\dagger$
            &    $8^\dagger$
            &    2.6927280047 &  2 \\
$8_\text{F}$&    &   &      &   &   & 2.7287513178 &  2 \\
\hline\hline
\end{tabular}
\caption{\label{table_summary_sq}
   Summary of qualitative results for the square-lattice 
   eigenvalue-crossing curves $\mathcal{B}$
   and for the isolated limiting points of zeros.
   For each cyclic square-lattice strip considered in this paper,
   we give the number of connected components of $\mathcal{B}$ (\# C),  
   the number of endpoints (\# E),
   the number of T points (\# T),
   the number of double points (\# D),
   and the number of enclosed regions (\# ER).
   We also give the value $q_0$ which is the largest real value where 
   $\mathcal{B}$ intersects the real axis, and the   
   number of real isolated limiting points of zeros (\# RI). 
   The symbol $^\dagger$ indicates uncertain results.   
}
\end{table}

\clearpage
%
%
\begin{table}
\centering
\scriptsize
\begin{tabular}{|l|l|l|}
\hline\hline
 Lattice & 3rd Zero  & 4th Zero  \\
\hline\hline
$ 3_{\rm F}\times 3_{\rm P}$ &  2   &  2.453397651516  \\
$ 3_{\rm F}\times 6_{\rm P}$ &  2.055981832687  &  2.096994387849  \\
$ 3_{\rm F}\times 9_{\rm P}$ &  2   &   \\
$ 3_{\rm F}\times 12_{\rm P}$ &  2.000122296702  &  2.293027675252  \\
$ 3_{\rm F}\times 15_{\rm P}$ &  2   &   \\
$ 3_{\rm F}\times 18_{\rm P}$ &  2.000001089940  &  2.317707700956  \\
$ 3_{\rm F}\times 21_{\rm P}$ &  2   &   \\
$ 3_{\rm F}\times 24_{\rm P}$ &  2.000000011921  &  2.325316414328  \\
$ 3_{\rm F}\times 27_{\rm P}$ &  2   &   \\
$ 3_{\rm F}\times 30_{\rm P}$ &  2.000000000143  &  2.328289813540  \\
\hline
$ 4_{\rm F}\times 4_{\rm P}$ &   &   \\
$ 4_{\rm F}\times 8_{\rm P}$ &  2.000043877962  &  2.370251603751  \\
$ 4_{\rm F}\times 12_{\rm P}$ &  2.000000074707  &  2.430973630621  \\
$ 4_{\rm F}\times 16_{\rm P}$ &  2.000000000188  &  2.453300957620  \\
$ 4_{\rm F}\times 20_{\rm P}$ &  2.000000000001  &  2.464208849270  \\
$ 4_{\rm F}\times 24_{\rm P}$ &  2.000000000000  &  2.470471774947  \\
$ 4_{\rm F}\times 28_{\rm P}$ &  2.000000000000  &  2.474474352391  \\
$ 4_{\rm F}\times 32_{\rm P}$ &  2.000000000000  &  2.477235439183  \\
$ 4_{\rm F}\times 36_{\rm P}$ &  2.000000000000  &  2.479250359727  \\
$ 4_{\rm F}\times 40_{\rm P}$ &  2.000000000000  &  2.480784447068  \\
\hline
$ 5_{\rm F}\times 5_{\rm P}$ &  2   &   \\
$ 5_{\rm F}\times 10_{\rm P}$ &  2.000000004274  &  2.485912024903  \\
$ 5_{\rm F}\times 15_{\rm P}$ &  2   &   \\
$ 5_{\rm F}\times 20_{\rm P}$ &  2.000000000000  &  2.546528797961  \\
$ 5_{\rm F}\times 25_{\rm P}$ &  2   &   \\
$ 5_{\rm F}\times 30_{\rm P}$ &  2.000000000000  &  2.560437990844  \\
$ 5_{\rm F}\times 35_{\rm P}$ &  2   &   \\
$ 5_{\rm F}\times 40_{\rm P}$ &  2.000000000000  &  2.566264491529  \\
$ 5_{\rm F}\times 45_{\rm P}$ &  2   &   \\
$ 5_{\rm F}\times 50_{\rm P}$ &  2.000000000000  &  2.569485040710  \\
\hline
$ 6_{\rm F}\times 6_{\rm P}$ &  2.000004484676  &  2.407498857052  \\
$ 6_{\rm F}\times 12_{\rm P}$ &  2.000000000000  &  2.559303044172  \\
$ 6_{\rm F}\times 18_{\rm P}$ &  2.000000000000  &  2.591819531465  \\
$ 6_{\rm F}\times 24_{\rm P}$ &  2.000000000000  &  2.604283491675  \\
$ 6_{\rm F}\times 30_{\rm P}$ &  2.000000000000  &  2.610364438319  \\
$ 6_{\rm F}\times 36_{\rm P}$ &  2.000000000000  &  2.613696502926  \\
$ 6_{\rm F}\times 42_{\rm P}$ &  2.000000000000  &  2.615612713769  \\
$ 6_{\rm F}\times 48_{\rm P}$ &  2.000000000000  &  2.616718339556  \\
$ 6_{\rm F}\times 54_{\rm P}$ &  2.000000000000  &  2.617340472059  \\
$ 6_{\rm F}\times 60_{\rm P}$ &  2.000000000000  &  2.617677872418  \\
\hline
$ 7_{\rm F}\times 7_{\rm P}$ &  2   &   \\
\hline
\hline
 Beraha &2   & 2.618033988750  \\
\hline
\end{tabular}
\caption{\label{zeros_sq_R}
  Real zeros of the chromatic polynomial $P_{m\times n}(q)$ of a square-lattice
  strip of width $m$ and lengths $n= k\times m$ ($k=1,\ldots,10$) with 
  cyclic boundary conditions. A blank means that the zero in question 
  is absent. The first two real zeros $q=0,1$ are exact on all lattices.
  For $m=2$ we have only found these two first zeros. ``Beraha'' indicates 
  the Beraha numbers $B_4=2$ and $B_5=(3+\sqrt{5})/2$. 
}
\end{table}

\clearpage
%
%
\begin{table}
\centering
\vspace*{-1cm}
\scriptsize
\begin{tabular}{|l|l|l|l|l|}
\hline\hline
 Lattice & 4th Zero  & 5th Zero  & 6th Zero  & 7th Zero  \\
\hline\hline
$ 2_{\rm F}\times 2_{\rm P}$ &  3   &   &   &   \\
$ 2_{\rm F}\times 4_{\rm P}$ &  2.515844688131  &  3   &  3.408771911807  &   \\
$ 2_{\rm F}\times 6_{\rm P}$ &  2.640265063616  &   &   &   \\
$ 2_{\rm F}\times 8_{\rm P}$ &  2.616262917475  &  2.751508037623  &  3   &   \\
$ 2_{\rm F}\times 10_{\rm P}$ &  2.614657987429  &  3   &  3.138323546170  &   \\
$ 2_{\rm F}\times 12_{\rm P}$ &  2.621439264722  &   &   &   \\
$ 2_{\rm F}\times 14_{\rm P}$ &  2.616560224418  &  2.887655738411  &  3   &   \\
$ 2_{\rm F}\times 16_{\rm P}$ &  2.618700179921  &  3   &  3.083264711515  &   \\
$ 2_{\rm F}\times 18_{\rm P}$ &  2.617797220618  &   &   &   \\
$ 2_{\rm F}\times 20_{\rm P}$ &  2.618107109614  &  2.926805351303  &  3   &   \\
\hline
$ 3_{\rm F}\times 3_{\rm P}$ &  2.546602348484  &   &   &   \\
$ 3_{\rm F}\times 6_{\rm P}$ &  2.617993627116  &   &   &   \\
$ 3_{\rm F}\times 9_{\rm P}$ &  2.618034110186  &  2.937234729089  &   &   \\
$ 3_{\rm F}\times 12_{\rm P}$ &  2.618033988011  &   &   &   \\
$ 3_{\rm F}\times 15_{\rm P}$ &  2.618033988754  &  2.965427896259  &   &   \\
$ 3_{\rm F}\times 18_{\rm P}$ &  2.618033988750  &   &   &   \\
$ 3_{\rm F}\times 21_{\rm P}$ &  2.618033988750  &  2.976132937129  &   &   \\
$ 3_{\rm F}\times 24_{\rm P}$ &  2.618033988750  &   &   &   \\
$ 3_{\rm F}\times 27_{\rm P}$ &  2.618033988750  &  2.981774802850  &   &   \\
$ 3_{\rm F}\times 30_{\rm P}$ &  2.618033988750  &   &   &   \\
\hline
$ 4_{\rm F}\times 4_{\rm P}$ &  2.617986010522  &  3   &  3.465246100723  &   \\
$ 4_{\rm F}\times 8_{\rm P}$ &  2.618033988761  &  3   &  3.230317951180  &   \\
$ 4_{\rm F}\times 12_{\rm P}$ &  2.618033988750  &  3.000090191719  &  3.165028046404  &   \\
$ 4_{\rm F}\times 16_{\rm P}$ &  2.618033988750  &  3   &  3.231342732991  &   \\
$ 4_{\rm F}\times 20_{\rm P}$ &  2.618033988750  &  3   &  3.219343048602  &   \\
$ 4_{\rm F}\times 24_{\rm P}$ &  2.618033988750  &  3.000000003255  &  3.200001385558  &   \\
$ 4_{\rm F}\times 28_{\rm P}$ &  2.618033988750  &  3   &  3.219217913659  &   \\
$ 4_{\rm F}\times 32_{\rm P}$ &  2.618033988750  &  3   &  3.217140972107  &   \\
$ 4_{\rm F}\times 36_{\rm P}$ &  2.618033988750  &  3.000000000000  &  3.210324007476  &   \\
$ 4_{\rm F}\times 40_{\rm P}$ &  2.618033988750  &  3   &  3.218488888204  &   \\
\hline
$ 5_{\rm F}\times 5_{\rm P}$ &  2.618033990394  &  3   &   &   \\
$ 5_{\rm F}\times 10_{\rm P}$ &  2.618033988750  &  3   &  3.246585484861  &   \\
$ 5_{\rm F}\times 15_{\rm P}$ &  2.618033988750  &  2.999999999573  &  3.250186825428  &  3.287751313682  \\
$ 5_{\rm F}\times 20_{\rm P}$ &  2.618033988750  &  3   &  3.246967522201  &   \\
$ 5_{\rm F}\times 25_{\rm P}$ &  2.618033988750  &  3   &  3.246982088945  &  3.332102289607  \\
$ 5_{\rm F}\times 30_{\rm P}$ &  2.618033988750  &  3.000000000000  &  3.246977286183  &   \\
$ 5_{\rm F}\times 35_{\rm P}$ &  2.618033988750  &  3   &  3.246979668997  &  3.329353767785  \\
$ 5_{\rm F}\times 40_{\rm P}$ &  2.618033988750  &  3   &  3.246979593498  &   \\
$ 5_{\rm F}\times 45_{\rm P}$ &  2.618033988750  &  3.000000000000  &  3.246979607399  &  3.327447532663  \\
$ 5_{\rm F}\times 50_{\rm P}$ &  2.618033988750  &  3   &  3.246979603508  &   \\
\hline
$ 6_{\rm F}\times 6_{\rm P}$ &  2.618033988750  &  3.001033705947  &  3.125892136302  &   \\
$ 6_{\rm F}\times 12_{\rm P}$ &  2.618033988750  &  3.000000000036  &  3.246874398154  &   \\
$ 6_{\rm F}\times 18_{\rm P}$ &  2.618033988750  &  3.000000000000  &  3.246979478144  &   \\
$ 6_{\rm F}\times 24_{\rm P}$ &  2.618033988750  &  3.000000000000  &  3.246979603417  &   \\
$ 6_{\rm F}\times 30_{\rm P}$ &  2.618033988750  &  3.000000000000  &  3.246979603717  &   \\
$ 6_{\rm F}\times 36_{\rm P}$ &  2.618033988750  &  3.000000000000  &  3.246979603717  &   \\
$ 6_{\rm F}\times 42_{\rm P}$ &  2.618033988750  &  3.000000000000  &  3.246979603717  &   \\
$ 6_{\rm F}\times 48_{\rm P}$ &  2.618033988750  &  3.000000000000  &  3.246979603717  &   \\
$ 6_{\rm F}\times 54_{\rm P}$ &  2.618033988750  &  3.000000000000  &  3.246979603717  &   \\
$ 6_{\rm F}\times 60_{\rm P}$ &  2.618033988750  &  3.000000000000  &  3.246979603717  &   \\
\hline
$ 7_{\rm F}\times 7_{\rm P}$ &  2.618033988750  &   3  & 3.247001348628  &  3.404690481534  \\
\hline
\hline
 Beraha &2.618033988750  & 3   & 3.246979603717  & 3.414213562373  \\
\hline
\end{tabular}
\caption{\label{zeros_tri_R}
  Real zeros of the chromatic polynomial $P_{m\times n}(q)$ of a 
  triangular-lattice
  strip of width $m$ and lengths $n= k\times m$ ($k=1,\ldots,10$) with 
  cyclic boundary conditions. A blank means that the zero in question 
  is absent. The first three real zeros $q=0,1,2$ are exact on all lattices.
  ``Beraha'' indicates the Beraha numbers $B_5=(3+\sqrt{5})/2$,
  $B_6=3$, $B_7$, and $B_8$. 
}
\end{table}

\clearpage
%
%
\begin{table}
\centering
\begin{tabular}{|r||c|c|c|c|c|l||c|}
\cline{2-8}
\multicolumn{1}{c||}{\mbox{}}&
\multicolumn{6}{|c||}{Eigenvalue-Crossing Curves $\mathcal{B}$} &
\multicolumn{1}{|c|}{Isolated Points}\\
\hline\hline
Lattice     & \# C & \# E & \# T & \# D & \# ER &\multicolumn{1}{|c||}{$q_0$}
            & \# RI \\
\hline\hline
$2_\text{R}$& 1 &  0 &\kk 0 &  1 &  2 & 3            &  2 \\
$3_\text{R}$& 1 &  0 &\kk 8 &  1 &  6 & 3            &  2 \\
$4_\text{R}$&    $1^\dagger$
            &    $0^\dagger$
            &\kk $8^\dagger$
            &    $0^\dagger$
            &    $5^\dagger$
            &     3.2281262261 &  2 \\
$5_\text{R}$&    $1^\dagger$
            &    $0^\dagger$
            &   $16^\dagger$
            &    $0^\dagger$
            &    $9^\dagger$
            &     3.3324469422 &  3 \\
$6_\text{R}$&    $1^\dagger$
            &    $2^\dagger$ 
            &   $10^\dagger$
            &    $0^\dagger$
            &    $5^\dagger$
            &     3.4062849655 &  3 \\
$7_\text{F}$&    $1^\dagger$
            &    $2^\dagger$
            &    $12^\dagger$ 
            &    $0^\dagger$
            &    $6^\dagger$
            &     3.4682618071 &  3 \\
$8_\text{F}$&   &    &      &    &    & 3.5134782609 &  3 \\
\hline\hline
\end{tabular}

\caption{\label{table_summary_tri}
   Summary of qualitative results for the triangular-lattice 
   eigenvalue-crossing curves $\mathcal{B}$
   and for the isolated limiting points of zeros.
   For each cyclic triangular-lattice strip considered in this paper,
   we give the number of connected components of $\mathcal{B}$ (\# C),  
   the number of endpoints (\# E),
   the number of T points (\# T),
   the number of double points (\# D),
   and the number of enclosed regions (\# ER).
   We also give the value $q_0$ which is the largest real value where 
   $\mathcal{B}$ intersects the real axis, and the   
   number of real isolated limiting points of zeros (\# RI). 
   The symbol $^\dagger$ indicates uncertain results.   
}
\end{table}

%
%
\begin{table}
\centering
\begin{tabular}{|c|lc|cr|}
\hline\hline
$q$   & Lattice    & $L$  & $\ell$ & \multicolumn{1}{c|}{$f$} \\
\hline
$B_5$ &  Square    & 6    &  0     & $0.28823784$ \\  
      &            & 7    &  0     & $0.28514284$ \\
      &            & 8    &  0     & $0.28291082$ \\
\hline
$B_5$ &  Triangular& 2    &  1     &$-0.24060591$ \\ 
      &            & 3    &  0     &$-0.35114882$ \\  
      &            & 4    &  1     &$-0.46727989$ \\
      &            & 5    &  0     &$-0.48121183$ \\
      &            & 6    &  1     &$-0.48248246$ \\
      &            & 7    &  0     &$-0.52290292$ \\
      &            & 8    &  1     &$-0.53899001$ \\
\hline
$B_7$ &  Triangular& 5    &  0     & $0.11980838$ \\ 
      &            & 6    &  0     & $0.11458681$ \\  
      &            & 7    &  0     & $0.11124861$ \\
      &            & 8    &  0     & $0.10886700$ \\
\hline
\end{tabular}
\caption{\label{table_F_qBn}
  Real free energy for $q=B_n$ with integer $n$. For several values
  of $q=B_n$ we list the free energy associated to the most relevant
  eigenvalue of the transfer matrix that gives a non-zero contribution to 
  the partition function. For each value of $q$, the type and width $L$ of 
  the lattice strip, we list the sector $\ell$ the eigenvalue belongs to
  and the free energy $f$. 
}
\end{table}

\clearpage
%
%
\begin{table}
\centering
\begin{tabular}{|r|r|r|r|r|r|}
\hline\hline
 $m$ & SqRing  
     & SqRingT 
     & \# blocks
     & $\max \dim$ 
     & SqRing$'$ \\ 
\hline\hline
  2  &     4 &     4 &   3 &    2 &   4 \\
  3  &    20 &    15 &   4 &    7 &  10 \\
  4  &    94 &    60 &   5 &   25 &  26 \\
  5  &   614 &   342 &   6 &  127 &  70 \\
\hline
\end{tabular}
\caption{\label{table_dim_T_cyclic_blocks_TB}
  Block structure of the square-lattice transfer matrix acting on the 
  top and bottom rows $\widehat{\mathsf{T}}$ \protect\reff{def_Tbis}. 
  For each strip width $m$, we quote the dimension of the standard 
  connectivity-state basis $\{\bm{w}\}$ SqRing and the dimension of the 
  $t$-reflection-invariant basis SqRingT. 
  We also give the number of blocks (\# blocks) the transfer matrix 
  $\widehat{\mathsf{T}}$ can be decomposed, 
  the dimension of the largest block ($\max\dim$),
  and the number of distinct eigenvalues we find SqRing$'$. 
}
\end{table}

\clearpage
%
%
%
%
\begin{figure}
   \setlength{\unitlength}{0.5pt}
   \centering
   \begin{picture}(300,300)
     \Thicklines
     \drawline(0,50) (300,50)
     \drawline(0,150)(300,150)
     \drawline(0,250)(300,250)

     \drawline(0,50)  (0,250)
     \drawline(100,50)(100,250)
     \drawline(200,50)(200,250)
     \drawline(300,50)(300,250)

     \put(0,50){\circle*{15}}
     \put(0,150){\circle*{15}}
     \put(0,250){\circle*{15}}
     \put(100,50){\circle*{15}}
     \put(100,150){\circle*{15}}
     \put(100,250){\circle*{15}}
     \put(200,50){\circle*{15}}
     \put(200,150){\circle*{15}}
     \put(200,250){\circle*{15}}
     \put(300,50){\circle*{15}}
     \put(300,150){\circle*{15}}
     \put(300,250){\circle*{15}}

     \put( -8, 10){\large \bf 1'}
     \put( 92, 10){\large \bf 2'}
     \put(192, 10){\large \bf 3'}
     \put(292, 10){\large \bf 4'}

     \put( -8,270){\large \bf 1}
     \put( 92,270){\large \bf 2}
     \put(192,270){\large \bf 3}
     \put(292,270){\large \bf 4}
   \end{picture}

   \caption{\label{figure_sq_cyclic_bc}
     Square lattice with cyclic boundary conditions of size $4\times 2$. 
     This lattice is obtained
     from a square lattice with free boundary conditions of size $4\times 3$
     by identifying the top and bottom rows 
     $i \leftrightarrow i'$ ($i=1,\ldots,4=m$). 
   }
\end{figure}
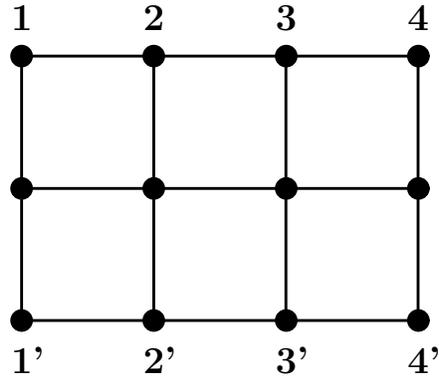

%
%

\clearpage
%
%

\clearpage
%
%

%
%

\clearpage
%
%

\clearpage
%
%

\clearpage
%
%

\clearpage
%
%

\clearpage
%
%

\clearpage
%
%

\clearpage
%
%

\clearpage
%
%
\end{document}